\documentstyle[12pt,psfig]{article}
\begin{document}

\pagenumbering{roman}

\pagestyle{empty}

\hfill PUPT-1740

\hfill hep-th/9711036

\vspace{0.4in}

\begin{center}

{\large\bf Sheaves on Toric Varieties }

{\large\bf for Physics }

\vspace{0.4in}

Allen Knutson \\
Mathematics Department \\
Brandeis University \\
Waltham, MA  02254 \\
{\tt allenk@alumni.caltech.edu} \\

$\:$  \\
$\:$  \\

Eric Sharpe \\
Physics Department \\
Princeton University \\
Princeton, NJ  08544 \\
{\tt ersharpe@puhep1.princeton.edu}

\vspace{0.2in}

\end{center}

In this paper we give an inherently toric description of 
a special class of sheaves (known as equivariant sheaves) 
over toric varieties, due in part to A. A. Klyachko. 
We apply this technology to heterotic compactifications, in 
particular to the (0,2) models of Distler, Kachru, 
and also discuss how knowledge of equivariant sheaves
can be used to reconstruct information about an entire
moduli space of sheaves.  Many results relevant to heterotic
compactifications previously known only to mathematicians
are collected here -- for example,
results concerning whether the restriction of a stable sheaf
to a Calabi-Yau hypersurface remains stable are stated.
We also describe 
substructure in the
K\"{a}hler cone, in which moduli spaces of sheaves are
independent of K\"{a}hler class only within any one subcone.
We study F theory compactifications in light of this fact,
and also discuss how it can be seen in the context of 
equivariant sheaves on toric varieties.    
Finally
we briefly speculate on the application of these 
results to (0,2) mirror symmetry.

\begin{flushleft}
November 1997  
\end{flushleft}

\newpage

\pagestyle{plain}

\tableofcontents

\newpage

\pagenumbering{arabic}

\section{Introduction}

Historically one of the biggest challenges facing anyone wishing to
study heterotic compactifications has been the lack of a good
description of vector bundles and, more generally, torsion-free
sheaves.  

This problem has been recently addressed for the special
case of vector bundles over elliptic varieties in 
\cite{wmf,wmf2,wmf3,bbundle,donagi,luest,me1,pauldave}.
However, their methods can neither be applied to more general
varieties nor to study more general sheaves.

Most Calabi-Yaus studied to date have been complete
intersections in toric varieties\footnote{For introductions to toric
varieties see for example \cite{fulton,oda,kempf,russ,cox}.}.  
A description of sheaves
relevant to all such Calabi-Yaus would be of tremendous
importance.

In \cite{old(02),(02)} attempts were made to study 
sheaves on such Calabi-Yaus by describing the sheaves with
short exact sequences and monads.  These methods are somewhat
clumsy and have not been thoroughly developed in the literature.
For example, to date these methods have not given a global
description of moduli spaces of sheaves, one of several
prerequisites needed to gain a solid understanding of heterotic
compactifications.

In this paper we present a radically different method of studying
sheaves over Calabi-Yaus realized as complete intersections
in toric varieties.  In particular, we present an inherently toric
method to describe certain sheaves on an ambient toric variety,
due largely to Alexander A. Klyachko.  Sheaves over a
complete intersection Calabi-Yau can be obtained by restricting
sheaves on the ambient space to the complete intersection.  
This does not allow
us to realize all possible sheaves over a Calabi-Yau,
but it does still give a tremendous amount of insight into moduli spaces.
In particular, it is closely analogous to the strategy of
studying the K\"{a}hler cone of a complete intersection Calabi-Yau
via the K\"{a}hler cone of the ambient toric variety, used to great
effect in the mirror symmetry literature.

Not any sheaf over a Calabi-Yau can be used
in a compactification -- one restriction is that the sheaf 
must be stable.  We are able to determine explicitly whether
a bundle on the ambient toric variety is stable,
but unfortunately the restriction of a stable sheaf to a hypersurface
is not, in general,
stable\footnote{with respect to restriction of the K\"{a}hler form}.
However, in special cases 
the restriction of a stable sheaf to a hypersurface
is stable; sufficient conditions will be discussed in 
section~\ref{stab}. 

The condition that a sheaf be stable implicitly depends upon the
choice of K\"{a}hler form.  It sometimes happens that a sheaf 
is stable with
respect to some, but not all, of the possible K\"ahler forms in the
K\"ahler cone.  In the more extreme versions of this phenomenon
the K\"ahler cone is subdivided into chambers, each associated
to distinct moduli spaces of sheaves.  This phenomenon will be
discussed in section~\ref{kcsubstruc}.

Although we have a lot to say about classical results, we largely
ignore quantum effects.
For example, we do not speak to worldsheet instanton corrections.
These are a very important aspect of heterotic compactifications,
but we do not at present have sufficient technology to deeply
understand their effects.
Perhaps the only exception to this are some speculations 
on the application of the results in this paper to (0,2)
mirror symmetry, in section~\ref{(02)mirrapp}. 
Our philosophy in writing this paper was to first try to understand
classical behavior, and leave quantum corrections to future work.

In this paper a great deal of mathematics is used which is 
unfamiliar to most physicists, even those who often think about
compactifications.  To help alleviate the resulting culture shock,
we have tried to loosely order this paper so that topics requiring
less mathematical sophistication appear near the beginning, and those
requiring more appear later.  
Throughout the paper, we have made an effort to explain nearly all of
the mathematics used, even to the extent of including lengthy elaborations
(where useful) of details that some would consider basic.
Extremely technical details, not elaborated upon in detail, are
usually banished to footnotes.
We have also provided several
appendices on the basics of relevant mathematical topics,
which ideally should help the reader understand not only this
paper but also parts of the mathematics literature cited.

In section~\ref{revhet} we begin with a short review of
constraints on consistent heterotic compactifications.
In section~\ref{genrmk} we make general remarks on sheaves
on toric varieties, which should help put the rest of the paper
in context.  In section~\ref{equivsheaves} we discuss
equivariant sheaves on toric varieties.  We discuss not only
how to construct equivariant bundles (of any structure group) 
and more general sheaves,
but also the calculation of invariants such as Chern classes, 
sheaf cohomology groups, and global Ext groups.  
We also discuss equivariant sheaves
on singular varieties.
In section~\ref{equivmodspace} we
explicitly construct moduli spaces of equivariant sheaves, and
discuss the underlying theory in some detail.  In 
section~\ref{kcsubstruc} we discuss substructure in the K\"ahler
cone, both in generality and in the special cases of elliptic
K3's with section (relevant to F theory) and to equivariant
sheaves on toric surfaces.   In section~\ref{genmodspace}
we discuss more general moduli spaces, of not-necessarily-equivariant
sheaves.  In particular, we discuss how knowledge of equivariant
sheaves can be used to reconstruct information about the rest
of the moduli space.  We also explicitly construct complete
moduli spaces of bundles on ${\bf P}^{2}$ (one of the few
cases that can be understood explicitly), and as an aside 
derive the ADHM construction.  Up to this point we have only discussed
sheaves on ambient toric varieties; in section~\ref{stab}
we discuss conditions under which the restriction of a stable
sheaf to a hypersurface is stable.  In section~\ref{(02)app}
we apply the technology above to the (0,2) models of Distler and
Kachru.  In section~\ref{(02)mirrapp} we apply the technology
above to make nontrivial statements about (0,2) mirror symmetry.
Finally, we conclude in section~\ref{concl} with speculations.
Four appendices review basic facts about
algebraic groups, sheaf theory, GIT quotients, and filtrations
commonly used in sheaf theory.

Much of this paper is devoted to reviewing highly relevant results known
to mathematicians but not physicists.  For example, an inherently
toric description of equivariant sheaves was originally worked out 
by A. A. Klyachko \cite{kl1,kl2,kl3,kl4}, but 
in order to make his results useful for physicists, we have had
to significantly extend them.  Since his work is also completely
unknown to physicists (despite its great relevance), we have taken
this opportunity to review it, and in places correct it.
Similarly, results on stability of the restriction of a stable
sheaf to a hypersurface exist in the mathematics literature
but are unknown to the physicists -- so we have presented known
results here.
In fact, for convenience we have collected here many results
highly relevant to heterotic compactifications that were previously
known only to mathematicians.

We should also mention a few sections can be read independently
of the rest of the paper.  Section~\ref{stab}, on stability
of restrictions of stable sheaves to hypersurfaces, can be
read apart from the rest.  The first two-thirds of section~\ref{kcsubstruc},
on K\"ahler cone substructure sensed by heterotic theories,
can be read independent of the rest of the paper.

A word of caution is in order concerning conventions used in this
paper.  We will assume the structure groups of all bundles are
reductive algebraic groups -- complexifications of compact Lie groups.
In particular, instead of working with $U(n)$, we shall
work with $GL(n, {\bf C})$, its complexification; similarly, instead of 
$SU(n)$, 
we work with $SL(n, {\bf C})$.

\section{A rapid review of heterotic compactifications} \label{revhet}

For a consistent perturbative compactification of either the
$E_{8}\times E_{8}$ or $Spin(32)/{\bf Z}_{2}$ heterotic string, in
addition to
specifying a Calabi-Yau $Z$ one must also specify 
a set of 
holomorphic vector bundles (or, more generally,
sheaves)
$V_{i}$.
These vector bundles must obey two constraints.  For $GL(n,{\bf C})$
bundles
one constraint\footnote{For example, for compactifications to four
dimensions, N=1 supersymmetry, on a Calabi-Yau $X$ 
one gets a D term in the low-energy
effective action proportional to $\langle X | \omega^{2} \cup c_{1}(V)
\rangle$.} can be written as
\begin{equation} \label{duy}
\omega^{n-1} \cup c_{1}( V_{i} ) \: = \: 0
\end{equation}
where $n$ is the complex dimension of the Calabi-Yau, and $\omega$ is
the K\"{a}hler form.  This constraint has a somewhat subtle implication.
In general, for any holomorphic bundle ${\cal E}$, if there exists
a Hermitian connection associated to ${\cal E}$ such that,
in every coordinate chart, the curvature $F$ satisfies
$F \wedge \omega^{n-1} = c I$, where $I$ is the identity
matrix and $c \in {\bf R}$ is a fixed chart-independent constant, 
then ${\cal E}$ is either properly Mumford-Takemoto stable 
\cite{okonek} or Mumford-Takemoto semistable and split
\cite{kobayashi,donald,uhlenyau}.
Thus, the constraint in equation~(\ref{duy}) implies that 
(but is not equivalent to the statement)
${\cal E}$
is either stable, or semistable and split.  In fact,
we can slightly simplify this statement.  Properly semistable sheaves
are grouped\footnote{More precisely, points on a moduli space
of sheaves that are properly semistable do not necessarily
correspond to unique semistable sheaves, but rather to $S$-equivalence
classes of properly semistable sheaves.  Points that are stable
do correspond to unique stable sheaves -- $S$-equivalence classes
are a phenomenon arising only for properly semistable objects.}
in $S$-equivalence classes, and each $S$-equivalence class contains
a unique split representative \cite[p. 23]{huybrechtslehn}.
Thus, constraint~(\ref{duy}) implies that ${\cal E}$ is 
Mumford-Takemoto semistable.

The other constraint is an anomaly-cancellation condition
which, 
if a single $GL(n,{\bf C})$ bundle $V_{i}$ is embedded in each $E_{8}$, 
is often\footnote{
On rare occasion it is possible for a perturbative compactification
to evade this condition.  See for example \cite{meqik}.} 
written as
\begin{displaymath}
\sum_{i} \, \left( c_{2}(V_{i}) \, - \, \frac{1}{2} c_{1}(V_{i})^{2} \,
\right) \: = \: c_{2}(TZ)
\end{displaymath} 

It was noted \cite{dmw} 
that the anomaly-cancellation conditions can be
modified slightly by the presence of five-branes in the heterotic
compactification.  Let $[W]$ denote the cohomology class of the
five-branes, then the second constraint above is modified to
\begin{displaymath}
\sum_{i} \, \left( c_{2}(V_{i}) \, - \, \frac{1}{2} c_{1}(V_{i})^{2} \,
\right) \: + \: [W] \: = \: c_{2}(TZ)
\end{displaymath}

Although the conditions above are necessary for a consistent heterotic
compactification, they are not sufficient -- quantum effects must
also be taken into account.  For example, it was once believed
that  
generic heterotic compactifications were destabilized by worldsheet instantons
\cite{xen}.  For the (0,2) models of \cite{(02)},
it has been shown that this is not the case \cite{eva(02)},
and in fact it has become fashionable to ignore this difficulty.
In addition, even for the (0,2) models of \cite{(02)}, there is
a more subtle and poorly-understood anomaly \cite{distanom} which
afflicts many potential heterotic compactifications.
In this paper we will speak to neither potential problem;
our philosophy is to first understand purely classical
behavior, and only then attempt to grasp quantum corrections.

Historically heterotic compactifications used only bundles, not
more general sheaves.  However, in \cite{dgm} it was shown that it
was possible to have consistent perturbative heterotic compactifications
involving sheaves (specifically, torsion-free sheaves) which are not bundles.
There is, however, a caveat.  On a smooth variety, all torsion-free sheaves
look like bundles up to codimension at least two, 
where the description as a bundle
breaks down.  Because of these bad points, it sometimes happens that the
conformal field theory breaks down -- in such cases the metric
degenerates and describes an infinite tube, sometimes loosely associated
with five-branes \cite{wit95,chs}.  In order to determine whether
a particular torsion-free sheaf suffers from this difficulty,
one writes down a linear sigma model describing the sheaf
and studies its properties as in \cite{dgm}.  Unfortunately at present
there is no invariant method to determine whether a given torsion-free
sheaf describes a singular conformal field theory -- one must study
an associated linear sigma model.  Most of the sheaves we shall
study in this paper are not obviously associated with linear sigma models,
so we have no way to determine whether a given (non-locally-free) sheaf
is associated with a singular conformal field theory.
We feel that the advantages of our new
approach to thinking about heterotic compactifications outweigh
such difficulties.

\section{General remarks on bundles on toric varieties}  \label{genrmk}

In this section we will make some general observations on
moduli spaces of bundles (of fixed Chern classes) on a toric variety $X$.  
(In fact, our
remarks will also hold for moduli spaces of reflexive,
and torsion-free, sheaves.)

First, we should mention a few basic facts about toric varieties
that will be used throughout this paper.  A toric variety
is a compactification\footnote{Not all compactifications of algebraic
tori are toric varieties -- toric varieties have additional nice
properties -- but the distinctions will not be relevant for
our discussion.} of some ``algebraic torus" 
$( {\bf C}^{\times})^n$, where ${\bf C}^{\times} = {\bf C} - \{ 0 \}$.  
For example, all projective spaces are toric varieties:
\begin{eqnarray*}
{\bf P}^{1} & = & {\bf C}^{\times} \cup \{ 0 \} \cup \{ \infty \} \\
{\bf P}^{2} & = & ({\bf C}^{\times})^{2} \cup  \{ x = 0 \}
\cup \{ y = 0 \} \cup \{ z = 0 \} 
\end{eqnarray*}
where $x$, $y$, and $z$ are homogeneous coordinates defining ${\bf P}^{2}$,
and so forth.

The codimension one subvarieties added to the algebraic torus to 
compactify it are known as the ``toric divisors."  For example,
the toric divisors of ${\bf P}^{1}$ are $\{ 0 \}$  and $\{ \infty \}$.
The toric divisors of ${\bf P}^{2}$ are $\{ x = 0 \}$,
$\{  y = 0 \}$, and $\{ z = 0 \}$.

Note that the algebraic torus $( {\bf C}^{\times} )^n = T$ 
underlying any toric variety $X$
has a natural action on $X$.  In the case of ${\bf P}^1$, this action
amounts to rotations about an axis and dilations.

A moduli space of bundles (or sheaves) ${\cal M}$ also has
a natural action of the algebraic torus $T$ defining the toric
variety $X$:  if ${\cal E}$ is some sheaf and $t: X \rightarrow X$
the action of an element $t \in T$, then we take ${\cal E} \rightarrow
t^{*} {\cal E}$.  In general,  ${\cal E} \not\cong t^{*} {\cal E}$.
The fixed-point locus of the $T$ action consists of sheaves ${\cal E}$ such
that $t^{*} {\cal E} \simeq {\cal E}$ for all 
elements\footnote{Some
toric varieties have automorphism groups larger than
the algebraic torus \cite{oda} which might, in principle, give
additional information.  We will not pursue this possibility
here.} 
$t \in T$.  For example, all line bundles on a smooth toric variety
have this property.  Such sheaves are known as equivariant\footnote{
In fact, we are being slightly sloppy.  In the mathematics
literature, an ``equivariant" sheaf would not only have the property
that it is mapped into itself by all algebraic torus actions, but in
addition would come with a fixed choice of ``equivariant structure" 
(a precise choice of sheaf involution).  Sheaves that are fixed
under all algebraic torus actions but do not have a fixed equivariant
structure should be called ``equivariantizable."  For reasons of
readability, we shall maliciously fail to distinguish ``equivariant"
from ``equivariantizable."  For more information on the relationship
between ``equivariant" and ``equivariantizable," 
see section~\ref{genmodspace}.} sheaves, or sometimes homogeneous
sheaves.  Such sheaves have an inherently toric description
\cite{kl1,kl2,kl3,kl4,kaneyama1,kaneyama2}, which we shall review
and extend in the first part of this paper.

This inherently toric description
of equivariant vector bundles and equivariant reflexive sheaves 
associates a filtration\footnote{
A filtration of a vector space $V$ is a sequence of nested subspaces
\begin{displaymath}
V \: = \: V_0 \: \supseteq \: V_1 \: \supseteq \: V_2 \: \supseteq \: \cdots 
\end{displaymath}
with a strictly increasing integer associated to each subspace.
In the filtration description of equivariant vector bundles,
the vector space we filter is precisely the fiber of the vector
bundle.  More general equivariant torsion-free sheaves are trivial
vector bundles over
the open $T$-orbit; the vector space we filter is that of the trivial
vector bundle.  This filtration description is valid for both bundles
and reflexive sheaves on both smooth and singular toric varieties,
but to aid the reader will will begin our discussion by restricting
to the special case of bundles on smooth toric varieties.}
of a vector space to each toric divisor.  Intuitively, equivariant sheaves
on an algebraic torus $( {\bf C}^{\times} )^n$ are trivial, so all
information is contained in their behavior near the toric divisors.
Not all choices of filtrations define a bundle -- to get a  bundle,
rather than merely a reflexive sheaf, there is a compatibility 
condition
that must be satisfied.  There is also a description of more general
equivariant torsion-free sheaves, 
but this description is more cumbersome and so is presented later.

In order to use the description of equivariant sheaves outlined
above, one must fix a precise action of the algebraic torus
$T$ on the equivariant sheaf -- it is not enough to know the fact
that the algebraic torus maps the sheaf into itself, we must
also specify precisely how the algebraic torus acts.  
Put another way, for any element $t$ of
the algebraic
torus $T$, there exists an action of $t$ on an equivariant sheaf 
${\cal E}$ which
makes the
following diagram commute:
\begin{displaymath}
\begin{array}{ccccc}
\, & {\cal E} & \: \stackrel{t}{\longrightarrow} \: & {\cal E} & \, \\
\pi & \downarrow & \: & \downarrow & \pi \\
\, & X & \: \stackrel{t}{\longrightarrow} \: & X & \,
\end{array}
\end{displaymath}
where $\pi: {\cal E} \rightarrow X$ is the projection.
This choice
of algebraic torus action, known as the choice of ``equivariant
structure,"
is not unique, and the filtration description outlined above
depends upon the precise choice, but for all that the choice 
is actually quite harmless -- it adds no continuous moduli, and
is under good control.

The ``inherently toric'' description of sheaves outlined
above only applies to equivariant sheaves, so what can we
say about the rest of a moduli space of sheaves, given knowledge
of only the equivariant ones?   In principle, quite a lot.
For any space with a torus action,
given information about only the fixed points of a torus action,
and the torus action on the normal bundle to the fixed points,
it is possible to determine a great deal of information about
the original space.  
In particular, even if we only know about equivariant sheaves
we can still determine a great deal of information about a moduli
space of not-necessarily-equivariant sheaves.
This perspective will be reviewed in
greater depth in section~\ref{genmodspace}.

Once we have discussed sheaves on ambient toric varieties in
great detail, we turn to sheaves on Calabi-Yau hypersurfaces
beginning in section~\ref{stab}. 

In most of the rest of this paper, we shall assume the
reader is well acquainted with toric varieties and their associated
machinery.  For introductions to toric varieties see
for example \cite{fulton,oda,kempf,russ,cox}.

\section{Equivariant sheaves}  \label{equivsheaves}

In this section we will review results on equivariant sheaves -- 
the sheaves located at the fixed points of the algebraic torus
action on the moduli space.

In section~\ref{evb} we begin by reviewing an inherently toric description of
equivariant vector bundles, due originally to Alexander A. Klyachko
\cite{kl1,kl2,kl3,kl4}.  In section~\ref{kaneyama} we compare
Klyachko's description to another, somewhat less useful, description due to
Kaneyama \cite{kaneyama1,kaneyama2}.  In section~\ref{appl} we
describe how to compute Chern classes and sheaf cohomology groups
using Klyachko's description.  In section~\ref{agg} we describe how
to modify Klyachko's description to describe bundles
with arbitrary gauge group.  Finally in section~\ref{etfs} we
describe how to generalize Klyachko's description to give
arbitrary torsion-free sheaves on arbitrary toric varieties.
We derive Klyachko's description from first principles, discuss
equivariant sheaves on singular varieties, give an efficient
description of reflexive sheaves on arbitrary varieties,
and on smooth toric varieties
we describe global Ext
calculations.

Many of the results in this section are due originally to
A. A. Klyachko \cite{kl1,kl2,kl3,kl4}, and are reviewed 
(and occasionally corrected)
for the reader's convenience.  In particular, much of sections~\ref{evb},
\ref{appl}, \ref{agg}, and \ref{tfsmooth} is due
to A. A. Klyachko.  In particular, A. A. Klyachko was concerned
with smooth toric varieties, but in order to make these methods
useful for physical applications we extended his work to singular
varieties, to global Ext calculations, and (in the next section)
to moduli space problems.

In sections~\ref{evb} through \ref{agg} we specialize to smooth
toric varieties and bundles on these varieties.  
More general toric varieties and more general sheaves are considered
in section~\ref{etfs}.

\subsection{Equivariant vector bundles} \label{evb}

In this section we will review work of A. A. Klyachko
\cite{kl1,kl2,kl3,kl4} describing equivariant bundles on smooth toric varieties
in an inherently toric fashion.  The basic idea is that an equivariant
bundle is completely determined by its behavior near the toric divisors.
Thus, Klyachko specifies bundles in terms of a family of 
filtrations of a vector
space, one filtration for each toric divisor.  We will first describe
the relevant technology, then afterwards try to give some intuitive
understanding.  Readers well-versed in algebraic geometry may
find the discussion of equivariant torsion-free sheaves in 
section~\ref{basicprin} somewhat more enlightening than the discussion 
in this section.  In section~\ref{etfs} we speak about more general sheaves
on toric varieties that are not necessarily smooth.

Klyachko describes equivariant vector bundles 
by associating to each toric divisor a filtration of the generic fiber.
In
order for these filtrations to yield a well-defined bundle, they must
satisfy a certain compatibility condition, to be defined shortly.
A set of (compatible) filtrations is sufficient to uniquely identify the
equivariant bundle.

In order to use this filtration prescription one must make a specific
choice of action of the algebraic torus on the bundle.
Since the bundle is equivariant, the algebraic torus action
maps the bundle into itself -- we need to be specific about the choice
of involution.
More precisely, for any element $t$ of
the algebraic
torus $T$, there exists an action of $t$ on an equivariant bundle 
${\cal E}$ which
makes the
following diagram commute:
\begin{displaymath}
\begin{array}{ccccc}
\, & {\cal E} & \: \stackrel{t}{\longrightarrow} \: & {\cal E} & \, \\
\pi & \downarrow & \: & \downarrow & \pi \\
\, & X & \: \stackrel{t}{\longrightarrow} \: & X & \,
\end{array}
\end{displaymath}
where $\pi: {\cal E} \rightarrow X$ is the projection.
This choice
of algebraic torus action, known as the choice of ``equivariant
structure,"
is not unique, and the filtration description outlined above
depends upon the precise choice, but for all that the choice
is actually quite harmless -- it adds no continuous moduli, and
is under good control.
In the case of line bundles, the choice of equivariant
structure is equivalent to a precise choice of $T$-invariant divisor.
For example, on ${\bf P}^{2}$ the line bundles
${\cal O}(D_{x})$, ${\cal O}(D_{y})$, and ${\cal O}(2D_{z}  - D_{y})$,
where $D_{x} = \{ x = 0 \}$ and $x$, $y$, $z$ are homogeneous
coordinates, are all equivalent as line bundles to ${\cal O}(1)$,
but have distinct equivariant structures.

Now we shall describe how to obtain a set of filtrations given some
bundle ${\cal E}$ over a toric variety $X$, following \cite{kl1,kl2,kl3,kl4}.  
Fix a point $p_{0}$ in the open torus orbit, and
define $E \, = \, {\cal E}(p_{0})$.  For each toric divisor $\alpha$,
let $p_{\alpha}$ be a generic point in the toric divisor.  We will
obtain a filtration for the toric divisor $\alpha$ by observing 
how the fiber $E$ changes
as we drag $p_{0} \rightarrow p_{\alpha}$.
In particular, let $f(p)$ be any rational function on $X$ with a pole of
order $i$ on the toric divisor $\alpha$, and $t$ a one-parameter
algebraic torus action dragging $p_0 \rightarrow p_{\alpha}$, then define
\begin{equation}
E^{\alpha}(i) \: = \: \left\{ \, e \in E \, | \, \lim_{t p_{0}
\rightarrow p_{\alpha}} \, f(t p_{0}) \cdot (t e) \mbox{ exists} \, \right\}
\end{equation}
These subspaces form a nonincreasing filtration:
\begin{displaymath}
\cdots \, \supseteq \, E^{\alpha}(i) \, \supseteq \, 
E^{\alpha}(i+1) \, \supseteq \, \cdots
\end{displaymath}
with limits 
\begin{displaymath}
\begin{array}{ll}
E^{\alpha}(i) \: = \: 0 & i \gg 0 \\
E^{\alpha}(i) \: = \: E & i \ll 0
\end{array}
\end{displaymath}

A random set of filtrations does not necessarily describe a vector
bundle; these filtrations are required to satisfy a compatibility
condition.  This condition is simply that, for any cone $\sigma \in
\Sigma$, the fan of $X$, the filtrations $E^{\alpha}(i)$, $\alpha \in
| \sigma |$, consist of coordinate subspaces of some basis of the space
$E$.   Put another way, on any open set corresponding to a maximal
cone, the vector bundle should split into a sum of line bundles.
Put yet another way, vector bundles on affine space are trivial.

This compatibility condition can be phrased more abstractly as follows.
For any toric divisor $\alpha$, define the parabolic subgroup $P^{\alpha}
\subseteq GL(n,{\bf C})$
by
\begin{equation} \label{paradefine}
P^{\alpha} \: = \: \left\{ \, g \in GL(n,{\bf C}) \, | \, g E^{\alpha}(i) \, = \,
E^{\alpha}(i) \, \forall i \, \right\}
\end{equation}
Then the compatibility condition says precisely that for all cones
$\sigma \in \Sigma$,
\begin{equation} \label{compatpara}
\bigcap_{\alpha \in | \sigma |} P^{\alpha} \mbox{ contains a maximal torus
of } GL(n,{\bf C})
\end{equation} 

Note that the bundle compatibility conditions are triangulation-dependent;
they are sensitive to more than just the edges of the fan.
In particular, a set of filtrations that defines a bundle in
one triangulation may fail to define a bundle after a flop
(or other birational transformation on the toric variety preserving
the edges of the fan).  In such a case, instead of defining a bundle
in the flopped triangulation, we would only get a reflexive sheaf.

Specifying a parabolic subgroup, as in equation~(\ref{paradefine}),
does not uniquely specify the filtration -- the indices at which
the filtration changes dimension also are meaningful.  We will see
later (section~\ref{agg}) that this information corresponds 
to a choice of ample
line bundle on the (partial) flag manifold $G/P^{\alpha}$, 
in addition to the choice of 
parabolic subgroup $P^{\alpha}$.

An equivariant vector bundle is defined uniquely by a set of filtrations
(one for each toric divisor) satisfying the compatibility condition
above,
up to a simultaneous rotation of all filtrations by an element of
$GL(n,{\bf C})$.

For example, let's discuss how line bundles are described in this
language.  Let $D \, = \, \sum a_{\alpha} D_{\alpha}$ be a Cartier
divisor on the toric variety, 
then the filtration on divisor $D_{\alpha}$ is given by
\begin{displaymath}
E^{\alpha}(i) \: = \: \left\{ \begin{array}{ll}
                              {\bf C} & i \leq a_{\alpha} \\
                              0 & i > a_{\alpha}
                              \end{array} 
                              \right. 
\end{displaymath}
(Recall that for line bundles the choice of equivariant structure
amounts to simply being specific about the choice of divisor.)
At this point we can see the necessity of the smoothness condition.
On a singular toric variety, not all divisors are Cartier, i.e., not
all divisors $D \, = \, \sum a_{\alpha} D_{\alpha}$ define line
bundles.  Equivariant sheaves on singular toric varieties will
be discussed in section~\ref{basicprin}.

For another example, consider the direct sum of two line bundles
${\cal O}(D_{1}) \oplus {\cal O}(D_{2})$.  Write $D_{1} \, = \, \sum
a_{1 \alpha} D_{\alpha}$, $D_{2} \, = \, \sum a_{2 \alpha} D_{\alpha}$
(where the $D_{\alpha}$ are the toric divisors),
then
\begin{displaymath}
E^{\alpha}(i) \: = \: \left\{ \begin{array}{ll}
                             {\bf C}^{2} & i \leq \mbox{min}(a_{1
\alpha}, a_{2 \alpha}) \\
                             {\bf C} & \mbox{min}(a_{1 \alpha}, a_{2
\alpha}) < i \leq \mbox{max}(a_{1 \alpha}, a_{2 \alpha}) \\
                             0 & i > \mbox{max}(a_{1 \alpha}, a_{2
\alpha})
                             \end{array}
                             \right.
\end{displaymath}
In particular, a vector bundle splits globally into a direct sum
of holomorphic line bundles precisely when all the filtrations
(not just associated to toric divisors in any one cone) satisfy
the compatibility condition~(\ref{compatpara}).

It may naively appear that all information is contained in the values
of $i$ at which a filtration changes dimension, but this is in fact
false.  We will see later that the precise vector subspaces appearing
also carry information -- for example, they partly determine most
of the Chern classes.

Note that we can rederive the (weaker) equivariant version of a famous
theorem of Grothendieck, which says that all holomorphic
vector bundles on ${\bf P}^{1}$ split into a direct sum of
holomorphic line bundles.  Since any two parabolic subgroups
intersect in a maximal torus (at least), and as ${\bf P}^{1}$ has only
two toric divisors, it should be obvious that all equivariant
vector bundles on ${\bf P}^{1}$ split as indicated.

\subsection{Kaneyama's approach} \label{kaneyama}

There is a closely related approach to equivariant bundles on toric 
varieties described in \cite{kaneyama1,kaneyama2}.  This approach
is more intuitively clear than Klyachko's, but computationally
far more cumbersome.  

The basic idea behind Kaneyama's approach is to study equivariant
vector bundles on each element of a cover of the base space.
In particular, the maximal cones defining coordinate charts on the
toric variety provide a suitable cover.  On each element of the
cover (affine spaces, by the assumption of smoothness),
the vector bundle trivializes 
into a sum of one-dimensional representations
of the algebraic torus $T$ defining the toric variety.
Put another way, for each maximal cone of the toric variety,
for a rank $r$ vector bundle ${\cal E}$ we can associate
$r$ characters of the algebraic torus (elements of the lattice $M$,
in the notation of \cite{fulton}).  This decomposition of the
vector space ${\bf C}^{r}$ describes precisely how the vector
space transforms (equivariantly) under the action of the algebraic
torus.

Given this set of $r$ characters associated with some
maximal cone, we can now associate
a set of integers with each toric divisor in the 
maximal cone.  (These integers are half the data Kaneyama needs to 
describe a vector bundle.) 
If we label the characters as $\chi_{1}, \ldots \chi_{r}$,
then the integers associated with toric divisor $\alpha$
are precisely $\langle \chi_{1}, \alpha \rangle, \ldots
\langle \chi_{r}, \alpha \rangle$.

In fact, this is precisely a generalization of the description of line
bundles in \cite{fulton}.  There, a line bundle on a compact toric
variety was specified by associating a character\footnote{
Technical fiends will note that each character is the 
location of the generator of the principal
fractional ideal associated to that cone.} to each maximal cone.
The characters on distinct cones need not agree, rather they can differ
by a character perpendicular to all vectors in the intersection of the
cones.  Thus, the characters themselves are not well-defined.  However,
it is possible to associate well-defined integers with each toric
divisor $\alpha$, as $\langle \chi, \alpha \rangle$ where $\chi$ is a
character associated with a maximal cone containing $\alpha$.  In
Kaneyama's generalization, we associate $r$ characters with each maximal
cone, and an element of $GL(r,{\bf C})$ with each overlap of cones.  
The characters themselves need not be well-defined; however the integers
$\langle \chi_{1}, \alpha \rangle$, $\langle \chi_{2}, \alpha \rangle$,
$\ldots$ are well-defined.

These integers can also be derived from Klyachko's filtration
description.  Let $E^{\alpha}(i)$ be a set of filtrations associated 
with the bundle ${\cal E}$, then define 
\begin{displaymath}
E^{[\alpha ]}(i) \: = \: \frac{E^{\alpha}(i)}{E^{\alpha}(i+1)}
\end{displaymath}
(recall that $E^{\alpha}(i+1) \subseteq E^{\alpha}(i)$).
Given this definition, the integers Kaneyama associates 
with a toric divisor $\alpha$ are precisely the $i$ for which
$E^{[\alpha ]}(i)$ is nonzero, counted with multiplicity equal
to $\mbox{dim } E^{[\alpha ]}(i)$.

An example should make this more clear.
Suppose we have a filtration defined by
\begin{displaymath}
E^{\alpha}(i) \: = \: \left\{ \begin{array}{ll}
                              {\bf C}^{3} & i \leq 5 \\
                              {\bf C}^2 & i = 6, 7 \\
                              0 & i > 7 
                              \end{array}
                      \right.
\end{displaymath}
then
\begin{displaymath}
E^{[\alpha ]}(i) \: = \: \left\{ \begin{array}{ll}
                                 {\bf C} & i = 5 \\
                                 {\bf C}^2   & i = 7 \\
                                 0           & \mbox{otherwise}
                                 \end{array}
                          \right.
\end{displaymath}
then in Kaneyama's description, to the toric divisor $\alpha$
we associate the integers $5$, $7$, and $7$.  

In fact, $\mbox{dim } E^{[\alpha ]}(i)$ is counting the multiplicities
with which characters of the algebraic torus $T$ are appearing
in maximal cones containing $\alpha$.  
The multiplicity with which a character $\chi$ appears
is equal to $\mbox{dim } E^{[\alpha ]}(\langle \chi, \alpha \rangle)$.

So far we have given only half of Kaneyama's description of vector
bundles.  To each toric divisor we have described how to associate
a set of integers, and shown how this is related to Klyachko's description of
vector bundles.  However, these integers are not sufficient to
completely describe the bundle, as we also need to describe the
transition functions on overlaps of coordinate charts. 

In Kaneyama's description the transition functions for a rank $r$ bundle
are given as elements of $GL(r,{\bf C})$ assigned to all pairs of
cones, satisfying the usual compatibility conditions.
How can such transition functions be derived from Klyachko's
description?
As was shown in \cite[p. 344]{kl2}, if ${\cal E}$, ${\cal F}$ are equivariant
bundles on an affine variety, then the space of toral homomorphisms
$\mbox{Hom}_{T}({\cal E},{\cal F})$ is isomorphic to the space of linear
operators $\phi: E \rightarrow F$ compatible with the filtrations as
$\phi( E^{\alpha}(i)) \subseteq F^{\alpha}(i) \: \forall \alpha, i$.
In particular, on overlaps between cones $\sigma_{1}$, $\sigma_{2}$, 
transition functions must map each filtration associated
to divisor $\alpha \in | \sigma_{1} \cap \sigma_{2} |$ back into itself. 
Thus, it should  be clear that for any cone $\sigma$,
Kaneyama's transition function must be an element of
\begin{displaymath}
\bigcap_{\alpha \in | \sigma | } P^{\alpha}
\end{displaymath}
Conversely, given the data Kaneyama specifies it is possible
to recreate Klyachko's description.

Note in particular that both descriptions implicitly rely on
equivariance of the bundles in question.  If a bundle is not
equivariant, then over any coordinate chart it trivializes and so of
course can be written as a direct sum of characters, but there need no
longer be any relation between characters of distinct coordinate charts,
and so one can no longer associate an invariant set of integers with
each toric divisor.  Similarly, Klyachko's description fails, as there
is no longer an invariant way to associate a filtration with each toric
divisor.

\subsection{Applications} \label{appl}

Here we will quote some results presented in \cite{kl1,kl2,kl3,kl4} on 
Chern classes and sheaf cohomology groups of equivariant bundles over
toric varieties.  We will also closely follow the notation of these
references, in order to make the connection more clear to the reader.

\subsubsection{Chern classes}

Before describing the Chern classes, we will first describe
a natural resolution of any bundle ${\cal E}$ by direct sums of
line bundles, as described in \cite{kl2}.  Given this resolution,
computing the Chern classes will then be quite straightforward.

Schematically, this resolution is an exact sequence
\begin{equation} \label{bundleres}
0 \rightarrow {\cal E} \rightarrow \bigoplus_{ {\it codim }\, \sigma = 0}
\sigma \otimes {\cal E} \rightarrow \bigoplus_{ {\it codim} \, \sigma = 1}
\sigma \otimes {\cal E} \rightarrow \cdots \rightarrow
\bigoplus_{ {\it codim } \, \sigma = n-1 } \sigma \otimes {\cal E}
\rightarrow \emptyset \otimes {\cal E} \rightarrow 0
\end{equation}
where each $\sigma \otimes {\cal E}$ is a holomorphic bundle,
of the same rank as ${\cal E}$, defined by the filtrations\footnote{The
equation shown corrects typos in equation~(3.3) of \cite{kl2}.}
\begin{displaymath}
(\sigma \otimes {\cal E})^{\alpha}(i) \: = \: \left\{
\begin{array}{ll}
E^{\alpha}(i) & \alpha \in | \sigma | \\
E             & \alpha \not\in | \sigma |, \, i \leq f(\alpha) \\
0             & \alpha \not\in | \sigma |, \, i >    f(\alpha)
\end{array} \right.
\end{displaymath}
and where $f(\alpha)$ is a function $f: D_{\alpha} \rightarrow {\bf Z}$
that assigns to each toric divisor an integer, the largest at which
the filtration is nonzero.  In other words, $E^{\alpha}(i) = 0$
precisely when $i > f(\alpha)$.
(We will see that the function $f(\alpha)$ cancels out of
Chern class computations.)  Note that the notation used above
is rather poor -- nothing is being tensored in $\sigma \otimes {\cal E}$,
rather $\sigma \otimes {\cal E}$ merely denotes an auxiliary bundle
associated to cone $\sigma$.
Each map in the exact sequence~(\ref{bundleres})
is the obvious inclusion between filtrations.

Each bundle $\sigma \otimes {\cal E}$
splits into a sum of line bundles, as we now show.  
By assumption, the filtrations
on the toric divisors of any single cone $\sigma$ are compatible,
in the sense of equation~(\ref{compatpara}), and as the filtrations
on all other toric divisors are trivial, all the filtrations are
compatible, so the bundle splits globally.

Note that the line bundles into which $\sigma \otimes {\cal E}$ splits
are in one-to-one correspondence with generators of $E$.  For
example, suppose ${\cal E}$ is rank two, defined on a toric surface.
Suppose for divisors we shall label 1,2,
\begin{displaymath}
\mbox{dim } E^{[\alpha]}(i) \: = \: \left\{ \begin{array}{ll}
1 & i=0 \\
1 & i=1 \\
0 & i \neq 0,1
\end{array} \right.
\end{displaymath}
Then if the two one-dimensional vector spaces at $i = 1$ coincide
as subspaces of ${\bf C}^{2}$, $c( \sigma \otimes {\cal E}) = [ 1 +
D_1 + D_2 ]$.  Otherwise, if those two one-dimensional vector
spaces are generic, we have $c( \sigma \otimes {\cal E}) =
[1 + D_1 ] [ 1 + D_2 ]$.  More generally, although it will turn out
that the first Chern class depends only on the dimensions of each
element of each filtration, to determine the higher Chern classes
more information about the filtrations is required.

Before using the resolution above to derive the Chern classes of
${\cal E}$, we shall define some notation.
Let ${\cal E}$ be an equivariant bundle on a toric variety $X$ with 
fan $\Sigma$, and let $E^{\alpha}(i)$ be a family of compatible
filtrations defining ${\cal E}$.
Define
\begin{equation}
E^{\sigma}(\chi) \: = \: \bigcap_{\alpha \in | \sigma |} 
E^{\alpha}( \langle \chi, \alpha \rangle )   \label{E(sigma)}
\end{equation}
\begin{equation}
E^{[\sigma ]}(\chi) \: = \: \frac{E^{\sigma}(\chi)}{\sum_{i_{\alpha}} 
\bigcap_{\alpha \in | \sigma |} E^{\alpha}(i_{\alpha}) }
\label{E[sigma]}
\end{equation}
where in equation~(\ref{E[sigma]}) the sum is taken over all 
$i_{\alpha} \in {\bf
Z}$, and the intersection is over $\alpha \in | \sigma |$ such that
$i_{\alpha} \geq \langle \chi, \alpha \rangle$ and for at least one
$\alpha \in | \sigma |$, $i_{\alpha} > \langle \chi, \alpha \rangle$.
In particular, this implies
\begin{displaymath}
E^{[\alpha]}(i) \: = \: \frac{ E^{\alpha}(i) }{ E^{\alpha}(i+1) }
\end{displaymath}
In fact, $E^{\sigma}$ is a freely-generated ${\bf C}[ \sigma^{\vee} ]$
module (see section~\ref{basicprin}), 
and $E^{[\sigma]}$ locates its generators.

Now, let us derive the Chern classes of ${\cal E}$, using the 
resolution~(\ref{bundleres}).
First consider the bundle $\alpha \otimes {\cal E}$, where $\alpha$
is an edge of the fan.  The first Chern class is clearly
\begin{displaymath}
c_{1}( \alpha \otimes {\cal E}) \: = \: \sum_{i} \, i \, \mbox{dim }
E^{[\alpha]}(i) \, D_{\alpha} \: + \: \sum_{\alpha' \neq \alpha}
\, f(\alpha') \, (\mbox{dim } E) \, D_{\alpha'}
\end{displaymath}
Using the splitting principle, we can derive the total Chern
class of $\alpha \otimes {\cal E}$:
\begin{displaymath}
c( \alpha \otimes {\cal E}) \: = \: \prod_{\chi \in \hat{T}_{\alpha} }
\left[ 1 + \sum_{\alpha' \in | \Sigma |} f^{\alpha}_{\chi}(\alpha')
D_{\alpha'} \right]^{ {\it dim } E^{[\alpha]}(\chi) }
\end{displaymath}
where 
\begin{displaymath}
f^{\sigma}_{\chi}(\alpha) \: = \: \left\{ \begin{array}{ll}
         f(\alpha) & \alpha \not\in | \sigma | \\
         \langle \chi, \alpha \rangle & \alpha \in | \sigma |
         \end{array} \right.
\end{displaymath}
and where $\hat{T}_{\sigma}$ is the weight lattice $M = \mbox{Hom }(T,
{\bf C}^{\times})$  
modulo the sublattice $\sigma^{\perp} \cap M$.
(More formally, define
$T_{\sigma}$ to be the subgroup of $T$ stabilizing the
subvariety corresponding to $\sigma$ (the subgroup of $T$ whose
Lie algebra is generated by $\sigma$ and $- \sigma$)
then $\hat{T}_{\sigma} = \mbox{Hom }( T_{\sigma}, {\bf C}^{\times} )$.)
In other words, if we took the product over $\chi \in M$ rather
than just $\chi \in \hat{T}_{\alpha}$, we would overcount.

It should now be clear that in general,
\begin{displaymath}
c( \sigma \otimes {\cal E}) \: = \: \prod_{\chi \in \hat{T}_{\sigma} }
\left[ 1 + \sum_{\alpha \in | \Sigma |} f^{\sigma}_{\chi}(\alpha)
D_{\alpha} \right]^{ {\it dim } E^{[\sigma]}(\chi) }
\end{displaymath}

Now, we should get the same result
for $c({\cal E})$ if we increase each value of $f:D_{\alpha} \rightarrow
{\bf Z}$, so the expression for $c({\cal E})$ must be
independent of the $f(\alpha)$'s, so we can simply set them to
zero and immediately recover the result
\begin{equation}
c({\cal E}) \: = \: \prod_{\sigma \in \Sigma, \chi \in \hat{T}_{\sigma} }
\left( 1 \, + \, \sum_{\alpha \in | \sigma |} \langle \chi, \alpha \rangle
D_{\alpha} \right)^{(-)^{{\it codim } \: \sigma } \,  {\it dim } \:
E^{[\sigma ]}(\chi)}
\label{c}
\end{equation}
(If the reasoning behind our omission of the $f(\alpha)$'s seemed
too loose, the reader is encouraged to check a few examples in
detail.)
Similarly, one can calculate the Chern character of ${\cal E}$
to be
\begin{equation}
ch({\cal E}) \: = \: \sum_{\sigma \in \Sigma, \chi \in \hat{T}_{\sigma} } 
(-)^{{\it codim } \: \sigma} \mbox{dim } E^{[\sigma ]}(\chi) 
\, \exp \left( \sum_{\alpha \in | \sigma |} \langle \chi, \alpha \rangle 
D_{\alpha} \right)
\label{ch}
\end{equation}
where $D_{\alpha}$ denotes the toric divisor associated with
$\alpha$.
Thus, for example,
\begin{displaymath}
c_{1}({\cal E}) \: = \: \sum_{i \in {\bf Z}, \alpha \in | \Sigma |}
i \: \mbox{dim } E^{[\alpha ]}(i) \, D_{\alpha}
\end{displaymath}

\subsubsection{Sheaf cohomology groups} \label{sheafcohom}

It is also straightforward to calculate sheaf cohomology groups
$H^{p}(X, {\cal E})$.  Since ${\cal E}$ is equivariant,
the sheaf
cohomology groups can be decomposed into groups each associated with 
an irreducible representation of
the algebraic torus\footnote{If ${\cal
E}$ is not equivariant, then the sheaf cohomology groups have no such
decomposition.}:
\begin{displaymath}
H^{p}(X, {\cal E}) \: = \: \oplus_{\chi} \, H^{p}(X, {\cal E})_{\chi}
\end{displaymath}
In particular, all irreducible representations of
the algebraic torus are one-dimensional, given precisely by the
characters $\chi$, and this decomposition of the sheaf cohomology
is known as an isotypic decomposition.
(Note that each $H^{p}(X, {\cal E})_{\chi}$ need not itself be 
one-dimensional; it is merely naturally associated with a (one-dimensional) 
irreducible representation.)

To calculate isotypic components of the sheaf cohomology groups 
$H^{p}(X,{\cal E})$ we
use the complex\footnote{In references \cite{kl1,kl2,kl3,kl4},
this complex is ordered in the opposite direction and
denoted $C_{*}({\cal E}, \chi)$.  We have chosen
different conventions for clarity.} $C^{*}({\cal E},\chi)$:
\begin{displaymath}
0 \: \rightarrow \: \bigoplus_{{\it dim } \: \sigma = n} E^{\sigma}(\chi)
\: \rightarrow \: \cdots \: \rightarrow \:
\bigoplus_{ {\it dim } \: \sigma = 2} E^{\sigma}(\chi) \: \rightarrow
\: \bigoplus_{ {\it dim } \: \sigma = 1} E^{\sigma}(\chi) \: \rightarrow
\: E \: \rightarrow \: 0
\end{displaymath}
where $n = \mbox{dim } X$, $E$ is the generic fiber of the vector
bundle.  We will observe
later (section~\ref{basicprin}) that $E^{\sigma}$ is a
graded ${\bf C}[\sigma^{\vee}]$-module,
equal to $\Gamma( U_{\sigma}, {\cal E})$ (where $U_{\sigma}$
is the open set in $X$ determined by cone $\sigma$),
so this is precisely a \u{C}ech cohomology calculation,
and the differential is then defined in the usual fashion.
Then, 
\begin{displaymath}
H^{p}(X, {\cal E})_{\chi} \: = \: H^{p}( C^{*}({\cal E},\chi) )
\end{displaymath}

It should be clear that
\begin{displaymath}
\sum_{i} (-)^{i} \mbox{dim } H^{i}( X, {\cal E})_{\chi} \: = \:
\sum_{\sigma \in \Sigma} (-)^{ {\it codim } \: \sigma} \mbox{dim }
E^{\sigma}(\chi)
\end{displaymath}
and also\footnote{The next result holds only for reflexive,
not for more general torsion-free sheaves.}
\begin{displaymath}
H^{0}( X, {\cal E})_{\chi} \: = \: \bigcap_{\alpha \in | \Sigma |}
E^{\alpha}(\chi)
\end{displaymath}

In principle, calculating the sheaf cohomology of the restriction
of ${\cal E}$ to a hypersurface in the toric variety is now
straightforward.  Suppose our hypersurface is specified by
some divisor D, then we have an exact sequence
\begin{displaymath}
0 \: \rightarrow \: {\cal E}(-D) \: \rightarrow \: {\cal E} \:
\rightarrow \: {\cal E} \otimes {\cal O}_{D} \: \rightarrow \: 0
\end{displaymath}
where ${\cal O}_{D}$ is a skyscraper sheaf with support on $D$.
From this short exact sequence we can derive a long exact sequence
describing sheaf cohomology of ${\cal E}$ restricted
to the hypersurface in terms of sheaf cohomology on the toric
variety.
Since ${\cal E}$ is equivariant, ${\cal E}(-D)$ is
also equivariant, so in principle this calculation can be performed
via the above technology.

\subsection{Alternate gauge groups} \label{agg}

So far in this paper we have discussed equivariant holomorphic vector
bundles whose structure group is $GL(n,{\bf C})$.
In principle, however, principal bundles with other (algebraic) 
gauge groups can be discussed 
equally easily in Klyachko's language, as was noted in \cite{kl4}.

When discussing $GL(n,{\bf C})$ bundles, we can speak in terms either
of filtrations of a vector space or in terms of parabolic subgroups
of $GL(n,{\bf C})$ which preserve the filtration (as in
equation~(\ref{paradefine})) (paired with ample line bundles, as 
mentioned in section~\ref{evb}).  For more general structure groups $G$
there is no filtration description, but the description in terms
of parabolics still holds.

In other words, to discuss principal bundles with arbitrary structure group
$G$, one associates a parabolic subgroup $P^{\alpha} \subset G$ 
to each toric divisor $\alpha$, together with
an ample line bundle $L_{\alpha}$
on $G/P^{\alpha}$.

The compatibility
condition on the parabolic subgroups associated to each toric divisor 
is very nearly the same as
equation~(\ref{compatpara}).  Rather than demand the parabolics
intersect in a maximal torus of $GL(n,{\bf C})$, one demands the
parabolics intersect in a maximal torus of $G$.
In other words, for all cones $\sigma \in \Sigma$, if
$P^{\alpha}$ denotes the parabolic associated with edge $\alpha$, then
\begin{equation} \label{Gcompatpara}
\bigcap_{\alpha \in | \sigma |} P^{\alpha}  \mbox{ contains a maximal
torus of } G
\end{equation}

Now, let us consider the ample line bundles $L_{\alpha}$
paired with the parabolics $P^{\alpha}$ more closely.
First, we shall describe a construction of vector bundles
on flag manifolds $G/P^{\alpha}$, then we shall describe
how the ample line bundles are derived for $GL(r,{\bf C})$
bundles.

There is a natural way to construct vector bundles on $G/P$.  First,
note that $G \rightarrow G/P$ is a principal $P$-bundle over $G/P$.
Then, there exists a family of vector bundles over $G/P$
(associated to the principal $P$-bundle $G \rightarrow G/P$),
corresponding to representations of $P$.  In other words,
given a representation of $P$, consisting of a vector space $V$
and a $P$-action $\lambda: P \times V \rightarrow V$,
we can construct a vector bundle associated to $G \rightarrow G/P$.
The total space of this vector bundle is just
$G \times_{P} V$.  In particular, line bundles constructed in this
fashion automatically come with a canonical $G$-linearization
(in fact, if $G$ is semisimple then the $G$-linearization is unique).
(This is the setup of the Bott-Borel-Weil theorem
\cite{bott}, which states -- among other things -- that the sections
of an ample line bundle associated to $G \rightarrow G/P$
form an irreducible representation of $G$.)

Now, we shall describe how to construct the ample line bundles
$L_{\alpha}$ for $G = GL(r,{\bf C})$ bundles.
First, associate $r = \mbox{rank }
{\cal E}$ integers to the filtration $E^{\alpha}$.  These integers
are the integers $i$ such that $E^{[\alpha]}(i)$ is nonzero,
counted with multiplicity equal to $\mbox{dim } E^{[\alpha]}(i)$.
(In other words, these are the same integers that Kaneyama
associates with a toric divisor, as described in section~\ref{kaneyama}.)
These integers define a point in the weight lattice of $G$.
If we put them in increasing order, then they correspond to
a weight in the fundamental Weyl chamber, and so describe
an ample line bundle.  Denote this line bundle on $G/P^{\alpha}$
by $L_{\alpha}$.  As promised in section~\ref{evb}, the ample
line bundles $L_{\alpha}$ encode information about the filtrations
that is missed by only specifying parabolics.

At this point we should mention an interesting subtlety.
Irreducible representations of a parabolic $P$ are in one-to-one
correspondence with irreducible representations of a Levi factor of the
parabolic, so when the parabolic is a Borel subgroup, any Levi
factor is a product of ${\bf C}^{\times}$'s, and so any set of
integers defines the dominant weight of a one-dimensional irreducible
representation of $P$ -- and in particular defines a line bundle
on $G/P$.  When $P$ is not a Borel, however, not any set of integers
will be the dominant weight for a one-dimensional irreducible
representation of $P$.  In such a case how can we be guaranteed of getting
a line bundle on $G/P$ (as opposed to some higher rank vector
bundle) in the prescription above?
For simplicity, consider the case that only two of the integers
associated to divisor $\alpha$ coincide.  Then, a Levi factor
of the corresponding parabolic will be the product of several
${\bf C}^{\times}$'s with a $GL(2,{\bf C})$.  The two coincident
integers define a weight of $GL(2,{\bf C})$.  However,
since the integers are identical, the semisimple part of
$GL(2,{\bf C})$, namely $SL(2,{\bf C})$, acts trivially on the
representation defined.  The only part of $GL(2,{\bf C})$ that
acts nontrivially is the overall ${\bf C}^{\times}$ factor,
and so we clearly have defined a one-dimensional irreducible representation
of $P$.  It should be clear in general that the prescription above
will always yield one-dimensional irreducible representations of
$P$, and in particular line bundles on $G/P$.

Principal $G$-bundles are uniquely identified by a set of parabolics
obeying the compatibility condition~(\ref{Gcompatpara}),
together with a set of ample line bundles, 
up to an overall simultaneous $G$
rotation of the parabolics.

The reader may wonder why vector bundles with
structure group $G$ do not have a description in terms of
filtrations analogous to that discussed earlier.  
Given some vector space $V$ acted on by a representation of $G$
inside some $GL(n,{\bf C})$, we can certainly associate parabolic
subgroups of $GL(n,{\bf C})$ to filtrations, and then 
we could intersect those parabolic subgroups of 
$GL(n,{\bf C})$ with the image of $G$.  Unfortunately those
intersections need not be parabolic subgroups, and in fact in general
to describe a vector bundle with structure group $G$ in this
fashion takes a great deal of work, dependent upon both $G$
and the representation chosen.  We shall not examine such constructions
in detail in this paper.

\subsection{Equivariant torsion-free sheaves} \label{etfs}

In this section we will discuss torsion-free sheaves
on arbitrary toric varieties.  (Note that our previous
discussions of bundles and reflexive sheaves were limited
to smooth varieties).  In section~\ref{basicprin} we discuss
the basic principles on which our description is based.
In section~\ref{tfsmooth} we discuss torsion-free sheaves
on smooth toric varieties.  
In section~\ref{glext} we discuss how global $\mbox{Ext}$ groups
can be calculated in principle for equivariant sheaves on
toric varieties.
In section~\ref{tfwp} 
we discuss torsion-free sheaves on the singular toric variety,
${\bf P}_{1,1,2}^{2}$.
Torsion-free sheaves are rather more complicated
to discuss on singular toric varieties, but this two example
should sufficiently illustrate the general method.
Finally, in section~\ref{cartierweil} we discuss the distinction
between Cartier and Weil divisors, and how it can be seen explicitly
in this framework.

Much of section~\ref{tfsmooth} was originally discussed in \cite{kl4},
however the rest of the material in this section is new.

\subsubsection{Basic principles} \label{basicprin}

In principle, 
coherent sheaves can be constructed over varieties by
associating
to each element $\mbox{Spec } R_{i}$ of an affine cover a finite-rank
$R_{i}$-module $M_{i}$, in a ``mutually
compatible'' way (see \cite[section II.5]{hartshorne}).  
Compatibility is defined by associating a module
to
every intersection of the affine cover, together with restriction maps.
In 
particular, to build a coherent sheaf on a toric variety, one must
associate a finite-rank module to each cone.

When the coherent sheaf is equivariant, we can refine the
description further.  The module associated to each cone $\sigma$ has
a weight space decomposition under the algebraic torus $T$ whose 
compactification
is the
toric variety.  For an $n$-dimensional toric variety, each
module is $M$-graded, where $M \cong {\bf Z}^{n}$ is the weight
lattice of $T$.  (This is essentially an isotypic decomposition
of the module under the action of the algebraic torus, which
holds only for the modules appearing in equivariant sheaves.)
Each (isotypic) component of the module is a vector space\footnote{
Each isotypic component of an $A$-module $M$, where both
$A$ and $M$ have $T$ actions compatible with the ring action,
is a module over the weight zero part of $A$.  In the cases
relevant here, the weight zero part of the ring will always
be the field ${\bf C}$.  A module over a field is precisely
a vector space, thus each isotypic component of the module
is a vector space.}, and in fact we shall see each component
is a vector subspace of some fixed vector space.  We shall
denote the component of a module $E$ associated to character
$\chi$ by $E(\chi)$.

If the module is torsion-free\footnote{Usually one states a sheaf
is torsion-free when each stalk is torsion-free, meaning, when each
localization of the module is torsion-free.  However, over the 
rings we shall encounter in toric varieties, torsion-free is a local
property, in the sense that if it is true for all localizations of
a module, then it is true for the entire module.  (See for
example \cite[exercise 3.13]{am}.)  The same statement is also
true of reflexivity and local freedom.} then the
action of the underlying monoid algebra ${\bf C} [ \sigma^{\vee} ]$ 
induces inclusion maps between
the graded components of the module.  
Let us work through this a little more carefully.
Let $E^{\sigma}$ denote the module associated to a cone $\sigma$,
then for $\mu \in \sigma^{\vee}$ and for all $\chi \in M$
we have maps $E^{\sigma}(\chi) \rightarrow E^{\sigma}(\chi + \mu)$.
Since the module is torsion-free, these maps are injective.
In fact, we can identify spaces with their images, and regard these maps
as inclusions.

In passing, note that for any $\rho$ in the interior of $\sigma^{\vee}$,
it is true that for all components of $E^{\sigma}$,
\begin{displaymath} 
E^{\sigma}(\chi) \: \subseteq \:
\lim_{n \rightarrow \infty} E^{\sigma}(n \rho)
\end{displaymath}
This is because for all $\chi \in M$, there exists
$N \in {\bf N}$ such that $\langle \chi + N \rho, \nu \rangle
> 0$ for all $\nu \in \sigma^{\vee}$, i.e., $\chi + N \rho \in 
\sigma^{\vee}$.
In particular, this means that each component $E^{\sigma}(\chi)$
is a subset of some fixed vector space, and in fact 
this vector space is independent of $\sigma$,
as expected.

To clarify these remarks,
consider the case that the affine space 
is of the form 
\begin{displaymath}
\mbox{Spec } {\bf C}[x^{a_{1 1}}_{1} x^{a_{2 1}}_{2}
\cdots
x^{a_{n 1}}_{n}, \cdots, x^{a_{1 k}}_{1} x^{a_{2 k}}_{2} \cdots
x^{a_{n k}}_{n}]
\end{displaymath}
then the inclusions are
\begin{eqnarray*}
E^{\sigma}(i_{1}, \cdots, i_{n}) & \hookrightarrow & E^{\sigma}(i_{1} +
a_{1 1}, i_{2} + a_{2 1},
\cdots, i_{n} + a_{n 1} ) \\
 \, & \hookrightarrow & E^{\sigma}(i_{1} + a_{1 2}, i_{2} + a_{2 2},
\cdots,
i_{n} + a_{n 2}) \\
\, & \cdots & \, \\
\, & \hookrightarrow & E^{\sigma}(i_{1} + a_{1 k}, i_{2} + a_{2 k},
\cdots,
i_{n} + a_{n k})
\end{eqnarray*}
In order to recover decreasing filtrations on smooth spaces,
the conventions used earlier in this paper, merely work on
the smooth space
\begin{displaymath}
\mbox{Spec } {\bf C}[ x_{1}^{-1}, x_{2}^{-1}, \cdots, x_{n}^{-1} ]
\end{displaymath}

To make this perspective more clear, let us consider 
some examples.  The structure sheaf on 
${\bf C}^2 = \mbox{Spec } {\bf C}[x_1, x_2]$ is equivariant, and
has bigraded module precisely ${\bf C}[x_1, x_2]$, meaning
\begin{displaymath}
E(i_1,i_2) \: = \: \left\{ \begin{array}{ll}
                     {\bf C} & i_1 \geq 0  \mbox{ and } i_2 \geq 0 \\
                     0 & \mbox{otherwise}
                     \end{array} \right.
\end{displaymath}
for the trivial choice of equivariant structure.
Now, consider
an ideal\footnote{For an introduction to ideal sheaves
see the first appendix of \cite{me1}.} 
sheaf on ${\bf C}^{2} = \mbox{Spec } {\bf C}[x_1, x_2]$.
In fact, consider the rank 1
ideal sheaf vanishing to order 1 at $x_1 = x_2 = 0$.
The module defining the sheaf is precisely the ideal
with generators $(x_1, x_2)$ inside ${\bf C}[x_1, x_2]$.
This ideal sheaf is equivariant\footnote{
Technical fiends will note that we can define equivariant Hilbert
schemes.},  
and
we have the bigraded module
\begin{displaymath}
E(i_1,i_2)  \: = \: \left\{ \begin{array}{ll}
                    {\bf C} & i_1 > 0 \mbox{ and } i_2 \geq 0 \\
                    {\bf C} & i_1 \geq 0 \mbox{ and } i_2 > 0 \\
                    0 &  \mbox{otherwise} 
                    \end{array} \right.
\end{displaymath}
As expected, the ideal sheaf sits naturally inside the structure sheaf.

Now, consider the case that a cone $\sigma$ is not maximal,
for example, that $\mbox{Spec } {\bf C}[\sigma^{\vee}] = {\bf C} \times
{\bf C}^{\times}$.  In such an example, we have inclusion maps
of the form
\begin{eqnarray*}
E^{\sigma}(i_1, i_2) & \hookrightarrow & E^{\sigma}(i_1 + 1, i_2) \\
 & \hookrightarrow & E^{\sigma}(i_1, i_2 + 1) \\
 & \hookrightarrow & E^{\sigma}(i_1, i_2 - 1)
\end{eqnarray*}
Clearly, $E^{\sigma}(i_1, i_2)$ is independent of $i_2$.
A more nearly invariant way to say this is that the module
associated to a cone $\sigma$ is nontrivial only over the lattice
$\hat{T}_{\sigma}$, where $\hat{T}_{\sigma}$ is simply the
weight lattice $M$ of the algebraic group modulo 
the subgroup $\sigma^{\perp}$ -- in other words,
$E^{\sigma}$ is constant along directions in $\sigma^{\perp}$.

So far we have described modules associated to any individual
cone; how are modules over distinct cones related?
If $\tau$ is a subcone of $\sigma$, then
by definition\footnote{$E^{\tau}$ is simply the restriction of the module
$E^{\sigma}$ associated to affine space $\mbox{Spec } {\bf C}[\sigma^{\vee}]$
to $\mbox{Spec } {\bf C}[\tau^{\vee}] \subset \mbox{Spec }
{\bf C}[\sigma^{\vee}]$.}
\begin{displaymath}
E^{\tau} \: = \: E^{\sigma} \otimes_{ {\bf C}[\sigma^{\vee}] }
{\bf C}[\tau^{\vee}]
\end{displaymath}

We can show that if $\tau$ is a subcone of $\sigma$
and $\rho$ is in the interior of $\tau^{\perp} \cap \sigma^{\vee}$,
then
\begin{displaymath}
E^{\tau}(\chi) \: = \: \lim_{n \rightarrow \infty} E^{\sigma}
(\chi + n \rho)
\end{displaymath}
for all $\chi \in M$. 
(In particular, this shows the maximal vector space $E$ is independent
of $\sigma$, as claimed earlier.)
The point is to show that for all $\mu \in \tau^{\vee}$,
there exists $N \in {\bf N}$ such that $\mu + N \rho \in \sigma^{\vee}$.
This follows from the definition of $\rho$:  
for any fixed $\mu \in \tau^{\vee}$, $\langle \mu, \nu \rangle$ 
is bounded from below as we vary
$\nu \in \sigma^{\vee}$, so as $\langle \rho, \nu \rangle \geq 0$
for all $\nu \in \sigma^{\vee}$, there exists $N \in {\bf N}$ such that
\begin{displaymath}
N \langle \rho, \nu \rangle \: + \: \langle \mu, \nu \rangle \:
\geq \: 0
\end{displaymath}
for all $\nu \in \sigma^{\vee}$.  Thus, for any $\mu \in \tau^{\vee}$,
there exists $N \in {\bf N}$ such that $N \rho + \mu \in \sigma^{\vee}$,
and so we have an inclusion map
\begin{displaymath}
E^{\sigma}(\chi - \mu)  \: \hookrightarrow \: E^{\sigma}(\chi + N \rho)
\end{displaymath}
for all $\chi \in M$, 
so in particular we find
\begin{displaymath}
E^{\tau}(\chi) \: = \: \lim_{n \rightarrow \infty} E^{\sigma}
(\chi + n \rho)
\end{displaymath}

We can summarize these results as follows.
Suppose $\tau$, $\sigma$ are both cones, and $\tau \subset \sigma$.
For any $\chi \in \hat{T}_{\sigma}$, we have
the inclusion
\begin{displaymath}
E^{\sigma}(\chi) \: \subseteq \: E^{\tau}(\chi)
\end{displaymath}
where on the right side of the equation above we can interpret
$\chi$ as an element of $\hat{T}_{\tau}$, using the natural
projection $\hat{T}_{\sigma} \rightarrow \hat{T}_{\tau}$.
Moreover, for any $\chi \in \hat{T}_{\tau}$,
\begin{displaymath}
E^{\tau}(\chi) \: = \: \bigcup_{\widetilde{\chi}} \, E^{\sigma}(
\widetilde{\chi} )
\end{displaymath}
where the union is over $\widetilde{\chi}$ that project to $\chi$
under $\hat{T}_{\sigma} \rightarrow \hat{T}_{\tau}$.

In the special case that a sheaf is reflexive, this description
can be simplified.  The module $E^{\sigma}$ associated to any cone
$\sigma$ is the intersection of the modules associated to the toric
divisors in the cone:
\begin{equation}  \label{rdfn}
E^{\sigma}(\chi) \: = \: \bigcap_{\alpha \in | \sigma | } E^{\alpha}(\chi)
\end{equation}
(This result is stated without proof in \cite{kl4} for the special case
of smooth varieties.)
In fact, somewhat more generally, for any equivariant torsion-free
sheaf with modules $\{ E^{\sigma} \}$, its bidual has modules
\begin{displaymath}
(E^{\vee \vee})^{\sigma}(\chi) \: = \: \bigcap_{\alpha \in | \sigma | } 
E^{\alpha}(\chi)
\end{displaymath}
The reader should find this result intuitively reasonable,
for the following reason.  For any torsion-free sheaf
${\cal E}$, there is an inclusion ${\cal E} \rightarrow {\cal E}^{\vee \vee}$.
In other words, any torsion-free sheaf should naturally map
into its bidual.  Now, for any set of modules $\{ E^{\sigma} \}$,
it is always true that
\begin{displaymath}
E^{\sigma}(\chi) \subseteq \bigcap_{\alpha \in | \sigma | } E^{\alpha}(\chi)
\end{displaymath}
so the expression given for $( E^{\vee \vee} )^{\sigma}$ should
seem extremely natural.

We shall now 
demonstrate the result~(\ref{rdfn}) for reflexive sheaves
explicitly.  In particular, we will show that for any torsion-free
sheaf ${\cal E}$, its dual ${\cal E}^{\vee}$ (which is always 
reflexive) has the property stated.
Let $X$ be an affine toric variety with principal
cone $\sigma$ and edges $\{ \tau_{\alpha} \}$.  
Let $A^{\sigma}$ be an equivariant
module over ${\bf C}[\sigma^{\vee}]$, and $\{ A^{\alpha} \}$ be the
modules associated to the toric divisors.  Let $\rho_{\alpha}$ be
in the interior of $\tau_{\alpha}^{\perp}
\cap \sigma^{\vee}$.
Let 
\begin{displaymath}
(A^{\vee})^{\sigma} \: = \: \mbox{Hom}_{{\bf C}[
\sigma^{\vee}]} \, ( A^{\sigma}, {\bf C}[\sigma^{\vee}])
\end{displaymath}
with associated
$T$-action.  
Now, consider $(A^{\vee})^{\sigma}(\chi)$, which is to say equivariant maps
$A^{\sigma} \rightarrow {\bf C}[\sigma^{\vee}]^{[- \chi]}$, where the
superscript indicates the grading is shifted.  For each
$\mu \in M$, $\nu \in \sigma^{\vee}$,
we have the commutative diagram
\begin{equation}  \label{commd}
\begin{array}{ccc}
A^{\sigma}(\mu) & \longrightarrow & A^{\sigma}(\mu + \nu) \\
\downarrow & \, & \downarrow \\
{\bf C}[\sigma^{\vee}](\mu + \chi) & \longrightarrow & 
{\bf C}[\sigma^{\vee}](\mu + \nu + \chi)
\end{array}
\end{equation}
so as $\nu$ gets infinitely deep in $\sigma^{\vee}$, the right edge converges
to a map $E \rightarrow {\bf C}$, where $E$ is the vector space in which
each component of the module $A^{\sigma}$ embeds. 
More generally for any $\mu$ we have the composition $A^{\sigma}(\mu)
\rightarrow E \rightarrow {\bf C}$   
-- thus, an element of $E^{*} = \mbox{Hom } (E, {\bf C})$ completely
determines a map $A^{\sigma} \rightarrow {\bf C}[\sigma^{\vee}]^{[-\chi]}$.
Note that not any random element of $E^{*}$ can be associated
with a map $A^{\sigma} \rightarrow {\bf C}[\sigma^{\vee}]^{[-\chi]}$:
if $\mu \not\in \sigma^{\vee} - \chi$, then the image of
$A^{\sigma}(\mu)$ must vanish,
because the lower left corner of diagram~(\ref{commd}) vanishes.
Thus, each component $(A^{\vee})^{\sigma}(\chi)$ of $(A^{\vee})^{\sigma}$ 
is identified
with a subspace of $E^{*}$ that kills certain components of
$A^{\sigma}$.
Now, this means that 
\begin{eqnarray}  \label{alp}
(A^{\vee})^{\sigma}(\chi) & = & \bigcap_{\mu \not\in \sigma^{\vee}} 
(A^{\sigma}(\mu - \chi))^{\perp} \\
 \, & = & \bigcap_{\alpha} \left[ \bigcap_{\mu \, s.t. \, \langle
\mu, \tau_{\alpha} \rangle < 0 } ( A^{\sigma}(\mu - \chi) )^{\perp} 
\right] 
\end{eqnarray}
where the elements on the right hand side should be thought of as
subspaces of $E^{*}$.
Since $A^{\sigma}(\mu - \chi) \subseteq A^{\alpha}(\mu - \chi)$,
we have that $(A^{\sigma}(\mu - \chi))^{\perp} \supseteq  
( A^{\alpha}(\mu  - \chi))^{\perp}$.  More precisely,
we have
$A^{\sigma}(\mu - \chi) \subseteq A^{\sigma}(\mu - \chi + m \rho_{\alpha})$,
and for sufficiently large $m$ we have $A^{\sigma}(\mu - \chi + m \rho_{\alpha}) = A^{\alpha}(\mu - \chi)$, so
\begin{equation} \label{almthere}
(A^{\vee})^{\sigma}(\chi) \: = \: \bigcap_{\alpha} \left[
\bigcap_{\mu \, s.t. \, \langle \mu, \tau_{\alpha} \rangle < 0 }
( A^{\alpha}(\mu - \chi) )^{\perp} \right]
\end{equation}
Now, we know 
\begin{eqnarray*}
(A^{\vee})^{\alpha}(\chi) & = &
(A^{\vee})^{\sigma}(\chi + m \rho_{\alpha}) \mbox{ for large $m$} \\
 & = & \bigcap_{\beta} \left[
\bigcap_{\mu \, s.t. \, \langle \mu, \tau_{\beta} \rangle <  0}
(A^{\beta}(\mu -  \chi - m \rho_{\alpha}))^{\perp} \right]
\end{eqnarray*}
Now, for $m$ sufficiently large, we can make
$\langle \tau_{\beta}, \mu  - \chi - m \rho_{\alpha}  \rangle$
arbitrarily negative if $\alpha \neq \beta$.
Thus, for sufficiently large $m$, if $\alpha \neq \beta$ then
$A^{\beta}(\mu - \chi - m \rho_{\alpha}) = 0$,
and so $ ( A^{\beta}(\mu  - \chi - m \rho_{\alpha}))^{\perp} = E^{*}$.
Thus, 
\begin{displaymath}
\bigcap_{\mu \, s.t. \, \langle \mu, \tau_{\alpha} \rangle < 0}
( A^{\alpha}(\mu - \chi))^{\perp} \: = \:
(A^{\vee})^{\alpha}(\chi)
\end{displaymath}
and so plugging into equation~(\ref{almthere}) we recover
\begin{displaymath}
(A^{\vee})^{\sigma}(\chi) \: = \: \bigcap_{\alpha} (A^{\vee})^{\alpha}(\chi)
\end{displaymath}
as was to be shown.

Note that the modules $E^{\alpha}$ associated to toric divisors
are all completely specified by filtrations.  This is because the
affine space associated to each toric divisor is of the form
${\bf C} \times ( {\bf C}^{\times} )^{k}$ for some $k$, so each
$\hat{T}_{\alpha}$ is completely specified by a single index.
Thus, even on a singular variety, reflexive sheaves are specified
by associating a filtration to each toric divisor.

To review, on both smooth and singular toric varieties,
we specify a reflexive sheaf (and also a $GL(r,{\bf C})$ bundle)
by associating a filtration to each toric divisor.  Modules associated
with larger cones are then obtained by intersecting the modules
associated to toric divisors.

Now, given some reflexive sheaf, how do we determine if it
is locally free?  On a smooth toric variety, we already
know the result -- the parabolic subgroups associated to each
filtration must satisfy equation~(\ref{compatpara}).
On a singular variety, this is necessary but not sufficient.
In order for a reflexive sheaf to be locally free, on each
element of an affine cover it must look like a direct sum of
line bundles.  (One line bundle for each generator of the associated
freely-generated module.)  On a smooth variety, any divisor will define
a line bundle, but on a singular toric variety only some divisors
(the so-called Cartier divisors) define line bundles.
Thus, on a singular toric variety, in addition to 
checking that equation~(\ref{compatpara}) is satisfied, one must
also check that the sheaf splits into a direct sum of line bundles
on each maximal cone.  The most efficient way to do this is to check
that, for any maximal cone, the integers associated to toric divisors
\`a la Kaneyama can be associated with a set of Cartier divisors.
(Cartier divisors will be studied in greater detail
in subsection~\ref{cartierweil}.)
 
In order to help clarify these remarks, let us
consider an example on ${\bf P}^{2}$.
Let the fan describing ${\bf P}^{2}$ as a toric variety
have edges $(1,0)$, $(0,1)$, and $(-1,-1)$,
and denote by cone $1$ the cone spanned by $(1,0)$, $(0,1)$,
cone 2 spanned by $(0,1)$, $(-1,-1)$,
and cone 3 spanned by $(-1,-1)$, $(1,0)$.
The coordinate charts corresponding to maximal cones
are as follows:
\begin{eqnarray*}
U_1 & = & \mbox{Spec } {\bf C}[x,y] \\
U_2 & = & \mbox{Spec } {\bf C}[x^{-1}, x^{-1} y ] \\
U_3 & = & \mbox{Spec } {\bf C}[y^{-1}, x y^{-1} ]
\end{eqnarray*}

Now, to be specific, let us describe an ideal sheaf, call it ${\cal I}$,
vanishing to order 1 at the origin of cone 1.
The module associated to cone 1, call it $M_1$, is the ideal
$(x,y) \subset {\bf C}[x,y]$:
\begin{center}
\begin{tabular}{c|cccccc}
 & -2 & -1 & 0 & 1 & 2 & 3  \\ \hline
 & & & $\vdots$ & & & \\
3 & & 0 & ${\bf C}$ & ${\bf C}$ & ${\bf C}$ & \\
2 & & 0 & ${\bf C}$ & ${\bf C}$ & ${\bf C}$ & \\
1 & $\cdots$ & 0 & ${\bf C}$ & ${\bf C}$ & ${\bf C}$ & $\cdots$ \\
0 & & 0 & 0 & ${\bf C}$ & ${\bf C}$ & \\
-1 & & 0 & 0 & 0 & 0 & \\
 & & & $\vdots$ & & & \\
\end{tabular}
\end{center}
Denote the module over $U_{2}$ by $M_{2}$.
As the ideal sheaf is trivial in cone 2, $M_{2}$ should
correspond to the structure sheaf on $U_{2}$  (meaning,
$M_2 = {\bf C}[x^{-1},x^{-1}y]$):
\begin{center}
\begin{tabular}{c|ccccccc}
 & -4 & -3 & -2 & -1 & 0 & 1 & 2 \\ \hline
 &    &    &    & $\vdots$ & & & \\
3 & & ${\bf C}$ & 0 & 0 & 0 & 0 & \\
2 & & ${\bf C}$ & ${\bf C}$ & 0 & 0 & 0 & \\
1 & $\cdots$ & ${\bf C}$ & ${\bf C}$ & ${\bf C}$ & 0 & 0 & $\cdots$ \\
0 & & ${\bf C}$ & ${\bf C}$ & ${\bf C}$ & ${\bf C}$ & 0 & \\
-1 & & 0 & 0 & 0 & 0 & 0 & \\
 & & & & $\vdots$ & & & \\
\end{tabular}
\end{center}
Denote the module over $U_{3}$ by $M_{3}$.
As for cone 2, the module $M_{3}$ should also correspond
to the structure sheaf on $U_{3}$ (meaning,
$M_3 = {\bf C}[y^{-1},xy^{-1}]$):
\begin{center}
\begin{tabular}{c|cccccc}
 & -2 & -1 & 0 & 1 & 2 & 3 \\ \hline
 &  & & $\vdots$ & & & \\
1 & & 0 & 0 & 0 & 0 &  \\
0 & $\cdots$ & 0 & ${\bf C}$ & 0 & 0 & $\cdots$ \\
-1 & & 0 & ${\bf C}$ & ${\bf C}$ & 0 & \\
-2 & & 0 & ${\bf C}$ & ${\bf C}$ & ${\bf C}$ & \\
 & & & $\vdots$ & & & \\
\end{tabular}
\end{center}

Now, the edge at the intersection of cones 1, 2
corresponds to the affine variety
\begin{displaymath}
U_{12} \: = \: \mbox{Spec } {\bf C}[x, x^{-1}, y] 
\end{displaymath}
and similarly for the other edges
\begin{eqnarray*}
U_{13} & = & \mbox{Spec } {\bf C}[y, y^{-1}, x ] \\
U_{23} & = & \mbox{Spec } {\bf C}[x^{-1}y, x y^{-1}, x^{-1} ] \\
\end{eqnarray*}

The module $M_{12}$ corresponding to $U_{12}$ is simply
\begin{center}
\begin{tabular}{c|ccccc} 
 & -2 & -1 & 0 & 1 & 2 \\ \hline
 & & & $\vdots$ & & \\
2 & & ${\bf C}$ & ${\bf C}$ & ${\bf C}$ & \\
1 & & ${\bf C}$ & ${\bf C}$ & ${\bf C}$ & \\
0 & $\cdots$ & ${\bf C}$ & ${\bf C}$ & ${\bf C}$ & $\cdots$ \\
-1 & & 0 & 0 & 0 & \\ 
 & & & $\vdots$ & & \\
\end{tabular}
\end{center}
where we have written it out over $M$ rather than $\hat{T}_{12}$.
(It should be clear that the module is constant along directions
in $v_{12}^{\perp}$, where $v_{12}$ is the edge of the fan bordering
cones 1,2.)

The module $M_{13}$ corresponding to $U_{13}$ is simply
\begin{center}
\begin{tabular}{c|ccccc}
 & -2 & -1 & 0 & 1 & 2 \\ \hline
 & & & $\vdots$ & & \\
1 & & 0 & ${\bf C}$ & ${\bf C}$ & \\
0 & $\cdots$ & 0 & ${\bf C}$ & ${\bf C}$ & $\cdots$ \\
-1 & & 0 & ${\bf C}$ & ${\bf C}$ & \\
 & & & $\vdots$ & & \\
\end{tabular}
\end{center}

The module $M_{23}$ corresponding to $U_{23}$ is simply
\begin{center}
\begin{tabular}{c|ccccccccc}
 & -4 & -3 & -2 & -1 & 0 & 1 & 2 & 3 & 4 \\ \hline
 & & & & & $\vdots$ & & & & \\
1 & & ${\bf C}$ & ${\bf C}$ & ${\bf C}$ & 0 & 0 & 0 & 0 & \\
0 & $\cdots$ & ${\bf C}$ & ${\bf C}$ & ${\bf C}$ & ${\bf C}$ & 0 & 0 & 0 & $\cdots$ \\
-1 & & ${\bf C}$ & ${\bf C}$ & ${\bf C}$ & ${\bf C}$ & ${\bf C}$ & 0 & 0 & \\
-2 & & ${\bf C}$ & ${\bf C}$ & ${\bf C}$ & ${\bf C}$ & ${\bf C}$ & ${\bf C}$
& 0 & \\
 & & & & & $\vdots$ & & & & \\
\end{tabular}
\end{center}

It is easy to check that the inclusion relations advertised are satisfied
in each case.  It is also easy to check that ${\cal I}^{\vee \vee} \cong
{\cal O}$, the structure sheaf, as expected.

In the case of a smooth compact toric variety, as above,
we can perform an $SL(2,{\bf Z})$ transformation on each
maximal cone to rotate them to all be of the same form.
Given cones in such a standard form, we can then work
somewhat more elegantly.  Such an approach is described
in the next subsection.

In later subsections we will speak to singular toric varieties
at greater length.

\subsubsection{Torsion-free sheaves on smooth toric varieties}
\label{tfsmooth}

In this section we will specialize to torsion-free sheaves
on smooth toric varieties, as has been discussed in \cite{kl4}.

According to \cite{kl4}, to describe a torsion-free sheaf, we generalize
the filtrations discussed earlier, to multifiltrations
\begin{displaymath}
E^{\sigma}(I) \: = \:
E^{(\alpha_{1}, \alpha_{2}, \ldots, \alpha_{k})}(i_{1}, i_{2}, \ldots,
i_{k})
\end{displaymath}
for $\sigma \: = \: (\alpha_{1}, \alpha_{2}, \ldots, \alpha_{k}) \: \in
\Sigma$ and $I \: = \: (i_{1}, i_{2}, \ldots, i_{k}), \: i \, \in \,
{\bf Z}$,.  These filtrations are nonincreasing:  for all $k$,
\begin{displaymath}
E^{\sigma}(i_{1}, \ldots, i_{k}, \ldots, i_{p}) \, \supseteq \,
E^{\sigma}(i_{1}, \ldots, i_{k} + 1, \ldots, i_{p})
\end{displaymath}
These multifiltrations must obey the compatibility condition that
for every pair of cones $\tau \subseteq \sigma$, $\tau \: = \:
(\alpha_{1}, \alpha_{2}, \ldots, \alpha_{p})$, $\sigma \: = \:
(\alpha_{1}, \alpha_{2}, \ldots, \alpha_{p}, \beta_{1}, \ldots,
\beta_{q})$,
\begin{equation} \label{tfcompat}
E^{\tau}(i_{1}, i_{2}, \ldots, i_{p}) \: = \: E^{\sigma}(i_{1}, i_{2},
\ldots, i_{p}, - \infty, \ldots, - \infty)
\end{equation}
This compatibility condition is not related 
to the compatibility condition for bundles, but rather is
the compatibility condition on inclusion morphisms usually used to
define sheaves.

Chern class computations, for example, can be carried out using
simple modifications of the formulae given earlier.
Instead of calculating $E^{\sigma}(i_{1}, \ldots i_{k})$ using
formula~(\ref{E(sigma)}), use the multifiltration
defining the torsion-free sheaf.  Formula~(\ref{E[sigma]}) must be
generalized to
\begin{displaymath}
E^{[\sigma ]}(i_{1}, i_{2}, \ldots, i_{p}) \: = \: \bigtriangleup_{1}
\bigtriangleup_{2} \cdots \bigtriangleup_{p} E^{\sigma}(i_{1}, i_{2},
\ldots, i_{p})
\end{displaymath}
where $\bigtriangleup_{k}$ is a difference operator yielding a formal
difference of vector spaces
\begin{displaymath}
\bigtriangleup_{k} E^{\sigma}(i_{1}, \ldots, i_{k}, \ldots, i_{p}) \: =
\: E^{\sigma}(i_{1}, \ldots, i_{k}, \ldots, i_{p}) \: - \:
E^{\sigma}(i_{1}, \ldots, i_{k} + 1, \ldots, i_{p}) 
\end{displaymath}
For example,
\begin{displaymath}
\mbox{dim } \bigtriangleup_{k} E^{\sigma}(i_{1}, \ldots, i_{k}, \ldots,
i_{p}) \: = \: \mbox{dim } E^{\sigma} (i_{1}, \ldots, i_{k}, \ldots,
i_{p}) \: - \: \mbox{dim } E^{\sigma} (i_{1}, \ldots, i_{k}+1, \ldots,
i_{p})
\end{displaymath}
In particular, the dimension of $E^{[\sigma ]}(I)$ may be negative.
Given these redefinitions, one may then use equations~(\ref{ch}), (\ref{c}) to
calculate the Chern character and total Chern class, respectively.

Sheaf cohomology can be calculated precisely as before,
with the note that each $E^{\sigma}$ should be reinterpreted
as the multifiltration associated to cone $\sigma$, rather
than the intersection of the filtrations on the toric divisors
bounding $\sigma$. 

How are reflexive sheaves described in this language?
According to \cite{kl4}, given a torsion-free sheaf ${\cal E}$ described by a
family of compatible multifiltrations $E^{\sigma}$, its bidual sheaf ${\cal
E}^{\vee \vee}$ is given by the family of multifiltrations
\begin{displaymath}
E^{\vee \vee (\alpha_{1}, \alpha_{2}, \ldots, \alpha_{p})}(i_{1}, i_{2},
\ldots, i_{p}) \: = \: E^{\alpha_{1}}(i_{1}) \cap \cdots
\cap E^{\alpha_{p}}(i_{p})
\end{displaymath}
In particular, if ${\cal E}$ is reflexive (${\cal E} \, = \, {\cal
E}^{\vee \vee}$), then the multifiltration is completely specified by a set of
ordinary filtrations.

How does a reflexive sheaf differ from a locally free sheaf (i.e., a
vector bundle)?  The filtrations defining a vector bundle must satisfy a
compatibility condition (not related to the compatibility condition for
torsion-free sheaves), whereas the filtrations defining a reflexive
sheaf are not required to satisfy any compatibility conditions -- any
random set of filtrations defines a reflexive sheaf.

In particular, note that a locally free sheaf is a special case of a
reflexive sheaf.

Also note that if we view a reflexive sheaf as a locally free sheaf with
singularities, then the singularities are located at
the intersections of toric divisors not satisfying the bundle
compatibility conditions.  As filtrations along any two toric divisors
automatically satisfy the bundle compatibility conditions, we implicitly
verify that singularities of a reflexive sheaf are located at
codimension $3$. 

As a check, note that we have verified the standard fact that reflexive 
sheaves on a smooth complex surface are locally
free, at least for equivariant sheaves on toric surfaces.  
In this language, this follows from the fact that if the maximal
cones in the fan are two-dimensional, then the compatibility condition
on the filtrations is obeyed trivially -- any pair of filtrations is
automatically compatible in the sense needed to define a vector bundle,
and so on a toric surface reflexive sheaves are defined by precisely the
same data as vector bundles.

In general, given a torsion free sheaf ${\cal E}$, there is a natural
map ${\cal E} \rightarrow {\cal E}^{\vee \vee}$.  How can this map be seen in
the present language?  First, note that the sheaf compatibility
conditions in equation~(\ref{tfcompat}) mean $\tau \subseteq \sigma$ implies
\begin{displaymath}
E^{\sigma}(i_{1}, \ldots, i_{p}, j_{1}, \ldots, j_{k}) \: \subseteq \:
E^{\tau}(i_{1}, \ldots, i_{p})
\end{displaymath}
and in particular, for all $k$
\begin{displaymath}
E^{(\alpha_{1}, \cdots, \alpha_{p})}(i_{1}, \ldots, i_{p}) \: \subseteq \:
E^{\alpha_{k}}(i_{k})
\end{displaymath}
so clearly
\begin{displaymath}
E^{(\alpha_{1}, \cdots, \alpha_{p})}(i_{1}, \ldots, i_{p}) \: \subseteq \:
E^{\alpha_{1}}(i_{1}) \cap \cdots \cap E^{\alpha_{p}}(i_{p}) \: = \:
E^{\vee \vee (\alpha_{1}, \cdots, \alpha_{p})}(i_{1}, \ldots, i_{p})
\end{displaymath}

This technology now allows us to gain a better grasp
of the sheaves appearing in extremal transitions
(previous studies of this matter have appeared in \cite{me1}).
Given a set of compatible filtrations defining a vector bundle, by
blowing down one of the toric divisors we are clearly left with a set
of filtrations, which no longer necessarily satisfy the usual
(bundle) compatibility conditions -- so we get a reflexive sheaf.
In other words, when blowing down a toric divisor, there exists a natural
transformation of (equivariant) bundles
into (equivariant) reflexive sheaves.

Note that the sheaves obtained in this manner are not the same sheaves
one obtains by pushforward along the blowdown morphism.
For example, consider an equivariant line bundle on the Hirzebruch surface
${\bf F}_{1}$, the blowup of ${\bf P}^{2}$ at a point.
In Klyachko's formalism, if we blowdown ${\bf F}_{1} \rightarrow {\bf
P}^{2}$, we recover another equivariant line bundle (as reflexive
sheaves on a surface are locally free).  By contrast, if we pushforward
the line bundle along the blowdown morphism, we will get a torsion-free
sheaf, which in general will not be locally free.

\subsubsection{Global Ext} \label{glext}

In principle it is possible to calculate global $\mbox{Ext}$ groups
for equivariant torsion-free sheaves on smooth toric varieties, 
with a calculation analogous
to the sheaf cohomology group calculation in section~\ref{appl}.
For completeness it is mentioned here.

The group $\mbox{(global) } \mbox{Ext}^{i}({\cal E}, {\cal F})$ is 
given by the
limit of the spectral sequence whose first-level terms are
\begin{displaymath}
E_{1}^{p,q} \: = \: \bigoplus_{{\em codim } \, \sigma = p}
\mbox{Ext}^{q} \left( {\cal E}|_{U_{\sigma}} , {\cal F}|_{U_{\sigma}} 
\right)
\end{displaymath}
where $\sigma$ denotes a cone,
and ${\cal E}|_{U_{\sigma}}$ denotes the module of
sections of ${\cal E}$ over the open set $U_{\sigma}$.
As noted earlier, when ${\cal E}$ is equivariant and
torsion-free, this module of sections is precisely the
graded module $E^{\sigma}$ defining the equivariant torsion-free
sheaf ${\cal E}$.

Like the sheaf cohomology groups, the groups $\mbox{Ext}^{i}({\cal E},
{\cal F})$ 
have an isotypic decomposition when ${\cal E}$, ${\cal F}$
are both equivariant:
\begin{displaymath}
\mbox{(global) } \mbox{Ext}^{i}( {\cal E}, {\cal F} )
\: = \: \oplus_{\chi} \mbox{Ext}^{i}( {\cal E}, {\cal F})_{\chi}
\end{displaymath}
If we let $\{ E^{\sigma} \}$, $\{ F^{\sigma} \}$
denote the corresponding sets of data, then
the group $\mbox{Ext}^{i}({\cal E}, {\cal F})_{\chi}$ is
given by the limit of the spectral sequence with first-level
terms
\begin{displaymath}
E_{1}^{p,q} \: = \: \bigoplus_{{\em codim} \, \sigma = p}
\mbox{Ext}^{q} \left( E^{\sigma}, F^{\sigma} \right)_{\chi}
\end{displaymath} 
Unfortunately, it is not clear at present how this can be
made more computationally effective.

As a check, note that when ${\cal E}$ is 
locally free, we recover the sheaf cohomology of
${\cal E}^{\vee} \otimes {\cal F}$.  When ${\cal E}$
is locally free, $E^{\sigma}$ is a freely generated module
for all $\sigma$, and so
\begin{displaymath}
\mbox{Ext}^{q} \left( E^{\sigma} , F^{\sigma} \right) 
\: = \: 0
\end{displaymath}
for all $q > 0$.  Recall that the differential 
\begin{displaymath}
d_{r}: \: E_{r}^{p,q} \: \rightarrow \: E_{r}^{p+r,q-r+1} 
\end{displaymath}
so in particular
\begin{displaymath}
d_{1}: \: E_{1}^{p,q} \: \rightarrow \: E_{1}^{p+1,q}
\end{displaymath}
and our spectral sequence degenerates to the complex
defining sheaf cohomology in section~\ref{sheafcohom}.

\subsubsection{An example on ${\bf P}_{1,1,2}^{2}$} \label{tfwp}

\begin{figure}
\centerline{\psfig{file=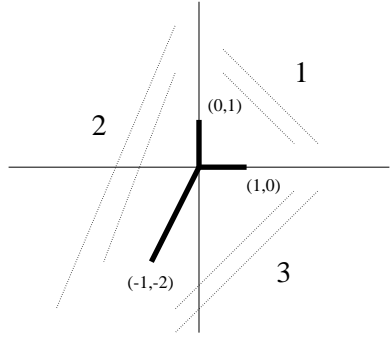,width=2.0in}}
\caption{\label{fanwp} A fan describing the weighted projective
space ${\bf P}_{1,1,2}^{2}$ as a toric variety.}
\end{figure}

First, we shall make some general remarks on the equivariant
structure of modules describing torsion-free sheaves on ${\bf P}_{1,1,2}^{2}$.
The fan for ${\bf P}_{1,1,2}^{2}$ can be taken to have edges
$(1,0)$, $(0,1)$, and $(-1,-2)$, as shown in figure~\ref{fanwp}.  
Denote as cone 1 the cone spanned
by $(1,0)$ and $(0,1)$.  Denote as cone 2 the cone spanned by
$(0,1)$ and $(-1,-2)$.  Denote as cone 3 the cone spanned by
$(-1,-2)$ and $(1,0)$.

To be specific, we shall describe an ideal sheaf on ${\bf P}^{2}_{1,1,2}$
which vanishes to order 1 at the origin of cone 1.

The affine space $U_{1}$ corresponding to cone 1 is 
$\mbox{Spec } {\bf C}[x,y]$,
and a torsion-free module $E^{\sigma_{1}}$ is a bifiltration with inclusions
\begin{eqnarray*}
E^{\sigma_{1}}(i_{1},i_{2}) & \hookrightarrow &
        E^{\sigma_{1}}(i_{1}+1, i_{2}) \\
  & \hookrightarrow & E^{\sigma_{1}}(i_{1},i_{2}+1)
\end{eqnarray*}
The module $M_{1}$ on $U_{1}$ for our example is the ideal
$(x,y) \subset {\bf C}[x,y]$:
\begin{center}
\begin{tabular}{c|cccccc}
 & -2 & -1 & 0 & 1 & 2 & 3 \\ \hline
 & & & $\vdots$ & & & \\
2 & & 0 & ${\bf C}$ & ${\bf C}$ & ${\bf C}$ & \\
1 & & 0 & ${\bf C}$ & ${\bf C}$ & ${\bf C}$ & \\
0 & $\cdots$ & 0 & 0 & ${\bf C}$ & ${\bf C}$ & $\cdots$ \\
-1 & & 0 & 0 & 0 & 0 & \\
 & & & $\vdots$ & & & \\
\end{tabular}
\end{center}

The affine space $U_{2}$ corresponding to cone 2 is 
$\mbox{Spec } {\bf C}[x^{-1},
x^{-2}y]$, and a torsion-free module $E^{\sigma_{2}}$ is a bifiltration
with inclusions 
\begin{eqnarray*}
E^{\sigma_{2}}(i_{1},i_{2}) & \hookrightarrow &
       E^{\sigma_{2}}(i_{1}-1,i_{2}) \\
  & \hookrightarrow & E^{\sigma_{2}}(i_{1}-2,i_{2}+1)
\end{eqnarray*}
The module $M_{2}$ on $U_{2}$ for our example should describe
the structure sheaf on $U_{2}$, and in fact $M_2 = {\bf C}[x^{-1},
x^{-2}y]$:
\begin{center}
\begin{tabular}{c|cccccccc}
 & -5 & -4 & -3 & -2 & -1 & 0 & 1 & 2 \\ \hline
 & & & & $\vdots$ & & & & \\
2 & & ${\bf C}$ & 0 & 0 & 0 & 0 & 0 & \\
1 & & ${\bf C}$ & ${\bf C}$ & ${\bf C}$ & 0 & 0 & 0 & \\
0 & $\cdots$ & ${\bf C}$ & ${\bf C}$ & ${\bf C}$ & ${\bf C}$ & ${\bf C}$
& 0 & $\cdots$ \\
-1 & & 0 & 0 & 0 & 0 & 0 & 0 & \\
 & & & & $\vdots$ & & & & \\
\end{tabular}
\end{center}

The affine space $U_{3}$ corresponding to cone 3 is 
$\mbox{Spec } {\bf C}[y^{-1},
xy^{-1}, x^{2}y^{-1}]$, which contains a ${\bf C}^{2}/{\bf Z}_{2}$
singularity.  A torsion-free module $E^{\sigma_{3}}$ is a bifiltration
with inclusions
\begin{eqnarray*}
E^{\sigma_{3}}(i_{1},i_{2}) & \hookrightarrow &
       E^{\sigma_{3}}(i_{1},i_{2}-1) \\
  & \hookrightarrow & E^{\sigma_{3}}(i_{1}+1,i_{2}-1) \\
  & \hookrightarrow & E^{\sigma_{3}}(i_{1}+2,i_{2}-1)
\end{eqnarray*}
The module $M_{3}$ on $U_{3}$ for our example should describe
the structure sheaf on $U_{3}$, and in fact $M_3 = {\bf C}[y^{-1},
xy^{-1},x^2y^{-1}]$: 
\begin{center}
\begin{tabular}{c|cccccccccc}
 & -2 & -1 & 0 & 1 & 2 & 3 & 4 & 5 & 6 & 7 \\ \hline
 & & & & & $\vdots$ & & & & & \\
1 & & 0 & 0 & 0 & 0 & 0 & 0 & 0 & 0 & \\
0 & & 0 & ${\bf C}$ & 0 & 0 & 0 & 0 & 0 & 0 & \\
-1 & $\cdots$ & 0 & ${\bf C}$ & ${\bf C}$ & ${\bf C}$ & 0 & 0 & 0 & 0 & 
$\cdots$ \\
-2 & & 0 & ${\bf C}$ & ${\bf C}$ & ${\bf C}$ & ${\bf C}$ & ${\bf C}$ & 0 
& 0 & \\
-3 & & 0 & ${\bf C}$ & ${\bf C}$ & ${\bf C}$ & ${\bf C}$ & ${\bf C}$ &
${\bf C}$ & ${\bf C}$ \\
 & & & & & $\vdots$ & & & & \\
\end{tabular}
\end{center}

How are these three bifiltrations glued together?
We shall examine modules over the intersections of the cones.

The affine space $U_{12}$ at the intersection of cones 1 and 2 is
$\mbox{Spec } {\bf C}[x,x^{-1},y]$, and a torsion-free module 
$E^{\sigma_{12}}$ over this space is a bifiltration with inclusions
\begin{eqnarray*}
E^{\sigma_{12}}(i_{1},i_{2}) & \hookrightarrow &
    E^{\sigma_{12}}(i_{1}+1,i_{2}) \\
   & \hookrightarrow & E^{\sigma_{12}}(i_{1}-1,i_{2}) \\
   & \hookrightarrow & E^{\sigma_{12}}(i_{1},i_{2}+1)
\end{eqnarray*}
For our example, the module $M_{12}$ associated to $U_{12}$ is
the ring ${\bf C}[x,x^{-1},y]$:
\begin{center}
\begin{tabular}{c|ccccc}
 & -2 & -1 & 0 & 1 & 2 \\ \hline
 & & & $\vdots$ & & \\
2 & & ${\bf C}$ & ${\bf C}$ & ${\bf C}$ & \\
1 & $\cdots$ & ${\bf C}$ & ${\bf C}$ & ${\bf C}$ & $\cdots$ \\
0 & & ${\bf C}$ & ${\bf C}$ & ${\bf C}$ & \\
-1 & & 0 & 0 & 0 & \\
 & & & $\vdots$ & & \\
\end{tabular}
\end{center}

The affine space $U_{13}$ at the intersection of cones 1 and 3 is
$\mbox{Spec } {\bf C}[x,y,y^{-1}]$, and a torsion-free module
$E^{\sigma_{13}}$ over this space is a bifiltration with inclusions
\begin{eqnarray*}
E^{\sigma_{13}}(i_{1},i_{2}) & \hookrightarrow &
      E^{\sigma_{13}}(i_{1}+1,i_{2}) \\
   & \hookrightarrow & E^{\sigma_{13}}(i_{1},i_{2}+1) \\
   & \hookrightarrow & E^{\sigma_{13}}(i_{1},i_{2}-1) 
\end{eqnarray*}
For our example, the module $M_{13}$ associated to $U_{13}$ is
the ring ${\bf C}[x,y,y^{-1}]$:
\begin{center}
\begin{tabular}{c|cccccc}
 & -2 & -1 & 0 & 1 & 2 & 3 \\ \hline
 & & & $\vdots$ & & & \\
1 & & 0 & ${\bf C}$ & ${\bf C}$ & ${\bf C}$ & \\
0 & $\cdots$ & 0 & ${\bf C}$ & ${\bf C}$ & ${\bf C}$ & $\cdots$ \\
-1 & & 0 & ${\bf C}$ & ${\bf C}$ & ${\bf C}$ & \\
 & & & $\vdots$ & & & \\
\end{tabular}
\end{center}

The affine space $U_{23}$ at the intersection of cones 2 and 3 is 
$\mbox{Spec } {\bf C}[x^{-2}y, x^{2}y^{-1}, xy^{-1}]$,
and a torsion-free module $E^{\sigma_{23}}$ over this space
is a bifiltration with inclusions
\begin{eqnarray*}
E^{\sigma_{23}}(i_1, i_2) & \hookrightarrow & 
        E^{\sigma_{23}}(i_1 - 2, i_2 + 1) \\
    & \hookrightarrow & E^{\sigma_{23}}(i_1 + 2, i_2 - 1) \\
    & \hookrightarrow & E^{\sigma_{23}}(i_1 + 1, i_2 - 1) 
\end{eqnarray*}
For our example, the module $M_{23}$ associated to $U_{23}$ is
the ring ${\bf C}[x^{-2}y,x^2y^{-1},xy^{-1}]$:
\begin{center}
\begin{tabular}{c|ccccccccc}
 & -5 & -4 & -3 & -2 & -1 & 0 & 1 & 2 & 3 \\ \hline
 & & & & & $\vdots$ & & & & \\ 
2 & & ${\bf C}$ & 0 & 0 & 0 & 0 & 0 & 0 & \\
1 & & ${\bf C}$ & ${\bf C}$ & ${\bf C}$ & 0 & 0 & 0 & 0 & \\
0 & $\cdots$ & ${\bf C}$ & ${\bf C}$ & ${\bf C}$ & ${\bf C}$ & ${\bf C}$ &
0 & 0 & $\cdots$ \\
-1 & & ${\bf C}$ & ${\bf C}$ & ${\bf C}$ & ${\bf C}$ & ${\bf C}$ & ${\bf C}$
& ${\bf C}$ & \\
 & & & & & $\vdots$ & & & & \\
\end{tabular}
\end{center}

In passing we should note that toric varieties can
have singularities worse than orbifold singularities -- for
example, an affine conifold singularity in three dimensions
is toric.  For completeness, we very briefly review
the conifold singularity below.

The affine conifold singularity in three dimensions
is the hypersurface $ad - bc = 0$ in ${\bf C}[a,b,c,d]$.
It can be described as a (noncompact) toric variety.
The fan has edges $(-1,0,1)$, $(0,-1,1)$, $(-1,1,0)$, and
$(1,-1,0)$, describing a single cone in three dimensions
(in the region $z \geq 0$, $x + z \geq 0$, $y + z \geq 0$,
and $x + y + z \geq 0$).  Its dual cone has the edges
$(0,0,1)$, $(1,0,1)$, $(0,1,1)$, and $(1,1,1)$.

Any torsion-free sheaf on this affine space will be described
in bulk by a ${\bf Z}^{3}$-graded module $E$, with inclusions
\begin{eqnarray*}
E(i_{1}, i_{2}, i_{3}) & \hookrightarrow & E(i_{1}, i_{2}, i_{3}+1) \\
   & \hookrightarrow & E(i_{1}+1, i_{2}, i_{3}+1) \\
   & \hookrightarrow & E(i_{1}, i_{2}+1, i_{3}+1) \\
   & \hookrightarrow & E(i_{1}+1, i_{2}+1, i_{3}+1)
\end{eqnarray*}

\subsubsection{Cartier divisors vs Weil divisors} \label{cartierweil}

On a smooth variety, for any divisor there exists a corresponding
line bundle, a standard fact well-known to physicists.
On singular varieties, not all divisors correspond to line
bundles.  In general, an arbitrary divisor is called a Weil
divisor, and a Weil divisor that happens to correspond to a line
bundle (rather than some other rank 1 sheaf) is called a 
Cartier divisor\footnote{Technical fiends will note we are deliberately
being sloppy, for reasons of readability.}.
Essentially, a divisor will not necessarily define a line bundle
when it intersects a singularity.

Put another way, on any normal variety (of which toric varieties
are examples) an arbitrary (Weil) divisor will yield a reflexive
rank 1 sheaf \cite{reid,cox2}.  In the special case that the reflexive
rank 1 sheaf is a line bundle, we say the Weil divisor is Cartier.
On a smooth variety, all reflexive rank 1 sheaves are line bundles,
thus on a smooth variety all Weil divisors are Cartier.

In brief, for a $T$-invariant divisor $D$ on an affine toric space 
$U_{\sigma}$ to be Cartier, 
there must exist a character\footnote{
Technical fiends will recognize that the character $\chi$
is the location of the generator of the principal fractional ideal.}
$\chi$ such that
\begin{displaymath}
D \: = \: \sum_{\alpha} \langle v_{\alpha}, \chi \rangle \, D_{\alpha}
\end{displaymath}
where $D_{\alpha}$ is the divisor associated to the edge of
the fan $v_{\alpha}$.

Let us work out a specific example in detail.
Consider the affine space $U = {\bf C}^{2}/{\bf Z}_{2}$ described
as a toric variety -- the corresponding fan has edges
\begin{eqnarray*}
v_1 & = & (2,-1) \\
v_2 & = & (0,1) 
\end{eqnarray*}
The affine space $U = \mbox{Spec } {\bf C}[x, xy, xy^{2} ] 
= \mbox{Spec } {\bf C}[x,y,z]/(xy = z^2 )$.
The affine space corresponding to divisor 1 is
$U_{1} = \mbox{Spec } {\bf C}[x y^2, x^{-1} y^{-2}, y^{-1} ]$,
and the affine space corresponding to divisor 2 is
$U_{2} = \mbox{Spec } {\bf C}[x, x^{-1}, y]$.

Now, let us consider the ideal sheaf defined by the principal
ideal $(x)$.  Since this ideal sheaf is defined by a principal
ideal, it should correspond to a Cartier divisor.  Indeed,
the corresponding module on $U$ is
\begin{center}
\begin{tabular}{c|cccccc}
 & -2 & -1 & 0 & 1 & 2 & 3 \\ \hline
 & & & $\vdots$ & & & \\
4 & & 0 & 0 & 0 & ${\bf C}$ & \\
3 & & 0 & 0 & 0 & ${\bf C}$ & \\
2 & & 0 & 0 & ${\bf C}$ & ${\bf C}$ & \\
1 & $\cdots$ & 0 & 0 & ${\bf C}$ & ${\bf C}$ & $\cdots$ \\
0 & & 0 & ${\bf C}$ & ${\bf C}$ & ${\bf C}$ & \\
-1 & & 0 & 0 & 0 & 0 & \\
 & & & $\vdots$ & & & \\
\end{tabular}
\end{center}
It is manifestly obvious that this module is freely
generated (and therefore a line bundle), 
with generator located at $i_1 = i_2 = 0$.

As this ideal sheaf is a line bundle, it should be
defined by some divisor on the toric variety.
Which divisor?
To determine the corresponding divisor, we first
find the modules $M_1$, $M_2$ over the affine spaces
$U_1$, $U_2$.  The module $M_1$ is given by
\begin{center}
\begin{tabular}{c|ccccccc} 
 & -2 & -1 & 0 & 1 & 2 & 3 & 4 \\ \hline
 & & & & $\vdots$ & & & \\
4 & & 0 & 0 & 0 & 0 & ${\bf C}$ & \\
3 & & 0 & 0 & 0 & 0 & ${\bf C}$ & \\
2 & & 0 & 0 & 0 & ${\bf C}$ & ${\bf C}$ & \\
1 & & 0 & 0 & 0 & ${\bf C}$ & ${\bf C}$ & \\
0 & $\cdots$ & 0 & 0 & ${\bf C}$ & ${\bf C}$ & ${\bf C}$ & $\cdots$ \\
-1 & & 0 & 0 & ${\bf C}$ & ${\bf C}$ & ${\bf C}$ & \\
-2 & & 0 & ${\bf C}$ & ${\bf C}$ & ${\bf C}$ & ${\bf C}$ & \\
-3 & & 0 & ${\bf C}$ & ${\bf C}$ & ${\bf C}$ & ${\bf C}$ & \\
-4 & & ${\bf C}$ & ${\bf C}$ & ${\bf C}$ & ${\bf C}$ & ${\bf C}$ & \\
 & & & & $\vdots$ & & & \\
\end{tabular}
\end{center}
and the module $M_2$ is given by
\begin{center}
\begin{tabular}{c|ccccc}
 & -2 & -1 & 0 & 1 & 2 \\ \hline
 & & & $\vdots$ & & \\
2 & & ${\bf C}$ & ${\bf C}$ & ${\bf C}$ & \\
1 & & ${\bf C}$ & ${\bf C}$ & ${\bf C}$ & \\
0 & $\cdots$ & ${\bf C}$ & ${\bf C}$ & ${\bf C}$ & $\cdots$ \\
-1 & & 0 & 0 & 0 & \\
 & & & $\vdots$ & & \\
\end{tabular}
\end{center}

Now, in each case, to determine the component of the divisor
along $D_{i}$ we compute $\langle v_{i}, \chi_{i} \rangle$
where $\chi_{i}$ is the location of the generator of $M_{i}$.
(Each generator is only defined up to an element of $v_{i}^{\perp}$,
but of course that will not matter for getting a number.)
It is easy to show that 
\begin{eqnarray*}
\langle v_{1}, \chi_1 \rangle & = & 2 \\
\langle v_2, \chi_2 \rangle & = & 0 
\end{eqnarray*}
so the ideal sheaf defined by the ideal $(x)$ 
is precisely the line bundle with divisor $ -2 D_1$.
(Note we can choose $\chi_{1} = \chi_{2}$,
in accordance with the fact this divisor is Cartier.)
Indeed, it is an exercise in \cite{fulton} that
$-2 D_1$ is a Cartier divisor in this example.

Now, in this example, the divisor $- D_1$ is not Cartier,
but merely Weil.  What sheaf does this correspond to?
We shall simply state the result that the
corresponding module on $U$ is
\begin{center}
\begin{tabular}{c|cccccc}
 & -1 & 0 & 1 & 2 & 3 & 4 \\ \hline
 & & & $\vdots$ & & & \\
4 & & 0 & 0 & 0 & ${\bf C}$ & \\
3 & & 0 & 0 & ${\bf C}$ & ${\bf C}$ & \\
2 & & 0 & 0 & ${\bf C}$ & ${\bf C}$ & \\
1 & $\cdots$ & 0 & ${\bf C}$ & ${\bf C}$ & ${\bf C}$ & $\cdots$ \\
0 & & 0 & ${\bf C}$ & ${\bf C}$ & ${\bf C}$ & \\
-1 & & 0 & 0 & 0 & 0 & \\
 & & & $\vdots$ & & & \\
\end{tabular}
\end{center}
This module is precisely the ideal $(x,xy)$.
It should be obvious that this module is not freely generated,
but rather is defined by two generators (at $(1,0)$ and $(1,1)$)
and one relation.  (Since the module is not freely generated,
the rank 1 sheaf does not correspond to a line bundle.) 
Although this module is not freely generated,
it is easy to check that the modules $M_{i}$ associated to either
toric divisor are freely generated.
Also, note that this module is reflexive.
On a smooth variety, a reflexive rank 1 sheaf
is precisely a line bundle -- here we see explicitly that 
on a singular variety
we can have reflexive rank 1 sheaves which are not bundles.

\section{Moduli spaces of equivariant sheaves}  \label{equivmodspace}

To construct a moduli space of, say, equivariant reflexive sheaves,
we should take the space of all filtrations of fixed Chern classes
and quotient by $GL(r,{\bf C})$ equivalences.  In doing so,
we would find that some of the sheaves should be omitted before
quotienting in order to get a well-behaved (for example, Hausdorff)
result.  This modified quotient is known formally as a GIT 
(Geometric Invariant Theory) quotient,
and the sheaves omitted before quotienting are labelled ``unstable."

In many ways GIT quotients\footnote{See appendix~\ref{gitap}.}
are the prototypes for moduli spaces, in that many features of moduli
space problems have analogues within GIT quotients.  In this section
we shall see that in the special case of moduli spaces of equivariant
sheaves, the moduli spaces are precisely GIT quotients.

In section~\ref{stability} we begin by describing Mumford-Takemoto
stability -- in other words, we describe the sheaves we wish to
include in any given moduli space.  In section~\ref{mers} we work
out the moduli space of equivariant reflexive sheaves, as a GIT
quotient of a closed subset of a product of flag manifolds,
in considerable detail.  In particular we show explicitly that
the GIT quotient realizes Mumford-Takemoto stability.  In 
section~\ref{meb} we describe moduli spaces of equivariant
bundles, and study orbifold singularities occurring in
the moduli space of $G$-bundles.  In section~\ref{metfs} we briefly
speak to moduli spaces of equivariant torsion-free sheaves,
and in section~\ref{rrtf} we remark on the interrelation of
the moduli spaces of equivariant reflexive and torsion-free sheaves. 

Much of section~\ref{geners} is a generalization of results
of A. A. Klyachko \cite{kl3}.

\subsection{Stability} \label{stability}

In order to get a well-behaved moduli space, one often must first
remove some nongeneric, badly-behaved objects.  This is true
for moduli spaces of sheaves, for example.  Sheaves are classified
as being stable, semistable, or unstable, and the unstable sheaves
are removed from consideration before forming the moduli space.

More precisely, when forming a moduli space we intuitively
want to quotient some space, call it ${\cal T}$, by the action
of some reductive algebraic group $G$.  In performing the quotient,
we often find that some (nongeneric) points are fixed by a subgroup
of $G$ of dimension greater than zero.  
The orbits of such points are in the closure of other orbits,
so if we included them when taking the quotient, the resulting
space would not be Hausdorff.
To get a well-behaved
moduli space we omit them.  As these points are typically nongeneric,
they are labelled ``unstable," and other points are either
``semistable" or just ``stable."  (Intuitively, a generic perturbation
of an unstable point will yield a semistable point, thus the notation.)
We shall define the relevant notion of stability 
(Mumford-Takemoto stability) for sheaves
momentarily, and then later in this section will demonstrate
explicitly that Mumford-Takemoto stability is realized in our
construction of moduli spaces.

Recall a torsion-free sheaf ${\cal E}$ on a K\"{a}hler variety is said to be
Mumford-Takemoto stable
\cite{okonek,takemoto} precisely when for any proper 
coherent subsheaf ${\cal F}
\subset {\cal E}$ such that $0 < \mbox{rank } {\cal F} < \mbox{rank }
{\cal E}$ and ${\cal E}/{\cal F}$ is torsion-free, we have
\begin{equation}
\frac{ c_{1}({\cal F}) \cup J^{n-1} }{ \mbox{rank } {\cal F} } <
\frac{ c_{1}({\cal E}) \cup J^{n-1} }{ \mbox{rank } {\cal E} }
\end{equation}
and semistable if
\begin{equation}
\frac{ c_{1}({\cal F}) \cup J^{n-1} }{ \mbox{rank } {\cal F} } \leq
\frac{ c_{1}({\cal E}) \cup J^{n-1} }{ \mbox{rank } {\cal E} }
\end{equation}
where $J$ is the K\"{a}hler form,
and $n$ is the complex dimension of the
K\"{a}hler variety.
The right side of the equation above is often called the slope of ${\cal
E}$, and is often denoted $\mu({\cal E})$. 

Note this means that if ${\cal E}$ is an $SL(r, {\bf C})$ bundle
($c_{1}({\cal E}) = 0$), then if ${\cal E}$ has sections it can be at
best semistable, not strictly stable.  This is because the section
defines a map ${\cal O} \rightarrow {\cal E}$, so we have a subbundle
${\cal F} = {\cal O}$ such that $\mu({\cal F}) = 0 = \mu({\cal E})$.

In addition to Mumford-Takemoto stability, there are other
(inequivalent) notions of stability for torsion-free coherent sheaves
(perhaps most prominently, Gieseker stability).  However, these other
notions are less relevant for physics 
than Mumford-Takemoto stability (see section~\ref{revhet}), 
so we will not discuss them
here.

\subsection{Moduli spaces of equivariant reflexive sheaves}
\label{mers}

The simplest moduli spaces to study are moduli spaces of
equivariant reflexive sheaves,
so that is where we shall begin.  After describing the
construction abstractly, we will work through a specific example
in great detail.  

In passing, we should clarify a point that may confuse the reader.
As the moduli spaces we construct here are moduli spaces of
equivariant sheaves (of fixed equivariant structure), the reader may wonder 
how these are related to moduli spaces of not-necessarily-equivariant
sheaves.  In fact, the forgetful map that ``forgets" the
equivariant structure is one-to-one on individual components
(of fixed equivariant structure) of a moduli space.  
More precisely \cite{kl2}, if we denote
a trivial line bundle with equivariant structure defined by a character $\chi$
by\footnote{The trivial bundle ${\cal O}(\chi)$ has
$T$-invariant divisor $D = \sum_{\alpha} \langle v_{\alpha}, \chi \rangle
\, D_{\alpha}$, where the $v_{\alpha}$ are the fan edges corresponding
to the toric divisors $D_{\alpha}$.}
${\cal O}(\chi)$, then if ${\cal E}$ is an indecomposable\footnote{
For a sheaf to be indecomposable means it cannot be written globally
as the direct sum of two subsheaves.} sheaf, all equivariant
structures on ${\cal E}$ can be obtained by tensoring with
characters as ${\cal E} \otimes {\cal O}(\chi)$.  (For
example, the line bundles ${\cal O}(D_{x})$ and ${\cal O}(D_{y})$
on ${\bf P}^2$ have distinct equivariant structures,
but we can get ${\cal O}(D_{y})$ from ${\cal O}(D_{x})$
by tensoring with ${\cal O}(D_{y} - D_{x})$,
which is simply the structure sheaf with a nontrivial 
equivariant structure.)  Thus, a moduli space
of equivariant sheaves of fixed equivariant structure sits naturally inside
a moduli space of not-necessarily-equivariant sheaves.

Subsection~\ref{geners} has been described
previously by A. A. Klyachko \cite{kl3} for the special case
of equivariant vector bundles on ${\bf P}^{2}$.

\subsubsection{Generalities} \label{geners}

By now the reader should be able to make a rough approximation
to the form of such moduli spaces.  One associates a parabolic
subgroup (paired with an ample line bundle on the corresponding
partial flag manifold) to each toric divisor, of a general form fixed by
the Chern classes.  The space of such parabolics on a single toric
divisor is a (partial) flag manifold, so the space of all the parabolics is
a product of (partial) flag manifolds (one for each toric divisor).
Define ${\cal T}_{r} \subset \prod_{\alpha} G/P^{\alpha}$,
associated to parabolics defining reflexive sheaves
of fixed Chern classes.  (Although any set of parabolics
will define a reflexive sheaf, for certain nongeneric flags the
Chern classes will change, and these we shall exclude.)
Then, recall that equivalent reflexive sheaves are defined
by sets of parabolics differing by an overall $G$-rotation,
so loosely speaking the moduli space should look like
the quotient of ${\cal T}_{r}$ by $G$.

This description is almost correct -- the only problem is that
we have been too loose in our description of the quotient.
The correct quotient to take is not an ordinary quotient, but
rather a GIT quotient\footnote{See appendix~\ref{gitap}.},
in which some ``bad" points are omitted.  
In particular, we do not want to consider all possible 
equivariant reflexive sheaves, but only those which are 
Mumford-Takemoto semistable. 

Note that to construct ${\cal T}_{r}$, we have excluded certain
nongeneric points (whose Chern classes differ) before taking the
GIT quotient.  The resulting moduli space of equivariant reflexive sheaves
is noncompact.

The description of Mumford-Takemoto stability in the equivariant language
presented earlier should be clear.
Given an equivariant reflexive sheaf ${\cal E}$ defined by 
filtrations $E^{\alpha}(i)$,
we say ${\cal E}$ is stable if for any nontrivial subspace\footnote{When
testing stability of a reflexive sheaf, it is sufficient to check
reflexive subsheaves \cite{rfprivate}, because if
$S$ is a destabilizing subsheaf of a reflexive sheaf ${\cal E}$, then
$S^{\vee \vee}$ is also a destabilizing subsheaf of ${\cal E}$, 
and $S^{\vee \vee}$
is reflexive.} $0 \neq
F \subset E$, we have
\begin{equation}
\frac{1}{ \mbox{dim } F} \sum_{i \in {\bf Z}, \alpha \in | \Sigma |}
\, i \, \mbox{dim } F^{[\alpha ]}(i) \, \left( D_{\alpha} \cup J^{n-1}
\right) <
\frac{1}{\mbox{dim } E} \sum_{i \in {\bf Z}, \alpha \in | \Sigma |}
\, i \, \mbox{dim } E^{[\alpha ]}(i) \, \left( D_{\alpha} \cup J^{n-1}
\right)
\end{equation}
and similarly for semistable reflexive sheaves.
We have defined $F^{\alpha}(i) \, = \, F \cap E^{\alpha}(i)$,
and
\begin{displaymath}
F^{[\alpha ]}(i) \: = \: \frac{ F^{\alpha}(i) }{ F^{\alpha}(i+1) }
\end{displaymath}

The reader may wonder why it is sufficient to check equivariant subsheaves,
rather than all subsheaves, in order to test stability.  The demonstration
of this is quite straightforward \cite{kl3}.  Let ${\cal E}$ be an
equivariant sheaf, then its Harder-Narasimhan filtration\footnote{See
appendix~\ref{hsfilt}.} is also equivariant (in the sense that
each subsheaf in the filtration is equivariant).  (Were this not the
case, then by acting on each element of the filtration with the
$({\bf C}^{\times})^{n}$ defining the toric variety, we could
produce a family of Harder-Narasimhan filtrations of ${\cal E}$,
but the Harder-Narasimhan filtration is unique.)  Now, the 
Harder-Narasimhan filtration is trivial precisely when ${\cal E}$ is
semistable, so as all its elements are equivariant, it clearly
suffices to check equivariant subsheaves when testing stability.

Now, we want to define a GIT quotient of ${\cal T}_{r}$
that acts by first restricting to the Mumford-Takemoto semistable 
sheaves in ${\cal T}_{r}$,
then mods out by $G$.  In fact, this is a straightforward generalization
of a standard example of GIT quotients (see \cite[section 4.4]{git}
or \cite[section 4.6]{newstead}).  The example in the references
cited corresponds to a GIT quotient of the space of reflexive
sheaves on ${\bf P}^{N-1}$, such that on each toric divisor $\alpha$,
$E^{[\alpha]}(i)$ is nontrivial only for $i = 0,1$, i.e., that
the parabolic associated to $\alpha$ is maximal.  In particular,
Mumford-Takemoto stability is realized naturally -- the notion of
stability defining the GIT quotient (see appendix~\ref{gitap}) 
is identical to Mumford-Takemoto stability.

The generalization of this example to describe moduli spaces
of other reflexive sheaves on arbitrary toric varieties is
straightforward.  Recall (appendix~\ref{gitap}) that to
construct a GIT quotient of a compact space we must specify
an ample, $G$-linearized line bundle on the space.
As ${\cal T}_{r} \subset \prod_{\alpha} G/P^{\alpha}$,
to construct an ample line bundle on ${\cal T}_{r}$ we will
first specify a line bundle on each factor.  
The relevant $G$-linearized ample line bundle on $G/P^{\alpha}$
is precisely the ample line bundle $L_{\alpha}$ which together
with $P^{\alpha}$ encodes all the information in the filtration
(as described in sections~\ref{evb}, \ref{agg}).

To take into
account the choice of K\"{a}hler form $J$, tensor together
pullbacks of the $L_{\alpha}$ to ${\cal T}_{r}$ weighted
by different factors.  By working in a dense
subset\footnote{More precisely, the subset intersecting
$\mbox{Num} \otimes_{{\bf Z}} {\bf Q}$ rather than
$\mbox{Num} \otimes_{{\bf Z}} {\bf R}$.} of the K\"{a}hler cone,
we can demand that there exist a fixed $\lambda \in {\bf R}$
such that for all $\alpha$, $D_{\alpha} \cup J^{n-1} = \lambda n_{\alpha}$,
where $n_{\alpha} \in {\bf Z}$.

We can now describe the ample $G$-linearized line bundle on ${\cal T}_{r}$
defining the relevant GIT quotient.  Let $\pi_{\alpha}: \prod_{\beta}
G/P^{\beta} \rightarrow G/P^{\alpha}$ be the canonical projection,
then the ample $G$-linearized line bundle $L$ on ${\cal T}_{r}$
is
\begin{displaymath}
L \: = \: \bigotimes_{\alpha} \, \pi_{\alpha}^{*} L_{\alpha}^{n_{\alpha}}
\end{displaymath}
Note that this is equivalent to composing the Segre embedding
of a project of projective spaces with the projective embeddings
defined by (tensor powers of) the ample line bundles $L_{\alpha}$.

\subsubsection{A simple example} \label{gitex}

Before going on, we shall work through a simple example
(\cite[section 4.4]{git}, \cite[section 4.6]{newstead}) in detail,
in which we shall demonstrate explicitly that the GIT quotient
described correctly reproduces Mumford-Takemoto stability.
(Our presentation closely follows that of both references.
As these methods are new to the physics literature, however,
we felt it appropriate to restate the (standard) derivation.)
Let us consider equivariant rank-$(n+1)$ reflexive sheaves 
on ${\bf P}^{N-1}$, with equivariant structure such that
for all toric divisors $\alpha$
\begin{displaymath}
\mbox{dim } E^{[\alpha]}(i) \: = \: \left\{
\begin{array}{ll}
q+1 & i = 1 \\
n-q & i = 0 \\
0   & i \neq 0,1
\end{array} \right.
\end{displaymath}
Clearly, each parabolic subgroup $P^{\alpha}$ is maximal,
and $G/P^{\alpha}$ is the Grassmannian
$G_{q+1}({\bf C}^{n+1})$ of $(q+1)$-dimensional
subspaces of ${\bf C}^{n+1}$ (or, equivalently, the space of
$q$-dimensional linear subspaces of ${\bf P}^{n}$).
The ample line bundle ${\cal L}$ defined over each $G/P^{\alpha}$
is in fact the very ample line bundle defining the Pl\"{u}cker 
embedding.
We shall assume that the K\"ahler form $J$ is such that 
$D_{\alpha} \cup J^{N-2} = 1$ for all toric divisors $\alpha$.

Now, we want to describe a GIT quotient that reproduces Mumford-Takemoto
stability, and in fact this will be quite straightforward.

First, recall the definition of the Pl\"{u}cker embedding
\begin{displaymath}
p: G_{q+1}({\bf C}^{n+1}) \rightarrow {\bf P}^{d}
\end{displaymath}
where
\begin{displaymath}
d \: = \:  \left( \begin{array}{c}
n + 1 \\ q+1  \end{array} \right) - 1
\end{displaymath}
Let $L$ be a $(q+1)$-plane in ${\bf C}^{n+1}$, spanned by vectors
$v_{0}, \cdots, v_{q} \in {\bf C}^{n+1}$, then the Pl\"{u}cker
embedding sends this plane to the multivector $v_{0} \wedge \cdots
\wedge v_{q}$.
Equivalently, if $(v_{j 0}, \cdots, v_{j q})$ are the
coordinates of the vectors defining $L$, forming a $(q+1) \times (n+1)$
matrix $\Lambda$, then $p_{i_{0} \cdots i_{q}}$
($ 0 \leq i_{0} < i_{1} < \cdots < i_{q} \leq n$)
is the determinant of the $(q+1) \times (q+1)$ submatrix of $\Lambda$
formed with columns $i_{0}, \cdots i_{q}$.  Note the $p_{i_0, \cdots,
i_q}$ are homogeneous coordinates of the projective space
in which the Grassmannian is embedded.

Now, in order to see explicitly that the choice of polarization\footnote{
Recall from appendix~\ref{gitap} that when computing the GIT
quotient of a variety ${\cal T}$ by an algebraic group $G$,
the ample line bundle (with $G$-linearization) on ${\cal T}$ 
defining the GIT quotient
is known as the polarization.}
above yields Mumford-Takemoto stability, we shall use the numerical
criterion for stability (appendix~\ref{gitap}).
Moreover, as ${\cal T}_{r} \subset \prod G_{q+1}({\bf C}^{n+1})$,
if $x_{\alpha} \in G_{q+1}({\bf C}^{n+1}) \, \forall \alpha$,
then we know (appendix~\ref{gitap}) 
\begin{displaymath}
\mu( \{ x_{\alpha} \}, \lambda) \: = \: \sum_{\alpha} \mu( x_{\alpha},
\lambda )
\end{displaymath}
so we shall first compute $\mu( x_{\alpha}, \lambda)$ for a single
Grassmannian.   (The reader should not confuse the $\mu$ appearing
in the numerical criterion of stability with the $\mu$ defining
the slope of a bundle.  The notation is unfortunate, but standard
in the mathematics literature, so we have decided to stay with
their conventions.)

The group whose action we wish to mod out is $GL(n+1,{\bf C})$;
however it should be clear that the overall ${\bf C}^{\times}$
breathing mode will simply act to multiply all the homogeneous
coordinates of any projective embedding by an overall factor,
which of course is irrelevant.  Thus, it is sufficient
to consider only $SL(n+1,{\bf C})$.

Let $\lambda$ be a one-parameter subgroup of $SL(n+1, {\bf C})$
defined by
\begin{displaymath}
\lambda(t) \: = \: \mbox{diag } ( t^{r_{0}}, t^{r_{1}}, \cdots
t^{r_{n}} )
\end{displaymath}
with $\sum r_{i} = 0$, and $r_0 \geq r_1 \geq \cdots \geq r_n$,
with not all the $r_i = 0$.
For given $L \in G_{q+1}({\bf C}^{n+1})$, by definition of $\mu$
we have
\begin{displaymath}
\mu(L, \lambda) \: = \: \mbox{max } \left\{
- \sum_{j = 0}^{q} r_{i_j} \, | \, p_{i_0 \cdots i_q}(L) \neq 0 \right\}
\end{displaymath}
(Note that this result for $\mu$ depends implicitly upon 
our choice of projective embedding (here, the Pl\"{u}cker embedding),
and so in particular $\mu$ depends on ${\cal L}$.)

In order to compute explicitly, we shall work in coordinates.
Let $e_0, \cdots, e_n$ be a basis for ${\bf C}^{n+1}$, and
define subspaces $L_{i} \in {\bf C}^{n+1}$ as 
\begin{eqnarray*}
L_{-1} & = & \emptyset \\
L_{i} & = & \mbox{subspace generated by images of $e_0, \cdots, e_i$
in } {\bf C}^{n+1} 
\end{eqnarray*}

Note that for any $L \in G_{q+1}({\bf C}^{n+1})$ and any
integer $j$, $0 \leq j \leq q$, there exists a unique integer
$\mu_{j}$ such that $\mbox{dim }(L \cap L_{\mu_j}) = j+1$
and $\mbox{dim }(L \cap L_{\mu_j -1}) = j$.
Clearly
\begin{displaymath}
0 \leq \mu_0 < \mu_1 < \cdots < \mu_q \leq n
\end{displaymath}
and $L$ is spanned by the rows of a matrix of the form
\begin{displaymath}
\left[
\begin{array}{cccccccccc}
a_{0 0} & \cdots & a_{0 \mu_0} & 0 & \cdots & & & & \cdots & 0 \\
a_{1 0} & \cdots & & \cdots & a_{1 \mu_1} & 0 & \cdots & & \cdots & 0 \\
\vdots & & & & & & & & & \vdots \\
a_{q 0} & \cdots & & & & \cdots & a_{q \mu_q} & 0 & \cdots & 0 
\end{array}
\right]
\end{displaymath}
with $a_{j \mu_j} \neq 0 \: \forall \, j$.
In this choice of coordinates, $p_{i_0 \cdots i_q}(L) = 0$
if $i_j > \mu_j$ for any value of $j$ (as, in that event,
the $j$th row would be a linear combination of the previous rows,
as all its entries past the $j$th column would be zero,
so the determinant must vanish).  It should also be
clear that $p_{\mu_0 \cdots \mu_q}(L) \neq 0$,
and in fact we can now compute
\begin{eqnarray*}
\mu(L, \lambda) & = & - \sum_{j=0}^{q} r_{\mu_j} \\
   & = & - \sum_{i=0}^{n} r_{i} \left[
\mbox{dim }(L \cap L_i) - \mbox{dim }(L \cap L_{i-1}) \right] \\
& = & - (q+1) r_n \: + \: \sum_{i=0}^{n-1} \left[
\mbox{dim }(L \cap L_i)  \right] ( r_{i+1} - r_i  )
\end{eqnarray*}

Now, recall that ${\cal T}_{r}$ is not a subset of a single Grassmannian,
but rather a product of them.  As a result, the $\mu$ 
we need to consider to apply the numerical criterion for stability is
\begin{eqnarray*}
\mu( \{ L^{(1)}, \cdots, L^{(N)} \}, \lambda) & = &
\sum_{j=1}^{N} \mu( L^{(j)}, \lambda ) \\
 & = & -N (q+1) r_n \: + \: \sum_{i=0}^{n-1} \left\{
\sum_{j=1}^{N} \left[ \mbox{dim }( L^{(j)} \cap L_i )  \right]
\right\} (r_{i+1} - r_{i})
\end{eqnarray*}

In order for a set of subspaces $\{ L^{(1)}, \cdots, L^{(N)} \}
\in {\cal T}_{r}$
to be stable, we must check that $\mu$ is positive.  
In fact, as $\mu$ is linear in the $r_i$, it will be positive
for all allowable $r_i$ (meaning, for all one-parameter subgroups
of $SL(n+1,{\bf C})$) precisely when it is positive for the
extreme cases
\begin{displaymath}
r_0 = \cdots = r_p = n-p, \: r_{p+1} = \cdots = r_n = -(p+1)
\end{displaymath}
for $0 \leq p \leq n+1$.

Thus, $\mu > 0$ for all one-parameter subgroups precisely when
\begin{displaymath}
N(q+1)(p+1) \: - \: (n+1) \sum_{j=1}^{N} \left[ \mbox{dim }(L^{(j)} \cap
L_p) \right] \: > \: 0
\end{displaymath}
for all $0 \leq p \leq n-1$.

Since every $(p+1)$-dimensional subspace of ${\bf C}^{n+1}$ is
equivalent under $SL(n+1, {\bf C})$ to $L_p$,
we have that a set of flags $\{ L^{(1)}, \cdots, L^{(N)} \}
\in {\cal T}_{r}$
is stable precisely when for all proper subspaces $L$ of
${\bf C}^{n+1}$,
\begin{displaymath}
\frac{1}{\mbox{dim } L} \sum_{j=1}^{N} \left[ \mbox{dim }(L^{(j)} \cap L) 
\right] \: < \: \frac{1}{n+1} (q+1) (N)
\end{displaymath}
and the reader should immediately recognize this as the statement
of Mumford-Takemoto stability for the example in question.

If the reader found this derivation overly technical, he may
wish to repeat it for the special case that the Grassmannians
are all projective spaces, in which case the Pl\"{u}cker embedding
is the identity map.

\subsection{Moduli spaces of equivariant bundles} \label{meb}

These moduli spaces are closely related to the moduli spaces
of equivariant reflexive sheaves.  Both bundles and reflexive
sheaves are described by associating
parabolic subgroups of $G$ (paired with ample line bundles on
corresponding partial flag manifolds) to each toric divisor.  
The difference is that in the case of bundles, there is a compatibility
condition
that must be satisfied
(equation~(\ref{compatpara}) for smooth toric varieties).

To construct a moduli space of equivariant bundles, therefore,
we proceed almost as in the previous section, except that
instead of performing a GIT quotient of 
${\cal T}_{r} \subset \prod G/P^{\alpha}$,
we perform a GIT quotient of the subspace ${\cal T}_{b} \subseteq
{\cal T}_{r}$ consisting of tuples of parabolic subgroups satisfying
the compatibility condition.  The ample $G$-linearized line
bundle defining the appropriate GIT quotient is then just the
restriction to ${\cal T}_{b}$ of the line bundle constructed
in the previous section.

In passing, we should note that in general, ${\cal T}_{b}$
will not even be of the same dimension as ${\cal T}_{r}$.
(In fact, ${\cal T}_{b}$ will never be a dense subset
of ${\cal T}_{r}$ except in the special case ${\cal T}_{b} = {\cal T}_{r}$.)
In many cases, equivariant bundles will be highly
nongeneric in a moduli space of equivariant reflexive sheaves.

Let us make a few observations on singularities in the moduli space of 
equivariant
$G$-bundles \cite{alleninprep}, relevant for generic\footnote{
For certain nongeneric K\"ahler forms, a moduli space of principal
$G$-bundles, for any $G$, (and for that matter a moduli space of
reflexive sheaves) will have singularities much worse than the
orbifold singularities described above.  The problem essentially
arises when, for certain nongeneric K\"ahler forms, one gets extra
strictly semistable sheaves.  For example, we will discuss in 
section~\ref{kcsubstruc} how moduli spaces of rank 2 sheaves on
surfaces vary depending upon which subcone of the K\"ahler cone
a K\"ahler form is in.  These moduli spaces are typically birational
to one another, and so are extremely singular on a chamber wall.}
K\"ahler forms.  When 
$G = GL(n, {\bf C})$, the moduli space
is a GIT quotient of a closed subset of a product of partial flag manifolds, 
and this will be
nonsingular.  
When $G$ is any
other reductive group, the corresponding GIT quotient will have
orbifold singularities.  
For example, when $G$ is (the
complexification of) $\mbox{Spin}(n)$ or $Sp(n)$, the moduli space will have
${\bf Z}_{2}$ orbifold singularities.  For $E_{6}$ or $G_2$, the moduli space
will have ${\bf Z}_{2}$ and ${\bf Z}_{3}$ singularities.  For 
$E_{7}$ and $F_4$
the moduli space will have ${\bf Z}_{2}$, ${\bf Z}_{3}$, and ${\bf Z}_{4}$
orbifold singularities.  For $E_{8}$ the moduli space will have
${\bf Z}_{2}$, ${\bf Z}_{3}$, ${\bf Z}_{4}$, ${\bf Z}_{5}$, and ${\bf Z}_{6}$
orbifold singularities.  More precisely, for each group the largest
singular subvarieties
will have the orbifold singularities indicated, but higher codimension
substrata may have worse orbifold singularities than indicated above.

In fact we have been rather sloppy.  The singularities described
above are singularities expected in a GIT quotient of a product of 
flag manifolds, but a moduli space of bundles is a subvariety of
such a GIT quotient.  Just as a hypersurface in a weighted projective
space need not have precisely the same orbifold singularities as 
the ambient weighted projective space, the singularities occurring
in a moduli space of bundles may not be precisely those indicated
above.  However, the singularities listed should be an excellent
first approximation.

How are these singularities derived?  We will not work through
the details (see instead \cite{alleninprep}), but the general 
idea is as follows.
In general, when forming the GIT quotient of a product of flag
manifolds $K/T$ by $K$, the stabilizers are intersections of tori.
Only finite intersections are semistable -- it can be shown 
the rest are unstable.  Furthermore, these finite intersections
turn out to be those subgroups equal to their double centralizers,
and it can be shown these are generated by elements corresponding
to vertices of the Weyl alcove, whose orders in the adjoint group
are the coefficients of the highest root in a basis of simple roots.

In particular, note the singularities above do not quite correspond
to singularities of
moduli spaces of bundles on elliptic curves \cite{looijenga1,looijenga2},
One might have expected that the singularities in a moduli space
of $G$-bundles on, for example, ${\bf P}^2$ would also appear
in any moduli space of $G$-bundles on a Calabi-Yau hypersurface
in ${\bf P}^2$, but here we see that might not necessarily happen.
In fact, there is no reason why it must -- in general,
the restriction of a stable sheaf to a hypersurface need not be stable.

\subsection{Moduli spaces of equivariant torsion-free sheaves}
\label{metfs}

To describe moduli spaces of equivariant torsion-free sheaves 
involves significantly more technical machinery, and so
this has been deferred to a future publication \cite{meworkinprogress}.

\subsection{Relations between reflexive, torsion-free moduli spaces}
\label{rrtf}

If the reader studies the possible moduli spaces, he will naively come to
an apparent contradiction.  Let us consider an example to spell out
this difficulty.  Consider a rank $2$ torsion-free sheaf
defined over ${\bf C}^{2}$.  This sheaf is defined
by a ${\bf Z}^{2}$-graded module, specified by some bifiltration.  
Suppose that the nontrivial
part of the bifiltration has the form 
\begin{equation} \label{tflimit1}
\begin{array}{c|ccccccc}
\, & & -1 & 0 & 1 & 2 & 3  & \, \\ \hline
\, & & & & \vdots & & \, \\
3 & & 0 & 0 & 0 & 0 & 0 & \, \\
2 & & 1 & 1 & 1^{*} & 0 & 0 & \, \\
1 & \cdots & 2 & 2 & 2 & 1^{*} & 0  & \cdots \\
0 & & 2 & 2 & 2 & 1 & 0  & \, \\
-1 & & 2 & 2 & 2 & 1 & 0  & \, \\
\, & & & & \vdots & & \, 
\end{array}
\end{equation}
The numbers on the outer edges are the indices of the bifiltration.
Those in the middle are the dimensions of elements of the bifiltration.
Those marked with a $(*)$ are generators of this module. 
The numbers indicated are the dimensions of elements of the bifiltration.
If the pair of one-dimensional vector spaces marked with a $(*)$ is generic
(inside the rank 2 vector space), then their intersection will vanish,
and so this bifiltration will correspond to the intersection of 
two ordinary filtrations, each of which has the form 
\begin{displaymath}
\mbox{dim } E(i) \: = \: \left\{ \begin{array}{ll}
                     2 & i \leq 1 \\
                     1 & i = 2 \\
                     0 & i \geq 3 
                     \end{array} \right.
\end{displaymath} 
In particular, the torsion-free sheaf will be reflexive, and
even a vector bundle.

Now consider the limit that the two one-dimensional vector subspaces
marked with a $(*)$
coincide.  In this limit the intersection of the two limit
filtrations is a bifiltration whose nontrivial components have
dimensions
\begin{equation}  \label{tflimit2}
\begin{array}{c|ccccccc}
\, & & -1 & 0 & 1 & 2 & 3  & \, \\ \hline
\, & & & & \vdots & & \, \\
3 & & 0 & 0 & 0 & 0 & 0 & \, \\
2 & & 1 & 1 & 1^{*} & 1^{\dagger} & 0 & \, \\
1 & \cdots & 2 & 2 & 2^{\dagger} & 1^{*} & 0  & \cdots \\
0 & & 2 & 2 & 2 & 1 & 0  & \, \\
-1 & & 2 & 2 & 2 & 1 & 0  & \, \\
\, & & & & \vdots & & \,
\end{array}
\end{equation}
Clearly, this bifiltration is not the same as the one indicated
in diagram~(\ref{tflimit1}).  For example, the (one-dimensional)
generators are now 
located in positions marked by $(\dagger)$.  In particular, 
in this limit the torsion-free sheaf 
indicated in diagram~(\ref{tflimit1}) is no longer reflexive.

The apparent problem arises when we now consider the same limit,
but only within the set of data defining a reflexive sheaf.
Reflexive sheaves are described only by a set of filtrations on
each toric divisor, and in particular in all such limits (in which
various vector spaces in filtrations coincide) one still gets a reflexive 
sheaf.  Yet we saw above that when we work in data defining torsion-free 
sheaves, we recover a sheaf that is torsion-free but specifically not 
reflexive in this limit!

Thus, the precise sheaf appearing in various limits, depends upon
whether we are describing sheaves using only reflexive data or more
general torsion-free data.

How can this possibly be consistent?
The answer is that in the limit that the two one-dimensional
vector subspaces coincide, the Chern classes change\footnote{
To be precise, if the label the toric divisors on ${\bf C}^2$ as
$D_1$, $D_2$, then the reflexive sheaf in diagram~(\ref{tflimit1})
is precisely ${\cal O}(D_1 + 2 D_2) \oplus {\cal O}(2 D_1 + D_2)$
for generic flags, whereas the reflexive sheaf in diagram~(\ref{tflimit2})
is precisely ${\cal O}(D_1 + D_2) \oplus {\cal O}(2 D_1 + 2 D_2)$.
The first Chern classes are identical, but the second Chern classes
differ.}.  Since
we only want to consider moduli spaces of fixed Chern classes,
such limiting cases should be omitted.  This point has been
made obliquely several times previously within this paper,
but we felt it sufficiently important to warrant repeating.

These limiting points of reflexive sheaves in which the Chern classes change
are excluded from the product of partial flag manifolds
before computing any moduli space ${\cal M}_{r}$ of equivariant
reflexive sheaves as a GIT
quotient.  As a result, any such moduli space ${\cal M}_{r}$
is noncompact.  To compactify the moduli space, we must add
torsion-free sheaves.

\section{Structure within the K\"{a}hler cone}  \label{kcsubstruc}

\subsection{Generalities}

Note that Mumford-Takemoto stability depends implicitly upon the 
choice of K\"{a}hler
form
\cite{qin1,qin2,qin3,qin4,friedmanqin,mw,hl,gottsche}.
This choice is extremely important -- sheaves that are stable with
respect to one K\"{a}hler form may not be stable with respect
to another.  In general, for fixed Chern classes a moduli space
of sheaves will not have the same form everywhere inside the
K\"{a}hler cone, but rather
will have walls along which extra sheaves become semistable.
When these walls are real codimension one in the classical K\"ahler cone,
they stratify the K\"ahler cone into subcones in each of
which the notion of stability is constant.  (In fact this phenomenon is
quite similar to
the behavior of GIT quotients under change of polarization, as discussed
in \cite{dolgachevhu,thaddeus}.)  

The most extreme case of this phenomenon, namely when
the walls are real codimension one and stratify the K\"ahler cone
into inequivalent subcones, is known to occur for the case
of rank $2$ sheaves on surfaces.  (Other cases are less well
understood at present.)  In this section we shall make some
general remarks on rank $r$ sheaves on algebraic surfaces
\cite[section 4.C]{huybrechtslehn}, and indicate when 
stronger statements can be made by restricting to $r = 2$.
Again, our remarks are specific to surfaces -- for varieties
of dimension not equal to two, there are nontrivial differences,
which are discussed to some extent in the literature.

Define the discriminant\footnote{
The discriminant is a special case of a ``logarithmic invariant"
of a coherent sheaf ${\cal E}$.  More generally, the logarithmic
invariants $\Delta_{i}({\cal E}) \in H^{2i}({\bf Z})$ are given by
\begin{displaymath}
\ln ch({\cal E}) \: = \: \ln r \: + \: \sum_{i=1} (-)^{i+1} 
\Delta_{i}({\cal E})
\end{displaymath} 
where $r$ is the rank of ${\cal E}$.} 
of a coherent sheaf ${\cal E}$ on an algebraic 
K\"{a}hler surface
to be
\begin{displaymath}
\Delta \: = \: 2 r \, c_{2}({\cal E}) \: - \: (r-1) c_{1}({\cal E})^{2}
\end{displaymath}
where $r$ is the rank of ${\cal E}$.  It can be shown 
\cite[section 3.4]{huybrechtslehn} that when ${\cal E}$
is semistable, $\Delta({\cal E}) \geq 0$.

Walls inside the K\"{a}hler cone $K$ are specified by divisors $\zeta$
such that $ - \frac{r^{2}}{4} \Delta \leq \zeta^{2} < 0$.
(Not all such divisors define walls along which 
extra sheaves become semistable; this is a necessary
but not sufficient condition.)  
For a divisor $\zeta$, the corresponding wall is
\begin{displaymath}
W_{\zeta} \: = \: \{ J \in K \, | \, \zeta \cdot J = 0 \}
\end{displaymath}  
(Note that by the Hodge index theorem, on an algebraic K\"{a}hler
surface
the positive definite part of $H^{1,1}$ is one-dimensional,
so the intersection form on $H^{1,1}$ has signature $(+,-, \cdots -)$.)
More precisely \cite[section 4.C]{huybrechtslehn}, 
if ${\cal E}$ is a rank $r$ torsion-free coherent
sheaf and ${\cal E}'$ is a subsheaf of rank $r'$, with $\mu_{J}({\cal
E}) =
\mu_{J}({\cal E}')$, then $J$ lies on a wall $W_{\zeta}$
defined\footnote{We need to check that $\zeta \cdot J = 0$ and
$ - \frac{r^{2}}{4} \Delta \leq \zeta^{2} < 0$.  The former is clear.
To show the latter, first define
${\cal E}'' = {\cal E}/{\cal E}'$.  Note ${\cal E}''$ is a semistable
(as all subsheaves of ${\cal E}$ descend to subsheaves of ${\cal
E}/{\cal E}'$)
torsion-free coherent sheaf, of rank $r'' = r - r'$.  Use the identity
\begin{displaymath}
\Delta({\cal E}) \: - \: \frac{r}{r'} \Delta({\cal E}') \: - \:
\frac{r}{r''} \Delta({\cal E}'') \: = \: - \frac{\zeta^{2}}{r' r''}
\end{displaymath}
and the fact that $\zeta \cdot J = 0$ implies $\zeta^{2} \leq 0$,
with equality only when $\zeta = 0$,
by the Hodge index theorem.}
by $\zeta = r c_{1}({\cal E}') - r' c_{1}({\cal E})$.
Again, not all divisors $\zeta$ such that $- \frac{r^2}{4} \Delta
\leq \zeta^2 < 0$ specify walls along which the moduli space changes.

In the special case $r=2$, the precise condition for a divisor
$\zeta$ to define a wall along which the moduli space changes
is known \cite{qin1,qin2,qin3,qin4}.  In addition to the
constraint $ - \frac{r^{2}}{4} \Delta \leq \zeta^2 < 0$,
one must also demand $\zeta - c_1 = 2F$ for some divisor $F$.
For $r=2$, these are the precise conditions for a divisor $\zeta$
to define a wall along which one finds additional strictly semistable sheaves. 
 
Moduli spaces associated to distinct chambers of a K\"{a}hler
cone are often, but not always, birational to one another.
For example it can be shown \cite[section 4.C]{huybrechtslehn}
that if $\Delta \gg 0$, then for any two polarizations $J$, $J'$,
the moduli spaces of sheaves of fixed rank $r$ and Chern classes
are birational.

In passing we should comment on the physics associated to this
phenomenon. 
First, we should warn that it is possible that this phenomenon
might not be seen in heterotic compactifications.  Stability
is a necessary but not sufficient condition for a consistent 
compactification, so it is
possible (though highly unlikely, in our opinion) that
sheaves that are stable in only part of a K\"ahler cone
can not be used in compactifications.
Assuming that this does not happen, then we should also
remark on K\"ahler forms in string theory. 
As is well-known, in string compactifications
the K\"ahler form is complexified, and we expect 
the imaginary part (the theta angle) will allow us to 
analytically continue from one moduli space to another,
effectively smoothing over the transition.

There is also a closely
related topological (rather than algebro-geometric) version
of this same phenomenon.  See \cite{frqin,mooreed} for
recent discussions.

\subsection{Application to F theory}

\begin{figure}
\centerline{\psfig{file=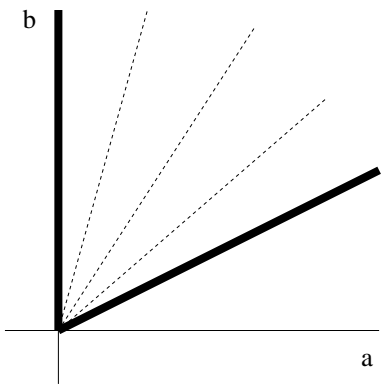,width=2.0in}}
\caption{\label{pickc} Schematically illustrated is a K\"{a}hler
cone of a generic elliptic K3 with section.  The outer boundaries of
the K\"{a}hler cone are shown in bold lines, and a few of the 
chamber boundaries
are shown as dotted lines.}
\end{figure}

Suppose we have an elliptic K3 with section.
Let $S$ denote the section, and $F$ the (elliptic) fiber,
so $S^{2} = -2$, $F^2 = 0$, and $S \cdot F = 1$.
Write a divisor $J = a S + b F$, then $J$ is ample precisely
when $a > 0$, $b > 2 a$.  See for example figure~\ref{pickc}.

Clearly by staying sufficiently close to the outer boundaries
of the K\"{a}hler cone it is possible to never cross a 
chamber wall.  For example, in \cite{wmf,wmf2,wmf3}
it is implicitly assumed that the fiber is very small compared to
the section.  In particular, this assumption means working close
to one of the outer boundaries of the K\"{a}hler cone, and 
so the authors of \cite{wmf,wmf2,wmf3} never had to worry about
crossing a chamber boundary.
For the other outer K\"{a}hler boundary, we can be slightly
more precise.
Consider the ample divisor $J = S + r F$.  According to \cite{qin1},
all such $J$ with $r > 1 + (4 c_2 - c_1^2)/2$ are in a single
chamber.

For example, consider 
rank $2$ sheaves of $c_1 = 0$ on an elliptic K3 with section.
For $c_2 = 4$ and $c_2 = 5$, there are two chambers, 
with a chamber wall defined
by $J = a S + b F$ with $b = 3 a$.

Although we can locate the chamber boundaries mathematically,
locating them in F theory is much more difficult.
First, F theory does not directly sense the distinction between
complex and K\"{a}hler structures on the K3, but only the choice
of Riemannian metric, which makes the problem of seeing the
K3 K\"{a}hler cone in F theory somewhat subtle.
Secondly, worldsheet instanton corrections will smooth over
transitions between chambers when $\alpha' > 0$, so in order
to clearly see a boundary one must go to an entirely classical
limit.

\subsection{K\"{a}hler cone substructure and equivariant moduli spaces}

Let us see explicitly how this phenomenon appears in moduli spaces of
equivariant reflexive sheaves.  These moduli spaces are
constructed as GIT quotients, in which essentially the same
phenomena occur under change of polarization.  In these moduli
spaces we will be able to see the chamber structure explicitly.

\begin{figure}
\centerline{\psfig{file=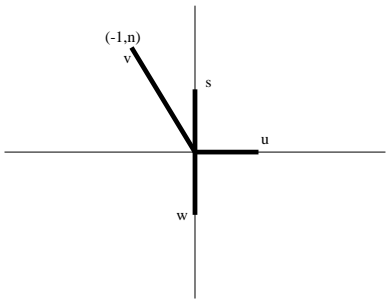,width=2.0in}}
\caption{\label{picfn} A fan describing the Hirzebruch surface
${\bf F}_{n}$ as a toric variety.}
\end{figure}

Let us consider a specific example, a variation on the example
considered in section~\ref{gitex}.
Let us work on a Hirzebruch surface ${\bf F}_{n}$.  A 
fan for such a surface is shown in figure~\ref{picfn}.
In our notation, $D_u = \{ u = 0 \}$,
for example, and there are the relations $D_u = D_v$, $D_w = n D_u + D_s$.
$H^{2}({\bf Z})$ is generated by $D_u$, $D_s$, and $H^4({\bf Z})$
is generated by $D_u \cdot D_s$, with  $D_u^2  = 0$ and $D_s^2 = -n D_u \cdot 
D_s$.

Our example is defined by filtrations associated to
each toric divisor of the form
\begin{displaymath}
\mbox{dim } E^{[\alpha]}(i) \: = \: \left\{
\begin{array}{ll}
1 & i = 1 \\
r-1 & i = 0 \\
0 & i \neq 0,1
\end{array} \right.
\end{displaymath}
For generic flags the Chern classes of this bundle are
$c_1 = 2 D_s + (n+2) D_u$ and $c_2 = 4 D_u \cdot D_s$, so
$\Delta = 8 D_u \cdot D_s$.  
Note that ${\cal T}_{r} \subset \prod {\bf P}^{r-1}$.
Expand the K\"{a}hler form\footnote{In $\mbox{Num}$ rather than
$\mbox{Num} \otimes {\bf R}$.} as $J = a D_u + b D_s$,
then the K\"{a}hler cone is given by $b > 0$, $a > n b$,
and we have the intersections with divisors
\begin{eqnarray*}
 D_u \cdot J & = & b \\
 D_v \cdot J & = & b \\
 D_w \cdot J & = & a \\
 D_s \cdot J & = & a - nb
\end{eqnarray*}

What chamber structure can we expect in the case $r = 2$ ?
Using the analysis described earlier, it is easy to show that
for $n=1$, the K\"ahler cone is split into two chambers, with
the chamber wall defined by $J = a D_u + b D_s$ with $2 a = 3 b$.
For $n=2$, $3$, there are no nontrivial chamber walls -- the
only chamber is the entire K\"ahler cone. 
We will momentarily check these results by an explicit
computation.

For reasons of notation, define $k_{\alpha} = D_{\alpha} \cdot J$.
Then, the GIT quotient is
defined by the ample $G$-linearized line bundle
\begin{displaymath}
\bigotimes_{\alpha} \pi_{\alpha}^{*} \left( {\cal O}_{{\bf P}^{n}}(k_{\alpha})
\right)
\end{displaymath}
on ${\cal T}_{r}$.   Note that this bundle is ample precisely
when all the $k_{\alpha} > 0$, which is true precisely because
$J$ lies in the ample cone.  In other words, the fact that $J$
defines a nondegenerate Kahler form is intimately connected to the
fact that the GIT quotient is well-defined.
Now, let $\{ L^{(\alpha)} \} \in {\cal T}_{r}$,
and it should be clear that this point is Mumford-Takemoto semistable
precisely when for all linear subspaces $W \subset {\bf P}^{r-1}$,
\begin{displaymath}
\frac{1}{\mbox{dim }W + 1} \sum_{\alpha \: s.t. \: L^{(\alpha)} \in W}
k_{\alpha} \: \leq \: \frac{1}{r} \sum_{\alpha} k_{\alpha}
\end{displaymath}
Now suppose there exists a subspace $W$ such that
\begin{displaymath}
\frac{1}{\mbox{dim }W + 1} \sum_{\alpha \: s.t. \: L^{(\alpha)} \in W}
k_{\alpha} \: = \: \frac{1}{r} \sum_{\alpha} k_{\alpha}
\end{displaymath}

Let us specialize to the case $r = 2$, and explicitly reproduce the chamber
structure described above.  (In fact our calculation is similar
to calculations presented in \cite{dolgachevhu}.) 
In this case, in order to get
the Chern classes described, one must be careful about the
filtrations chosen.  In particular, the only times a pair
of $L^{(\alpha)}$ can coincide as subspaces of ${\bf C}^{2}$
is when the toric divisors do not intersect.
In particular, this means that the pair $L^{(u)}$, $L^{(v)}$
can coincide, as well as the pair $L^{(w)}$, $L^{(s)}$,
but no others.

On ${\bf F}_1$, the condition for strict semistability is
that there exists a subspace $W \subset {\bf P}^1$ such that
\begin{eqnarray*}
\sum_{\alpha \: s.t. \: L^{(\alpha)} \in W} k_{\alpha} & = & 
\frac{1}{2} \sum_{\alpha} k_{\alpha} \\
\: & = & a \: + \: \frac{1}{2} b 
\end{eqnarray*}
In the case that $W$ contains only a single $L^{(\alpha)}$,
we get no information.
Now, the only time a subspace $W$ can contain more than a single
$L^{(\alpha)}$ is when the toric divisors are the pair $D_u$, $D_v$,
or the pair $D_w$, $D_s$.
In the first case,
\begin{displaymath}
\sum_{\alpha \: s.t. \: L^{(\alpha)} \in W} k_{\alpha} \: = \: 2b 
\end{displaymath}
In the second case,
\begin{displaymath}
\sum_{\alpha \: s.t. \: L^{(\alpha)} \in W} k_{\alpha} \: = \: 2a - b
\end{displaymath}
and in either event, for strict semistability we immediately
derive 
\begin{displaymath}
a \: = \: \frac{3}{2} b
\end{displaymath}
This is precisely the chamber wall expected.
Similarly, for ${\bf F}_2$ and ${\bf F}_3$, it is easy to show
that there are no chamber walls inside the K\"ahler cone,
as claimed earlier.  Thus, we can see explicitly the chamber structure
predicted by the more abstract analysis given earlier.

\section{Moduli spaces of not-necessarily-equivariant sheaves}
\label{genmodspace}

So far in this paper we have primarily studied only equivariant
sheaves and their moduli spaces.  The reader may well wonder
what can be said about moduli spaces of not-necessarily-equivariant
sheaves, and how much information about the general case can be
obtained by studying only equivariant objects.

In fact, a great deal of information about more general moduli spaces
can be gained by
studying only the equivariant objects.  Full moduli spaces
have natural stratifications, the Bia\l ynicki-Birula stratifications,
in which each stratum is a vector bundle over a 
moduli space of equivariant sheaves.  This is discussed in more
detail in subsection~\ref{bb}.  (This has also been discussed 
previously in \cite{kl1,kl2,kl3,kl4}.)

In addition, the reader may wonder if any examples of 
not-necessarily-equivariant moduli spaces are known explicitly.
In fact, a few examples are indeed known explicitly, and
we examine one in detail in subsection~\ref{monadstuff}.

Finally, as an aside, in subsection~\ref{adhm} we show
how the ADHM construction can be recovered given the
results in subsection~\ref{monadstuff}, following
\cite{donaldsongit}.

The material in this section is not used in other sections,
so on a first reading the reader may want to skip to the
next section.

In passing, we should clarify a few points that may confuse the reader.
First, we have been somewhat sloppy in our use of the word
``equivariant.''  We have implicitly assumed throughout that
if a sheaf is equivariant, in our sense of the word, then
we are necessarily able to give it an equivariant structure.
In fact, this is not quite correct, there is a minor subtlety.
Consider for example the tautological bundle on ${\bf P}^1$.
The space ${\bf P}^1$ can be realized as the coset
$SL(2,{\bf C})/B$, $B$ a Borel subgroup of $SL(2,{\bf C})$,
so ${\bf P}^1$ has a natural $SL(2,{\bf C})$ action.
The center of $SL(2,{\bf C})$ acts trivially on ${\bf P}^1$ -- the
algebraic torus of which ${\bf P}^1$ is a compactification
is actually
\begin{displaymath}
\frac{ \mbox{maximal torus of } SL(2,{\bf C}) }{ {\bf Z}_2 }
\end{displaymath}
The center of $SL(2,{\bf C})$ does not act trivially on the
total space of the tautological bundle, however -- it acts as
$(-)$ on the fibers.  Thus, the algebraic torus action on ${\bf P}^1$
does not lift to an action on the tautological bundle,
but a finite extension of it does.  Throughout this paper,
we have assumed that any such extensions are performed whenever
needed.  In other words, our definition of equivariant sheaf
is not rigorous, and in a more mathematical treatment should
be replaced with a definition that specifies a lift of the
algebraic torus action.  

As the moduli spaces we construct here are moduli spaces of
equivariant sheaves (of fixed equivariant structure), the reader may wonder
how these are related to moduli spaces of not-necessarily-equivariant
sheaves.  In fact, the forgetful map that ``forgets" the
equivariant structure is one-to-one on individual components
(of fixed equivariant structure) of a moduli space.
More precisely \cite{kl2}, if we denote
a trivial line bundle with equivariant structure defined by a character $\chi$
by\footnote{In other words,
${\cal O}(\chi)$ is the trivial line bundle with equivariant
structure defined by the $T$-invariant divisor $D = \sum_{\alpha}
\langle v_{\alpha}, \chi \rangle \, D_{\alpha}$ where $v_{\alpha}$ is
the edge of the fan associated to toric divisor $D_{\alpha}$.}
${\cal O}(\chi)$, then if ${\cal E}$ is an indecomposable\footnote{
For a sheaf to be indecomposable means it cannot be written globally
as the direct sum of two subsheaves.} sheaf, all equivariant
structures on ${\cal E}$ can be obtained by tensoring with
characters as ${\cal E} \otimes {\cal O}(\chi)$.  (For
example, the line bundles ${\cal O}(D_{x})$ and ${\cal O}(D_{y})$
on ${\bf P}^2$ have distinct equivariant structures,
but we can get ${\cal O}(D_{y})$ from ${\cal O}(D_{x})$
by tensoring with ${\cal O}(D_{y} - D_{x})$,
which is simply the structure sheaf with a nontrivial
equivariant structure.)  Thus, a moduli space
of equivariant sheaves of fixed equivariant structure sits naturally inside
a moduli space of not-necessarily-equivariant sheaves.

\subsection{Bia\l ynicki-Birula stratifications} \label{bb}

Any compact variety with a ${\bf C}^{\times}$ action has a canonical
stratification, the Bia\l ynicki-Birula stratification \cite{bb}.
We explain the B-B stratification first and then apply it to moduli spaces
of sheaves on toric varieties.

Let $X$ be a projective complex variety with a ${\bf C}^{\times}$ action.
Let $F$ be the set of fixed-point components of this action.
For $f \in F$, define the stratum $X^{0}_{f}$ corresponding to $f$ as 
\begin{displaymath}
X^{0}_{f} = \{ x \in X | \lim_{z \rightarrow 0} z \cdot x \in f \}
\end{displaymath}
This limit exists precisely because $X$ is projective.
This plainly contains all the points of $f$ and has a natural retraction
onto $f$.  If $X$ is smooth, then $X^{0}_{f}$ is in fact a vector bundle
over $f$ \cite{basshaboush,kraft,ginzburg}.
The fixed-point components contain much of the information of the
variety.  For example, the birational geometry and all the Hodge numbers
can be computed given those of the fixed-point components \cite{ginzburg}.

In many cases this stratification can also be understood in terms of 
Morse theory.  
If the action of the $U(1)$ inside ${\bf C}^{\times}$ is Hamiltonian --
in particular, if $X$ is smooth (or even just normal) and projective --
then 
the critical points 
for the Hamiltonian are exactly the fixed points, and the B-B stratum 
$X^{0}_{f}$ is precisely the points of $X$ that limit to $f$ under
gradient\footnote{with respect to a $U(1)$-invariant
metric} flow (known as the ``unstable manifold" of $f$ in mathematics
circles).

It is straightforward to determine the rank of a bundle describing a
stratum over a component $f$ of ${\cal M}^{T}$.  Let $\lambda$ be a
one-parameter subgroup determining the stratification, then
the rank of the bundle over $f$ equals the
number of positive weights in the representation of $\lambda$ in a
tangent space to any point of $f$.

For example, consider the Riemann sphere 
${\bf P}^{1} = {\bf C} \cup \{ \infty \}$.  This space is a toric
variety
and so has a natural action (by multiplication) of ${\bf C}^{\times}$,
of which it is a
compactification.  The ${\bf C}^{\times}$ action has two fixed points:
$0$ and $\infty$.  
The action on the tangent space at $0$ has weight $1$, and on the tangent
space 
at $\infty$ has weight $-1$.
We can then view ${\bf P}^{1}$
as a disjoint union of $\{ 0 \} \times {\bf C}$ and $\{ \infty \}$ --
so, as
promised,
${\bf P}^{1}$ looks like a disjoint union of vector bundles over the
components of the fixed-point locus.

In the case of a torus action, in principle there are many B-B
stratifications, one
for each one-dimensional subgroup of the torus.  (In fact, even for a
one-dimensional torus, there are actually two distinct stratifications,
depending upon the
isomorphism chosen with ${\bf C}^{\times}$.) 
However,
there
are only finitely many distinct stratifications
\cite{hu,dolgachevhu,thaddeus}.
These distinct stratifications can be described pictorially
be a fan of some compact toric variety, and we shall denote this fan
the stratification fan.

To describe the stratification fan, first we must discuss a fan
naturally associated to the moduli space ${\cal M}$ of
not-necessarily-equivariant
bundles.  As is well-known \cite{atiyah,gs} the image of a moment map
associated with a set of commuting circle-generating Hamiltonian
actions is a convex polytope.  As the algebraic torus acts on ${\cal
M}$, we can restrict to $S^{1}$'s to find a set of commuting
circle-generating Hamiltonian actions, and so (for a fixed
projective embedding of ${\cal M}$) we can naturally associate
a convex polyhedron ${\cal P}$ with ${\cal M}$.

In principle \cite{atiyah,gs} this polytope ${\cal P}$ is the 
convex hull of the
images (under the moment map) of the fixed points of the circle
actions.  (Note that sometimes some of the fixed points will
lie in the interior of the polytope, but typically they will
be vertices of the polytope.)  Such polytopes ${\cal P}$ may
be of use in understanding (0,2) mirror symmetry, as will be
discussed later.

Now, given the convex polyhedron ${\cal P}$ derived above, it can be
shown \cite{hu} that the fan describing distinct Bia\l ynicki-Birula
stratifications of ${\cal M}$, is a refinement of the fan associated to
${\cal P}$.  In particular, we can obtain the stratification fan by
adding edges to the fan associated with ${\cal P}$, so that $v$ is an
edge precisely when $-v$ is an edge.  
Intuitively this should be sensible.  A one-parameter subgroup
$\lambda$ borders two stratifications when
it has more fixed points than do generic subgroups. 
In particular, if a one-parameter subgroup
$\lambda$ borders two stratifications, then the same should be true
of $\lambda^{-1}$.  Thus, it should seem intuitively reasonable
that $v$ is an edge of the fan precisely when $-v$ is an edge.

We now apply these ideas to normal, projectively embedded 
moduli spaces of torsion-free sheaves on toric varieties.
This should be relatively clear.
Recall the torus action on the toric variety induces a torus
action on any moduli space of sheaves, as ${\cal E} \mapsto t^{*}
{\cal E}$ for a torus action $t$.  The fixed points of this
torus action are precisely the equivariant sheaves, which we
have studied in great detail earlier in this paper.

\subsection{Monads and bundles on projective spaces} \label{monadstuff}

In this section, we will describe moduli spaces of $SL(r,{\bf C})$
bundles on ${\bf P}^{2}$.  These moduli spaces
have been studied in great detail since the early 1970s,
and are quite well understood.  

The reader who is interested in the $c_{1} \neq 0$ case
may wish to consult \cite{okonek,hulekodd}.

To what extent can arbitrary bundles on toric varieties be described?
It is known \cite{horrocks1,barth} that any vector bundle ${\cal E}$ 
on ${\bf P}^{n}$ can be
obtained as the cohomology of a monad\footnote{A short complex,
such that the first and last pairs of maps are exact.}
\begin{displaymath}
0 \: \rightarrow \: {\cal A} \: \rightarrow \: {\cal B} \: \rightarrow
\: {\cal C} \: \rightarrow \: 0
\end{displaymath}
where ${\cal A}$, ${\cal C}$ are direct sums of line bundles and
${\cal B}$ is a vector bundle such that
\begin{displaymath}
\begin{array}{ll}
H^{1}({\cal B}(k)) \: = \: H^{n-1}({\cal B}(k)) \: = \: 0 , & k \in {\bf Z} \\
H^{i}({\cal B}(k)) \: = \: H^{i}({\cal E}(k)) , & k \in {\bf Z}, 1 < i < n-1
\end{array}
\end{displaymath}
Unfortunately, except for bundles over ${\bf P}^{2}$ and ${\bf P}^{3}$
the classification of such monads is not well-understood, and so
this is not as useful for describing moduli spaces as one would
wish.

To begin our discussion of $SL(r, {\bf C})$ bundles on ${\bf P}^2$, 
we will describe a result due to Beilinson \cite{beilinson}.
Let ${\cal E}$ be a holomorphic rank $r$ bundle on ${\bf P}^{n}$.
There are spectral sequences with $E_{1}$ terms
\begin{eqnarray*}  
E_{1}^{p,q} & = & H^{q} \left( {\bf P}^{n}, {\cal E}(p) \right)
\otimes \Omega^{-p}(-p) \\
E_{1}^{' p,q} & = & H^{q} \left( {\bf P}^{n}, {\cal E} \otimes
\Omega^{-p}(-p) \right) \otimes {\cal O}(p)
\end{eqnarray*}
both of which converge to
\begin{displaymath}
{\cal E}^{i} \: = \: \left\{ \begin{array}{ll}
                             {\cal E} & i = 0 \\
                             0 & \mbox{otherwise}
                             \end{array} 
                             \right.
\end{displaymath}
In other words, $E_{\infty}^{p,q} = 0$ when $p+q \neq 0$,
and $\oplus_{p} E_{\infty}^{-p,p}$ is the associated graded sheaf of a 
filtration of ${\cal E}$, for both spectral sequences.
In this notation $\Omega^{p}$ denotes $\wedge^{p} T^{*} {\bf P}^{n}$. 

In other words, just as Klyachko \cite{kl1,kl2,kl3,kl4} describes
bundles by associating filtrations to divisors, Beilinson 
describes bundles in terms of spectral sequences.  We shall
see shortly that in special cases, the spectral sequence simplifies
greatly.

Now, let us use Beilinson's result to derive a monad describing $SL(r,{\bf C})$
bundles on ${\bf P}^{2}$, following \cite[section II.3.2]{okonek}.
Let ${\cal E}$ be a (properly) stable holomorphic $r$-bundle 
over ${\bf P}^{2}$.

Since ${\cal E}$ is properly stable, we can derive a few constraints
on certain sheaf cohomology groups.  In particular, for $k \geq 0$,
$h^{0}({\bf P}^{2}, {\cal E}(-k)) = 0$.  If this were not the case,
then a section $s$ of ${\cal E}(-k)$ would define a map ${\cal O}
\rightarrow {\cal E}(-k)$, and so we would have a map
${\cal O}(k) \rightarrow {\cal E}$.  But then ${\cal O}(k)$ would
be a subbundle of ${\cal E}$, with $\mu( {\cal O}(k) ) = k > \mu({\cal
E}) = 0$, violating the stability condition.  Also, by Serre duality,
this means $h^{0}({\bf P}^{2}, {\cal E}^{*}(k)) = 0$ for $k \geq -3$.

We can also derive a constraint on $c_{2}({\cal E})$ for properly stable
bundles on ${\bf P}^{2}$.  Using Hirzebruch-Riemann-Roch it is easy
to show $h^{1}({\bf P}^{2}, {\cal E}) = c_{2}({\cal E}) - r$,
so clearly $c_{2}({\cal E}) \geq r$.

Apply one of Beilinson's spectral sequences to ${\cal E}(-1)$:
\begin{displaymath}
E_{1}^{p,q} \: = \: H^{q} \left( {\bf P}^{2}, {\cal E}(-1) \otimes
\Omega^{-p}(-p) \right) \otimes {\cal O}(p)
\end{displaymath}

We can show
\begin{displaymath}
E_{1}^{0,q} \: = \: \left\{ \begin{array}{ll}
                            H^{1}( {\bf P}^{2}, {\cal E}(-1) ) 
\otimes {\cal O} & q = 1 \\
                            0 & q = 0,2 
                            \end{array} 
                            \right.
\end{displaymath}

By tensoring the short exact sequence
\begin{displaymath}
0 \: \rightarrow \: \Omega^{1}(1) \: \rightarrow \: {\cal O}^{3} \:
\rightarrow \: {\cal O}(1) \: \rightarrow \: 0
\end{displaymath}
with ${\cal E}(-1)$, we can show $h^{0}( {\bf P}^{2}, {\cal E}\otimes
\Omega^{1}) = 0$, and similarly $h^{2}( {\bf P}^{2}, {\cal E}\otimes
\Omega^{1}) = 0$.

Using these results, one can then show
\begin{eqnarray*}
E_{1}^{-1,q} & = & \left\{ \begin{array}{ll}
                           H^{1}( {\bf P}^{2}, {\cal E} \otimes
\Omega^{1}) \otimes {\cal O}(-1) & q = 1 \\
                           0 & q = 0,2
                           \end{array}
                           \right. \\
E_{1}^{-2,q} & = & \left\{ \begin{array}{ll}
                           H^{1}( {\bf P}^{2}, {\cal E}(-2))
\otimes {\cal O}(-2) & q = 1 \\
                           0 & q = 0,2 
                           \end{array}
                           \right.
\end{eqnarray*}
so $E_{\infty}^{p,q} = E_{2}^{p,q}$, and ${\cal E}$ is the
cohomology of the monad
\begin{equation} \label{Emonad}
0 \: \rightarrow \: H \otimes {\cal O}(-1) \: \rightarrow \: K \otimes
{\cal O} \: \rightarrow \: L \otimes {\cal O}(1) \: \rightarrow \: 0
\end{equation}
where
\begin{eqnarray*}
H & = & H^{1}( {\bf P}^{2}, {\cal E}(-2) ) \\
K & = & H^{1}( {\bf P}^{2}, {\cal E}\otimes \Omega^{1}) \\
L & = & H^{1}( {\bf P}^{2}, {\cal E}(-1) ) 
\end{eqnarray*}

For notational convenience define $n = c_{2}({\cal E})$.
Using Hirzebruch-Riemann-Roch it is easy to show $\mbox{dim } H =
n = \mbox{dim } L$, and $\mbox{dim } K = 2 n 
+ r$.

It can be shown \cite[p. 291]{okonek} that two monads of the 
form of equation~(\ref{Emonad})
define the same bundle ${\cal E}$ precisely when they differ by the
action of $GL(n,{\bf C})$, $GL(2n+r,{\bf C})$, and $GL(n,{\bf C})$
on $H$, $K$, and $L$, respectively.

The moduli space of rank $r$ bundles on ${\bf P}^{2}$ can be
characterized a little more precisely by the use of Kronecker modules
\cite{hulek,barthinvmath,maruyamats}, but we will not do so here. 

More precise characterizations of the moduli space of rank $r$ bundles
on ${\bf P}^{2}$ are somewhat difficult to construct except in special
cases.  For example \cite{okonek}, 
the moduli space of stable rank $2$ bundles on 
${\bf P}^{2}$ of $c_{1} = 0$, $c_{2} = 2$ is ${\bf P} H^{0}({\bf P}^{2 *},
{\cal O}(2)) = {\bf P}^{5}$ (the space of conics on the dual\footnote{
If $V$ is a rank 3 vector space, so ${\bf P}^2 = {\bf P} V$,
then the dual projective plane is ${\bf P}^{2 *} = {\bf P}
\wedge^2 V$.} projective plane), modulo automorphisms
of ${\bf P}^{2}$.
(In fact almost any such conic can be related to almost any other 
by an automorphism
of ${\bf P}^{2}$.)

In passing, one might have hoped that the moduli space of bundles on a
Calabi-Yau hypersurface were related to the moduli space of bundles
on the ambient space, whose top Chern class is minimal.  Here, however,
we see that this is wrong, as the moduli space of $SL(2, {\bf C})$ bundles on
$T^{2}$ is $T^{2}$ \cite{atiyahbund}.  (For that matter,
the moduli space of flat $SU(2)$ connections on $T^2$ is ${\bf P}^1$.)
${\bf P}^{1}$.  We will see in section~\ref{stab}
that all stable $SL(2, {\bf C})$ bundles on ${\bf P}^{2}$
restrict to semistable bundles on $T^{2}$, but there is no map of
${\bf P}^{5}$ onto $T^2$ (or ${\bf P}^{1}$), 
so clearly
there are semistable bundles on $T^{2}$ not obtained by
restriction from ${\bf P}^{2}$.

\subsection{An aside:  the ADHM construction}  \label{adhm}

The ADHM construction of instantons on ${\bf R}^{4}$ can be derived 
from the moduli space of bundles on ${\bf P}^{2}$ \cite{donaldsongit}. 
Intuitively this should be clear:  ${\bf P}^{2}$
looks like ${\bf C}^{2} = {\bf R}^{4}$ with a ${\bf P}^{1}$
at infinity.  To derive the ADHM construction, one simply
restricts to those bundles on ${\bf P}^{2}$ which are trivial
on the line at infinity.  (In principle, one can also
derive the ADHM construction via Penrose transformations
\cite{atiyahward,drinfeldmanin} from bundles on 
${\bf P}^{3}$, but we shall not do so here.)

For completeness, we briefly review the derivation of the ADHM 
construction.  In equation~(\ref{Emonad}), label the map $H \otimes
{\cal O}(-1) \rightarrow K \otimes {\cal O}$ by $A$, and the map
$K \otimes {\cal O} \rightarrow L \otimes {\cal O}(1)$ by $B$,
and expand
\begin{eqnarray*}
A & = & A_{x} x + A_{y} y + A_{z} z \\
B & = & B_{x} x + B_{y} y + B_{z} z
\end{eqnarray*}
where $x$, $y$, and $z$ are homogeneous coordinates on ${\bf P}^{2}$.

Now, demanding that the bundle ${\cal E}$ be trivial on the line
$z = 0$ is equivalent to saying that over $z = 0$,
$\mbox{ker } B / \mbox{im } A$ is a fixed $r$-dimensional
subspace.  Given that $A$ is one-to-one and $B$ is onto,
we can choose bases such that
\begin{displaymath}
A_{x} = \left( \begin{array}{c}
               I_{n \times n} \\
               0_{n \times n} \\
               0_{r \times n} 
               \end{array}
               \right), \:
A_{y} = \left( \begin{array}{c}
               0_{n \times n} \\
               I_{n \times n} \\
               0_{r \times n}
               \end{array}
               \right), \:
A_{z} = \left( \begin{array}{c}
               \alpha_{1} \\
               \alpha_{2} \\
               a
               \end{array}
               \right)
\end{displaymath}
where $I_{n \times n}$ denotes the identity $n \times n$ matrix, and 
$0_{r \times n}$
the zero $r \times n$ matrix, and
\begin{displaymath}
B_{x} = (0_{n \times n}, I_{n \times n}, 0_{n \times r}) , \:
B_{y} = (- I_{n\times n}, 0_{n \times n}, 0_{n \times r} ), \:
B_{z} = (- \alpha_{2}, \alpha_{1}, b)
\end{displaymath}

The choice of bases above reduces the original $GL(n,{\bf C})$,
$GL(2n+r,{\bf C})$, $GL(n,{\bf C})$ symmetries to
$GL(n,{\bf C})$ and $GL(r,{\bf C})$.  However, the action
of $GL(r,{\bf C})$ acts as a gauge transformation on the
${\bf P}^{1}$ at infinity, and in constructing the moduli space
of instantons on ${\bf R}^{4}$ we restrict to gauge transformations
which are trivial at infinity, so for the purposes of deriving
the ADHM construction we have only a $GL(n,{\bf C})$ symmetry
left to mod out.

The constraint $BA = 0$ reduces to $B_{z} A_{z} = 0 \Leftrightarrow [
\alpha_{1}, \alpha_{2} ] + ba = 0$.

In particular, the moduli space of instantons on ${\bf R}^{4}$ can now
be described as the quotient of the set of matrices $(\alpha_{1},
\alpha_{2}, a, b)$ satisfying
\begin{eqnarray*}
& (1) & [ \alpha_{1}, \alpha_{2} ] + ba = 0 \\
& (2) & A \mbox{ is one-to-one, } B \mbox{ is onto}
\end{eqnarray*}
by the action of $GL(n,{\bf C})$ given by
\begin{eqnarray*}
\alpha_{i} & \rightarrow & p \alpha_{i} p^{-1} \\
a & \rightarrow & a p^{-1} \\
b & \rightarrow & pb 
\end{eqnarray*}
where $p \in GL(n, {\bf C})$.

Even more succinctly, the moduli space of instantons on ${\bf R}^{4}$
is the GIT quotient of the set of matrices $(\alpha_{1}, \alpha_{2},
a, b)$ satisfying
\begin{displaymath}
[ \alpha_{1}, \alpha_{2} ]\:  + \:  ba \: = \: 0
\end{displaymath}
by the action of $GL(n,{\bf C})$ indicated.
With additional work, it can be shown this is equivalent to
the symplectic quotient usually mentioned in the physics literature.

\section{Restriction to hypersurfaces \label{stab}}

So far in this paper we have discussed sheaves on 
toric varieties.  However, in order to get valid heterotic
compactifications, we must restrict to Calabi-Yau hypersurfaces
(or, more generally, Calabi-Yau complete intersections) 
in these toric varieties.

In general, the restriction of a (semi)stable sheaf on a toric
variety to a Calabi-Yau hypersurface
need not be (semi)stable (with respect to the restriction of the
K\"ahler form).  However, there do exist conditions under
which
the restriction is known to necessarily be (semi)stable. 
There are two sets of results in this
matter known in the mathematics literature, which we shall
outline below.

Let $X$ be an $n$-dimensional normal projective variety,
$L$ a very ample line bundle on $X$ defining Mumford-Takemoto
stability, and let ${\cal E}$ be a semistable torsion-free
sheaf of rank $r$ on $X$.  Consider restricting the sheaf
${\cal E}$ to a generic hypersurface $H$ in the linear system
$| L^{d} |$, with $d > 0$.

The first result, due originally to Maruyama 
\cite{maruyamaboundedness,mehtalec,barthmathann},
is that the restriction of ${\cal E}$ to $H$
is semistable for arbitrary $d$ if $r < n$.

The second result, due to Flenner \cite{flenner}, says that
the restriction of ${\cal E}$ to $H$ is semistable if
\begin{equation} \label{flenn}
\frac{  \left( \begin{array}{c}
               n + d \\
               d 
               \end{array}
               \right) - d - 1 }{ d } \: > \: \mu( L ) 
                              \mbox{ max } \left\{ \frac{r^2-1}{4}, 1
\right\}
\end{equation}
where $\mu( L )$ is the Mumford-Takemoto slope of $L$,
namely $c_1 (L)^{n}$.

For example, Flenner's result~(\ref{flenn}) says that
for a Calabi-Yau hypersurface in ${\bf P}^{2}$, with
very ample line bundle $L = {\cal O}(1)$, the restriction
of ${\cal E}$ is semistable if $r \leq 2$.
For Calabi-Yau hypersurfaces in ${\bf P}^{3}$, the restriction
of ${\cal E}$ is semistable if $r \leq 5$.
For Calabi-Yau hypersurfaces in ${\bf P}^{4}$, the restriction
of ${\cal E}$ is semistable if $r \leq 9$.

Flenner's result is a refinement of a slightly older result often
cited in the mathematics literature, which we shall describe
here for completeness.  The older result is due to Mehta and
Ramanathan \cite{mehtaramanathan}, and says that for fixed 
semistable ${\cal E}$, there exists an integer $d_{0}$ such
that for generic hyperplanes $H \in | L^{d} |$, $d \geq d_{0}$,
the restriction of ${\cal E}$ to $H$ is semistable.  Unfortunately,
the integer $d_{0}$ depends very much upon the precise sheaf ${\cal E}$,
not just numerical invariants such as its Chern classes.  (A very
slight refinement of their result also exists \cite{mehtaramanathan2},
which concerns whether the restriction of a stable sheaf is stable.) 

Unfortunately all of these results are only valid for very
specific K\"ahler forms.  It would be very interesting to have
analogous results for more general K\"ahler forms, but to our knowledge
the mathematics community has not yet been so obliging.

\section{Distler, Kachru's (0,2) models}  \label{(02)app}

\subsection{Review}

In addition to the recent description of vector bundles given by
Friedman, Morgan, Witten \cite{wmf,wmf2,bbundle,donagi},
there is another description of vector bundles\footnote{Actually,
Distler, Kachru \cite{(02)} are more general and describe
torsion-free sheaves, not just bundles. }
due originally to Maruyama \cite{maruyama2}
and generalized in work of Distler, Greene \cite{old(02)} and Distler,
Kachru
\cite{(02)}.

A Distler-Kachru model is a conformal field theory with global (0,2)
supersymmetry constructed as the IR limit of
a gauged linear sigma model \cite{phases}.  The linear sigma model in
question
describes a Calabi-Yau hypersurface in a toric variety.  
A sheaf ${\cal E}$ lies
over the ambient toric variety, 
and ${\cal E}$ is often\footnote{Either by a short exact sequence as shown, or
as the
cohomology of a monad,
which is a short complex such that the first and last pairs of maps are
exact.  The monads of \cite{(02)} are typically only defined 
over complete intersections in 
toric varieties, not over the entire ambient space. 
We will not consider such monads in this paper.} specified by 
a short exact sequence of the form
\begin{displaymath}
0 \: \rightarrow \: {\cal E} \: \rightarrow \: \oplus \, {\cal O}(n_{a}) \:
\rightarrow \:
\oplus \, {\cal O}(m_{i}) \: \rightarrow \: 0
\end{displaymath}
The sheaf over the hypersurface is simply the restriction of
${\cal E}$.\footnote{
In fact, we should be slightly more careful.  Let $\iota: Y \hookrightarrow
X$ denote the inclusion morphism mapping a Calabi-Yau hypersurface $Y$
into the ambient toric variety $X$.  Strictly speaking, given a sheaf 
${\cal E}$ defined over $X$ by the short exact sequence shown, when we restrict
to the Calabi-Yau $Y$ we recover
\begin{displaymath}
\cdots \: \rightarrow {\em Tor}_{1}^{{\cal O}_{Y}} \left( \oplus {\cal
O}(m_{i}), {\cal O}_{Y} \right) \: \rightarrow \: \iota^{*} {\cal E} \:
\rightarrow \: \iota^{*} \oplus {\cal O}(n_{a}) \: \rightarrow \:
\iota^{*} \oplus {\cal O}(m_{i}) \:  \rightarrow \: 0
\end{displaymath}
In reasonably nice cases, however, ${\em Tor}_{1}^{{\cal O}_{Y}} \left(
\oplus {\cal O}(m_{i}) , {\cal O}_{Y} \right) \, = \, 0$, so we recover
the naive short exact sequence.}

Some sheaves over the hypersurface in the toric 
variety -- even some line bundles -- may be unobtainable as restrictions
of such sheaves over the
toric ambient space.  This is essentially because a single toric
divisor can sometimes intersect a hypersurface more than once,
so a single divisor in the ambient space can become multiple divisors
on a complete intersection \cite{agm}.
Moreover, it is not obvious that all bundles over the ambient toric variety
can be described as kernels of short exact sequences.
(Admittedly though for sheaves on projective toric varieties,
the results described at the beginning of section~\ref{monadstuff}
make this seem not completely unreasonable.)

Typically Distler-Kachru 
models are assumed to describe bundles whose structure
group is $GL(n,{\bf C})$, though it has been observed in \cite{me2} that more
general structure groups can be obtained in principle, as special cases
of $GL(n,{\bf C})$ structure groups.  One of the advantages to Klyachko's
approach is that we can describe bundles of arbitrary structure
group with equal facility.

\subsection{Application to Distler-Kachru models}

Recall in \cite{(02)} sheaves ${\cal E}$ over toric varieties
are often specified as the kernel of a short exact sequence
\begin{displaymath}
0 \: \rightarrow \: {\cal E} \: \rightarrow \: \oplus {\cal O}(n_{a}) \:
\rightarrow \: \oplus {\cal O}(m_{i}) \: \rightarrow \: 0
\end{displaymath}
Note that such sheaves ${\cal E}$ are not usually equivariant (as the
maps are not).

As noted earlier, the moduli space ${\cal M}$ of torsion-free sheaves
has a stratification, in which each stratum looks like a vector bundle
over a component of ${\cal M}^{T}$.  Thus, we shall begin by
describing one way to construct an equivariant sheaf associated with
a family of sheaves specified as above.  Unfortunately it will
usually be the case that our recipe will yield a badly behaved
result -- it seems at least naively that most (0,2) models of Distler, Kachru
do not have equivariant limits.

First, for each line bundle, pick a specific equivariant structure --
in other words, write ${\cal O}(D) \, = \, {\cal O}( \sum a_{\alpha}
D_{\alpha})$, for some $\sum a_{\alpha} D_{\alpha}$ in the linear
equivalence class of $D$.  (In this notation, each $D_{\alpha}$ is a toric
divisor.)  In fact, each equivariant line bundle is 
uniquely written as such a sum,
if an equivariant structure is taken into account \cite{brion}.
Note that divisors differing by linear
equivalence define line bundles with distinct equivariant structures,
so it is necessary to fix specific divisors.

Let $x_{\alpha}$ denote the homogeneous coordinate corresponding to the
toric divisor $D_{\alpha}$ \cite{cox}.  Then the unique (up to rescalings)
equivariant map ${\cal O}( \sum a_{\alpha} D_{\alpha} ) \rightarrow
{\cal O}( \sum b_{\alpha} D_{\alpha} )$ consists of multiplying
sections of ${\cal O}( \sum a_{\alpha} D_{\alpha} )$ by
the monomial $x_{0}^{b_{0} - a_{0}} x_{1}^{b_{1} - a_{1}} \cdots
x_{k}^{b_{k} - a_{k}}$, up to a scale factor.

This constraint can be made somewhat more intuitive as follows.
In terms of a homogeneous coordinate description of a toric
variety, the algebraic torus appears as rescalings of the individual
homogeneous coordinates by ${\bf C}^{\times}$'s, modulo the
${\bf C}^{\times}$'s one mods out to recover the toric variety.
Suppose that one of the maps ${\cal O}( \sum a_{\alpha} D_{\alpha} )
\rightarrow {\cal O}( \sum b_{\alpha} D_{\alpha})$ was a polynomial
rather than a monomial, then it should be clear that under the
algebraic torus action the coefficients of each monomial in the
polynomial would vary, independently of one another.  Clearly,
when the maps are polynomials, the algebraic torus action
deforms ${\cal E}$ -- a necessary (but not sufficient) condition
for ${\cal E}$ to be invariant under the algebraic torus is
for each polynomial to be a monomial.

The sheaf ${\cal E}$ defined by these choices of equivariant structures and
maps is an equivariant torsion-free sheaf.  

Note that we can deform ${\cal E}$ equivariantly by varying scale
factors in front of the monomials defining the maps $\oplus {\cal
O}(n_{a}) \rightarrow \oplus {\cal O}(m_{i})$.  These scale factors
maps out a Grassmannian, after removing points at which the scale
factors no longer define an exact sequence (i.e., if the map drops rank)
and modding out by rescalings of the homogeneous coordinates.
This projective space is, in general, a subspace of the moduli space of
equivariant torsion-free sheaves.

\begin{figure}
\centerline{\psfig{file=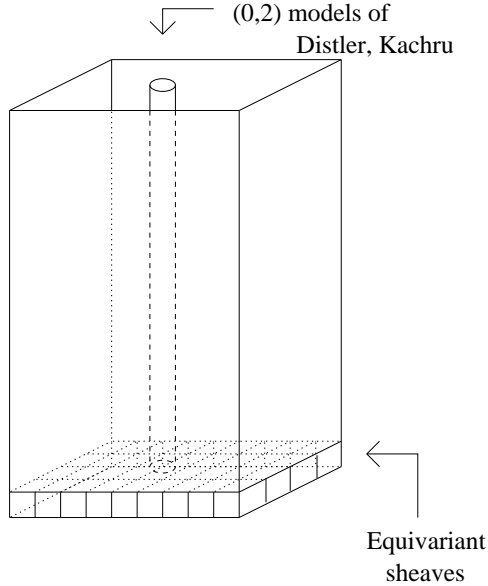,width=2.5in}}
\caption{\label{pic02} Schematically illustrated is a stratum of
a moduli space of sheaves of fixed Chern classes on a toric variety.
Equivariant sheaves are located in the shaded region at the bottom,
and the region described by a Distler-Kachru model is
contained in the cylindrical region.  }
\end{figure}

In figure~\ref{pic02} we have schematically drawn a picture
of a stratum of a moduli space of sheaves of fixed Chern
classes on some toric variety.  The point it is meant to illustrate
is that Distler-Kachru models and the equivariant
sheaves discussed here scan different parts of the moduli space,
and neither is contained within the other.  In fact, the figure
shows the two subspaces as intersecting, though in fact that seems
to be quite rare -- most Distler-Kachru models do not
seem to have well-behaved equivariant limits.

By deforming the maps $\oplus {\cal O}(n_{a}) \rightarrow \oplus {\cal
O}(m_{i})$ to more generic polynomials, we can also study non-equivariant
sheaves.

In passing, we feel we should make an important point 
which has long been overlooked in the (physics) literature on
heterotic compactifications.  For some coherent sheaf ${\cal E}$,
its deformations are parametrized by elements of (global)
$\mbox{Ext}^{1}({\cal E}, {\cal E})$, which gets contributions
from both\footnote{${\em Hom }({\cal E},{\cal E}) = {\em End } {\cal E} 
\oplus {\cal O}$, so on a simply-connected Calabi-Yau, 
$H^1({\em Hom}({\cal E},{\cal E})) = H^1({\em End }{\cal E})$ 
Here ${\em End}$ refers to traceless endomorphisms, as is standard
in the physics literature.}
$H^{1}({\em Hom } ({\cal E},{\cal E}))$ and also
$H^{0}( {\em Ext}^{1} ({\cal E}, {\cal E}))$.  In the special
case that ${\cal E}$ is locally free (meaning, is a bundle),
it is true that (global) $\mbox{Ext}^{1}({\cal E},{\cal E})
= H^{1}({\em Hom }({\cal E},{\cal E}))$, but more generally the
contribution from $H^{0}({\em Ext}^{1}({\cal E},{\cal E}))$
will be nonzero.  This contribution has long been overlooked.
Assuming that the additional directions in the moduli space
are not lifted by quantum corrections, this should be a source
of additional marginal operators overlooked in 
heterotic nonlinear sigma models.  (In non-geometric models,
such as heterotic Landau-Ginzburg orbifolds,
marginal operators morally corresponding to elements 
of $H^{0}({\em Ext}^{1}(
{\cal E},{\cal E}))$ presumably appear in twisted sectors.)
In addition, other massless modes, traditionally counted by
sheaf cohomology groups, are correctly counted with
$\mbox{Ext}$ groups \cite{meqik,ralph2}.

\section{(0,2) mirror symmetry}  \label{(02)mirrapp}

For type II compactifications there is a well-known worldsheet
duality called mirror symmetry.  In mirror symmetry two distinct
Calabi-Yaus (a ``mirror pair") are described by the same conformal
field theory.

Many people have conjectured that there exists a heterotic
version of  the  same phenomenon, known as ``(0,2) mirror symmetry."
The most conservative potential (0,2) mirror symmetry is that
there exist pairs of Calabi-Yaus and bundles that are described
by the same conformal field theory.  In passing, however, we should
note that (0,2) mirror symmetry may do far more.  Unlike ordinary
mirror symmetry, (0,2) mirror symmetry is not constrained to
be a single ${\bf Z}_{2}$ involution of the moduli space.
For example, there may be multiple distinct (0,2) mirror symmetries
acting on the moduli space.  Some of them might, for example,
say that triples of heterotic compactifications are described
by the same conformal field theory.

To date, the only hypothesized examples of (0,2) mirrors
have been constructed either as orbifolds \cite{blumsethi,blumflohr,blumetc}
of Distler-Kachru models, or via WZW models.  No analogue of
Batyrev's construction \cite{batyrev} is yet known for (0,2) mirror symmetry.

One might ask whether equivariant sheaves are relevant to 
the (0,2) generalization of mirror symmetry.  
In fact, we shall conjecture that there exists a (0,2) mirror
symmetry that at least sometimes carries (restrictions to Calabi-Yaus of)
equivariant sheaves to other (restrictions to Calabi-Yaus of) equivariant
sheaves. 

A priori, this would not be expected.
In principle, given some sheaf ${\cal E}$ over a Calabi-Yau 
hypersurface, it should be extremely difficult to determine
if ${\cal E}$ is the restriction of an equivariant sheaf to 
the hypersurface.  Thus, a priori, one might be surprised if
some (0,2) generalization of mirror symmetry relates
sheaves obtained by restriction of equivariant sheaves.
Although such a result is naively surprising, it is completely
consistent with what we know about ordinary mirror symmetry,
in which analogous surprises are well-known to occur.
We will give two rather weak bits of evidence for our conjecture.

First, for at least some (0,2) mirrors constructed
as orbifolds of Distler-Kachru models,
it is easy to show that formally the mirror of the restriction of
an equivariant sheaf, is the restriction of another equivariant
sheaf. 

We shall consider a very simple example
just to illustrate the point.  Our example is 
an equivariant sheaf on ${\bf P}^{2}$ constructed as the kernel
of a short exact sequence
\begin{displaymath}
0 \: \longrightarrow \: {\cal E} \: \longrightarrow \:
{\cal O}(D_{x}) \oplus {\cal O}(D_{y}) \oplus {\cal O}(D_{z})
\: \stackrel{ F_{a} }{\longrightarrow} \: {\cal O}(D_{x} + D_{y} + D_{z}) \:
\longrightarrow \: 0 
\end{displaymath} 
where
\begin{eqnarray*}
F_{x} & \propto & yz \\
F_{y} & \propto & xz \\
F_{z} & \propto & xy
\end{eqnarray*}
(In our notation, $D_{x} = \{ x = 0 \}$.  Note
we have explicitly stated the equivariant structure on each
line bundle above.) 
The (restriction of the) sheaf ${\cal E}$ defined above is 
closely related to a deformation
of the tangent bundle of $T^{2}$.  As discussed in \cite{distanom}
such a Distler-Kachru model suffers from certain poorly understood anomalies
that prevent it from being used as a string compactification;
however, it will serve admirably to illustrate the general idea. 

To construct the orbifold mirror to this equivariant Distler-Kachru model,
we mod out by a ${\bf Z}_{3}$ symmetry that acts on the Calabi-Yau as
\begin{equation}  \label{orbsymm}
x \rightarrow x, \: y \rightarrow \alpha y, \: z \rightarrow \alpha^2 z
\end{equation}
where $\alpha$ is a cube root of unity, and on the ${\cal O}(1)^3$
as
\begin{displaymath}
\Lambda_1 \rightarrow \Lambda_1, \: \Lambda_2  \rightarrow \alpha \Lambda_2,
\: \Lambda_3 \rightarrow \alpha^2 \Lambda_3 
\end{displaymath}
(specifying an action on ${\cal O}(1)^3$ is sufficient to specify
an action on the bundle).
Now, given a Distler-Kachru model defined by a sheaf ${\cal E}$ of the
form
\begin{displaymath}
0 \: \longrightarrow \: {\cal E} \: \longrightarrow \:
\bigoplus_{a} {\cal O}(n_{a}) \: \stackrel{ F_{a} }{ \longrightarrow }
{\cal O}(m) \: \longrightarrow \: 0
\end{displaymath}
when orbifolding the (0,2) model we must restrict to 
maps $F_{a}$ such that the lagrangian of the (0,2) model is invariant 
under the
orbifold symmetry.  It is straightforward to check that in our
example above the maps $F_{x}$, $F_{y}$, and $F_{z}$ 
(together with an appropriate choice of complex structure for
the hypersurface)
yield a Lagrangian invariant under the symmetry in equation~(\ref{orbsymm}).

Thus, in our example, the orbifold construction of (0,2) mirror
symmetry formally appears to carry 
(restrictions to hypersurfaces of) equivariant
sheaves to (restrictions to hypersurfaces of) equivariant sheaves.

Our second bit of evidence for our conjecture comes from studying
a special limit of ordinary mirror symmetry.  Consider a limit
of a mirror pair of Calabi-Yaus $X$, $Y$ in which all worldsheet
instanton corrections on either side are suppressed to zero -- 
in other words, consider the large K\"ahler modulus and large
complex structure limit.  In this limit the Calabi-Yau degenerates,
typically in such a way that its tangent bundle becomes (stably equivalent to)
the restriction of an equivariant sheaf.  Ordinary mirror symmetry
takes such a degenerate Calabi-Yau, into another similarly
degenerate Calabi-Yau.  In other words, in this special limit,
ordinary mirror symmetry maps (restrictions of) equvariant
sheaves to (restrictions of) equivariant sheaves.

If our conjecture is correct, it
may give us information about the action
of (0,2) mirror symmetry on a moduli space of 
not-necessarily-equivariant sheaves.  In particular,
if we associate a polygon to each such moduli space as in
section~\ref{bb}, then this suggests that (0,2) mirror
symmetry may map the vertices of the polygons into each other.

If our conjecture is correct -- if there exists a (0,2) mirror
symmetry mapping equivariant sheaves to equivariant sheaves -- 
then in principle it should be possible to at least find a mirror map
precisely relating moduli of equivariant sheaves.  Unfortunately
we have not yet been able to do so.

\section{Conclusions} \label{concl}

In this paper we have presented an inherently toric description
of a special class of sheaves on toric varieties, something that
has long  been considered extremely desirable.  We have stated 
relevant results on when the restriction of sheaves to Calabi-Yau
hypersurfaces are stable, a necessary (but not sufficient) condition
for a consistent heterotic compactification.  We have described
nontrivial substructure in the classical K\"ahler cone.  Finally,
we have made some basic attempts to apply this technology
to the study of (0,2) mirror symmetry.

We believe that one of the most promising future directions
for this work lies in building a nontrivial understanding
of the (0,2) generalization of mirror symmetry.
Such insights would vastly improve our knowledge of heterotic
compactifications, and surely lead to improvements in our
understanding of nonperturbative effects in heterotic string theory.

In passing we should mention two other recent developments
that may also lead to an understanding of (0,2) mirror symmetry.
First, in addition to the ``traditional" construction of
mirror symmetric pairs \`a la Batyrev \cite{batyrev}, there has
recently been an attempt to understand mirror symmetry
from a very different perspective \cite{syz,morrsyz,grosswilson,gross1},
as a sort of T-duality on the fibers of a special Lagrangian
fibration of a Calabi-Yau.  At the moment this approach to
understanding ordinary mirror symmetry is still in its
infancy, but it holds great promise for giving insight
into ordinary mirror symmetry and perhaps also (0,2) mirror symmetry.

There have also recently been some very interesting 
attempts by mathematicians to apply technology originally designed for
real 3- and 4-manifolds to Calabi-Yau 3- and 4-folds with
stable holomorphic sheaves \cite{figfarr,donaldsonthomas,thomasthesis}.
Their holomorphic versions of the Casson, Floer, and Donaldson
invariants may well turn out to be very important in
understanding heterotic compactifications.

Finally, as this paper was being finished, a paper appeared
\cite{vafaleung} which claims to give an intuitive understanding
of ordinary mirror symmetry as a particular form of T-duality.
It may be possible to apply their ideas to our technology
to gain insight into (0,2) mirror symmetry; work on this
is in progress \cite{meworkinprogress}.

\section{Acknowledgements}

The authors would like to thank R. Friedman, T. Gomez, S. Kachru, J. Morgan,
E. Silverstein, K. Uhlenbeck,
and E. Witten for useful discussions.  We would also like
to thank A. A. Klyachko for supplying us with the
preprints \cite{kl3,kl4} and for creating a beautiful mathematical
theory.

\appendix

\section{Algebraic Groups}

An algebraic group is a group that also happens to be a variety,
just as a Lie group is a group that also happens to be a manifold.
Put another way, algebraic groups are to algebraic geometry
what Lie groups are to differential geometry.
In what follows we shall only
discuss connected, complex subgroups of $GL(n,{\bf C})$.
For more information, see for example \cite{humphreys,borel,jantzen}.

Reductive algebraic groups are algebraic groups which are the
complexification of a compact Lie group.  
Put another way, a reductive algebraic group is an algebraic group such
that 
its maximal compact subgroup has half the ${\bf R}$ dimension of the
algebraic group itself.
Not all algebraic
groups are reductive:  a basic example is ${\bf C}$ under $+$.
An algebraic group is called solvable if its maximal compact
subgroup is a torus.  
An algebraic group is called unipotent if its maximal compact
subgroup is a point (meaning, the group is contractible).
Unipotent algebraic groups have the property that their Lie
algebras can be represented by nilpotent matrices.
A Borel subgroup of a reductive group $G$ is a maximal
solvable subgroup.  All Borel subgroups of $G$ are
conjugate.  A parabolic subgroup is a subgroup 
containing a Borel subgroup.

For example, for the algebraic group $GL(n,{\bf C})$, the subgroup consisting
of all upper-triangular matrices is a Borel subgroup.
A subgroup consisting of block-upper-triangular matrices is a parabolic
subgroup.  Parabolic subgroups of an algebraic group
(other than the group itself)
are not reductive. 

Borel and parabolic subgroups can be understood efficiently in terms of
the Lie algebra of an algebraic group $G$.  
For some choice of Cartan subalgebra and Weyl chamber, 
a Borel subgroup
is a subgroup generated by the Cartan subalgebra and all the
positive roots.  To get a parabolic subgroup, add some of the negative
simple roots and what they generate.  In our conventions, both a Borel and $G$
itself are considered special cases of parabolic subgroups.  Maximal
parabolic subgroups ($\neq G$) are in one-to-one correspondence with
points on the Dynkin diagram.

Parabolic subgroups have a (non-canonical) splitting known as the
Levi decomposition.  More precisely, any parabolic subgroup $P$
can be written $P = L U$ where $U$ is a unipotent subgroup,
and $L$ (known as the Levi factor) is reductive.
($U$ is unique, but $L$ is only unique  up to conjugation by elements
of $U$.)

A useful fact is that the intersection of any two parabolic subgroups
contains a maximal torus of $G$.  (Remember that any two parabolic
subgroups both contain a Borel subgroup, but not necessarily
the same Borel subgroup.)

(Partial) flag manifolds can be obtained as  
quotients of 
$GL(n,{\bf C})$ by parabolic subgroups.  This is closely related
to a description of flag manifolds in terms of Lie groups.
For example, 
the
Grassmannian of $k$ planes in ${\bf C}^{n}$ is
\begin{eqnarray*}
G_{k}({\bf C}^{n}) & = & GL(n, {\bf C}) / P_{k}  \\
                   & = & \frac{ U(n) }{ U(k) \times U(n-k) }  
\end{eqnarray*}
where the $P_{k}$ are maximal parabolics of $GL(n,{\bf C})$.
The $P_{k}$ are very closely related to $U(k) \times U(n-k)$
 -- the Levi factors of the $P_{k}$ are $GL(k,{\bf C}) \times GL(n-k,{\bf C})$,
the complexification of $U(k) \times U(n-k)$. 
One can define ``generalized flag manifolds," which are coset spaces
$G/P$, $P$ a parabolic subgroup of $G$, for arbitrary $G$.

Fix a Borel subgroup $B$ of $G$, and let $T$ be a maximal torus in 
$B \subset G$.
Define 
$N = N_{G}(T) = \{ x \in G \, | \, x T x^{-1} = T \}$, the normalizer of
$T$ in $G$.  The Weyl group is $W = N/T$.
The Bruhat decomposition of $G$ is then
\begin{displaymath}
G \: = \: \bigcup_{\sigma \in W} \, B \widetilde{\sigma} B
\end{displaymath}
where the union is a disjoint union, and each $\widetilde{\sigma}$ is
a representative of a coset $\sigma \in N/T$.  In other words,
for each element $g \in G$ there is a unique $\sigma \in W$ such that
$g$ lies inside $B \widetilde{\sigma} B$.  
The reader may enjoy proving this for 
$GL(n,{\bf C})$, taking the $\{ \widetilde{\sigma} \}$ to be
permutation matrices (a single 1 in each row and column).
In this case the Bruhat decomposition says that any invertible
matrix can be brought to a unique permutation matrix using upward
row and rightward column operations.

\section{Sheaf Classifications}

A few general remarks are in order concerning sheaves.
By a locally free sheaf we mean
a sheaf associated to a vector bundle -- the sheaf
of local sections of the vector bundle.  A torsion-free sheaf
is a sheaf which, on a smooth variety, is locally free except along 
a codimension two subvariety -- so (on a smooth variety)
a torsion-free sheaf can be thought of as a vector bundle with singularities
at codimension two.  (For example, a vector bundle in which an
instanton has become small is associated with a torsion-free sheaf.)

Given a sheaf ${\cal E}$, we can define its dual ${\cal E}^{\vee}
= {\em Hom}({\cal E}, {\cal O})$.  
A reflexive sheaf ${\cal E}$ is a sheaf isomorphic to its 
double dual ${\cal E}^{\vee \vee}$, i.e., ${\cal E} = {\cal E}^{\vee \vee}$.
In general, for any torsion-free sheaf ${\cal E}$,
${\cal E}^{\vee \vee}$ and even ${\cal E}^{\vee}$ are reflexive,
regardless of whether ${\cal E}$ was reflexive.
On a smooth variety,
reflexive sheaves are 
locally free except on a codimension three subvariety.

Thus, a reflexive sheaf is also torsion-free, and a locally free
sheaf is both reflexive and torsion-free.

On a smooth variety, a reflexive rank 1 sheaf is precisely
a line bundle.
 
Over singular varieties this description is sometimes misleading.
For example, on a singular curve it is possible to have 
sheaves which are torsion-free but
not locally free.  Also, on a singular variety it is possible
to have reflexive rank 1 sheaves which are not line bundles
(see for example section~\ref{cartierweil}).

The nomenclature ``locally free,'' ``torsion-free,'' and ``reflexive''
refers to properties of the stalks of the sheaf.
For a sheaf to be locally free means precisely each stalk is
a freely generated module, torsion-free means each stalk is torsion-free,
and reflexive means each stalk is reflexive.  (These definitions
hold even on singular varieties.)

Deformations of a vector bundle ${\cal E}$ are classified by
elements of $H^{1}(\mbox{End } {\cal E})$.  When ${\cal E}$ is
a more general sheaf, its deformations are classified by
(global) $\mbox{Ext}^{1}({\cal E}, {\cal E})$ \cite{maruyama2}.  
(As the reader
may guess, on a Calabi-Yau surface the Yoneda pairing
$\mbox{Ext}^{1}({\cal E}, {\cal F}) \, \times \, \mbox{Ext}^{1}({\cal F},
{\cal G}) \, \rightarrow \, \mbox{Ext}^{2}({\cal E}, {\cal G})$
defines a symplectic structure on the moduli space of sheaves \cite{mukai}.)

For more information on sheaf theory and sheaf-theoretic homological
algebra, see for example \cite{okonek,hartshorne} and
also \cite{me1,meqik}.

\section{\label{gitap} GIT quotients}

Geometric Invariant Theory quotients are described in more detail 
in \cite{git,newstead,maruyama1} and \cite[section 4.2]{huybrechtslehn}.

A GIT quotient ${\cal T}//G$ should be thought of as closely related
to the quotient space ${\cal T}/G$.  The difference is that before
taking the quotient, some subset (the ``unstable points") 
of ${\cal T}$ is removed.  Less often, some properly semistable
points are identified.

An elementary example should be useful.
The space ${\bf P}^{n}$ can be described as a GIT quotient
of the space ${\bf C}^{n+1}$ by ${\bf C}^{\times}$.
More precisely,
\begin{displaymath}
{\bf P}^{n} \: = \: \frac{ \left( {\bf C}^{n+1} \, - \, 0 \right) }{
{\bf C}^{\times} }
\end{displaymath}

To specify the unstable points of ${\cal T}$ (the points to be omitted
before quotienting) one must first specify an ample line bundle ${\cal
L}$, called the polarization.  
Then,
\begin{displaymath}
{\cal T}//G \: = \: \mbox{Proj } \left( \bigoplus_{n \geq 0} H^{0}( {\cal T}, {\cal L}^{n}
)^{G} \right)
\end{displaymath}
Points of ${\cal T}$ are classified as unstable, semistable, or stable.
The unstable points are precisely the
points omitted from ${\cal T}$ before quotienting by $G$.
These unstable points are points $x$ such that for all $G$-invariant 
sections $\sigma$ of
${\cal L}^{n}$, for all $n > 0$, $\sigma(x) = 0$.
Semistable points of ${\cal T}$ are points $x \in {\cal T}$ such that
there exists a $G$-invariant section $\sigma \in H^{0}({\cal T}, {\cal
L}^{n})$ for some $n$ such that $\sigma(x) \neq 0$.
Stable points of ${\cal T}$ are precisely semistable points with finite
stabilizer and the property that in the set of
semistable points, their $G$-orbit is closed. 

Intuitively, why is this definition sensible?  Suppose that
${\cal L}$ is a very ample line bundle on ${\cal T}$, then more or
less by definition of very ample, 
\begin{displaymath}
{\cal T} \: \cong \: \mbox{Proj } \bigoplus_{n /geq 0} H^0( {\cal T},
{\cal L}^n )
\end{displaymath}
If $G$ acts freely on ${\cal T}$, then
\begin{displaymath}
{\cal T}/G \: \cong \: \mbox{Proj } \left( \bigoplus_{n \geq 0} H^0({\cal T},
{\cal L}^n)^G \right)
\end{displaymath}
It should now seem quite reasonable that when the action of $G$
is not so well-behaved, one recovers the GIT quotient outlined above.

Note that the language defined above coincides with the language
used to define moduli spaces of sheaves.  In both cases, one speaks
of stable, semistable, and unstable objects, and in particular
before forming the moduli space one removes the unstable objects.
This is not a coincidence.  Historically, GIT quotients were originally
developed in large part to study moduli spaces of sheaves on curves.
In particular, in the text we show that the moduli spaces of
equivariant sheaves on a toric variety are precisely GIT quotients.
This partially justifies the terrible notation, in which points with
big (positive-dimensional) stabilizers in the group are the least
``stable" !

In fact we have been somewhat sloppy.  We must also specify a
(linearized) action of $G$ on the ample line bundle ${\cal L}$.
Let $\pi: {\cal L} \rightarrow {\cal T}$ be the bundle projection,
then we say \cite[p. 81]{newstead} a linearized action of $G$ is 
an action such that ${\cal L}$ is $G$-equivariant, meaning
\begin{displaymath}
\pi( gy ) \: = \: g \pi(y) \mbox{ for all } y \in {\cal L}, g \in G
\end{displaymath}
and in addition to being $G$-equivariant,
for all $x \in {\cal T}, g \in G$, the map
\begin{displaymath}
{\cal L}_{x} \rightarrow {\cal L}_{g x} : y \mapsto gy
\end{displaymath}
must be linear.  The second condition is equivalent to saying
that $G$ acts linearly on the homogeneous coordinates of the
projective space in which ${\cal T}$ is embedded via ${\cal L}$
(or a tensor power of ${\cal L}$, if it is not very ample).

In practice there is a more nearly straightforward way to
test stability of a point of ${\cal T}$, known as the numerical
criterion for stability (see \cite[section 4.2]{newstead}
or \cite[section 2.1]{git}). 
Let $x \in {\cal T}$, and $\hat{x} \in {\bf P}^{n}$
the image of $x$ under the projective embedding defined by
${\cal L}$.  Write $\hat{x}$ in terms of homogeneous coordinates as
$\hat{x} = (\hat{x}_{0}, \hat{x}_{1}, \cdots, \hat{x}_{n})$.
Let $\lambda$ be a one-parameter subgroup of $G$, whose
action on $x$ is determined by its action on $\hat{x}$ as
\begin{displaymath}
\lambda(t) \hat{x} \: = \: ( t^{r_0} \hat{x}_{0}, t^{r_{1}} \hat{x}_{1},
\cdots, t^{r_{n}} \hat{x}_{n} )
\end{displaymath}
then define 
\begin{eqnarray*}
\mu(x, \lambda) & = & \mbox{unique integer $\mu$ such that }
\lim_{t \rightarrow 0} \, t^{\mu} \, \lambda(t) \, \hat{x} \mbox{ exists and
is nonzero} \\
 & = & \mbox{max } \left\{ - r_{i} \, | \, \hat{x}_{i} \neq 0 \right\}
\end{eqnarray*}
(The reader should not confuse the $\mu$ above with the $\mu$
used to denote the slope of a sheaf.  Unfortunately it is standard
in the mathematics literature to use $\mu$ to denote both, and we
have chosen to abide by their conventions.)
Then, it can be shown that $x$ is semistable precisely when
$\mu( x, \lambda) \geq 0$ for all one-parameter subgroups $\lambda$
of $G$, and $x$ is stable precisely when $\mu(x,\lambda) > 0$
for all one-parameter subgroups $\lambda$ of $G$.

Intuitively, why should the numerical criterion for stability
be correct?  First, it should be intuitively clear\footnote{In
fact, it is rigorously correct.}
that $x$ is semistable precisely when the closure of the $G$-orbit
of $\hat{x}$ does not intersect 0.  
Then, note that if $0$ is not in
the closure of the $G$-orbit of $\hat{x}$, then for all
one-parameter subgroups $\lambda(t)$ of $G$, it must be the case that
\begin{displaymath}
\lim_{t \rightarrow 0} \, \lambda(t) \hat{x} \: \neq \: 0
\end{displaymath}
and from this the numerical criterion for stability follows
immediately.

A useful fact is that if we have specified several identical spaces
${\cal T}$, with a projective embedding defined by composing
the projective embeddings defined by the ${\cal L}$'s over each
${\cal T}$ with a Segre embedding,
then 
\begin{displaymath}
\mu( (x_{1}, \cdots, x_{N}), \lambda) \: = \: \sum_{i=1}^{N} 
\mu( x_{i}, \lambda )
\end{displaymath}
for all one-parameter subgroups $\lambda$ of $G$ (\cite[lemma 4.12]{newstead}).
This is relatively straightforward to demonstrate, as we shall
now outline.  Let $\hat{x}_{1}$, $\hat{x}_{2}$ denote sets of
homogeneous coordinate for either of two projective embeddings,
and suppose we have a one-parameter subgroup action $\lambda(t)$ defined by
\begin{eqnarray*}
\lambda(t) \hat{x}_{1} & = &  \left( t^{r_{1 0}} \hat{x}_{10},
t^{r_{1 1}} \hat{x}_{1 1}, \cdots, t^{r_{1 n}} \hat{x}_{1 n} \right) \\
\lambda(t) \hat{x}_{2} & = & \left( t^{r_{2 0}} \hat{x}_{2 0},
t^{r_{2 1}} \hat{x}_{2 1}, \cdots, t^{r_{2 m}} \hat{x}_{2 m} \right)
\end{eqnarray*}
The action of $\lambda$ on the homogeneous coordinates of the
Segre embedding is clear:
\begin{displaymath}
\lambda(t) \left( \hat{x}_{1 i} \hat{x}_{2 j} \right)
\: = \: t^{r_{1 i} + r_{2 j}} \, \hat{x}_{1 i} \hat{x}_{2 j}
\end{displaymath}
so we can now compute
\begin{eqnarray*}
\mu( x, \lambda ) & = & \mbox{max } \left\{
- r_{1 i} - r_{2 j} \, | \, \hat{x}_{1 i} \hat{x}_{2 j} \neq 0 \right\} \\
 & = & \mu( x_{1}, \lambda ) \: + \: \mu( x_{2}, \lambda )
\end{eqnarray*}
just as stated.

The resulting GIT quotient depends not only upon the choice of ample
line bundle ${\cal L}$, but also upon the choice of $G$-linearization
\cite{dolgachevhu,thaddeus} -- different choices of either the
ample line bundle or $G$-linearization may yield distinct results.   

For example, let us consider a simple example of a flop.
Let $a_{1}, a_{2}, b_{1}, b_{2}$ be coordinates on ${\bf C}^{4}$.
Define a ${\bf C}^{\times}$ action $\lambda$ on ${\bf C}^{4}$ by
\begin{eqnarray*}
a_{1,2} & \rightarrow & \lambda a_{1,2} \\
b_{1,2} & \rightarrow & \lambda^{-1} b_{1,2}
\end{eqnarray*}
and define 
\begin{eqnarray*}
x & = & a_1 b_1 \\
y & = & a_2 b_2 \\
z & = & a_1 b_2 \\
t & = & a_2 b_1 
\end{eqnarray*}
We fix the action of $G = {\bf C}^{\times}$ on ${\bf C}^4$
and vary the $G$-linearization of ${\cal O}$; 
we shall see shortly that this reproduces the expected distinct
quotients. 

For convenience, we shall define some notation.
Let $R = H^0 ({\bf C}^{4}, {\cal O})$, and give it a grading
determined by weight under the action of $G = {\bf C}^{\times}$.
Then,
\begin{eqnarray*}
R_0 & = & H^0 ({\bf C}^{4}, {\cal O})^G \\
  & = & {\bf C}[x,y,z,t] /  (xy-zt) 
\end{eqnarray*}
and also define the graded rings
\begin{eqnarray*}
R_{\geq 0} & = & \bigoplus_{n \geq 0} R_n \\
    & = & {\bf C}[x,y,z,t][a_1 , a_2 ] / ( a_1 y - a_2 z , a_1 t - a_2 x ) \\
R_{\leq 0} & = & \bigoplus_{n \geq 0} R_{-n} \\
 & = & {\bf C}[x,y,z,t][b_1 , b_2 ] / ( b_1 z - b_2 x , b_1 y - b_2 t )
\end{eqnarray*}
where $a_1 , a_2 $ should be considered as homogeneous variables
defining the grading of $R_{\geq 0}$, and $b_1 , b_2 $ defining
the grading of $R_{\leq 0}$.

Now, although as line bundles ${\cal O}^{n}$ and ${\cal O}$ are
identical for all $n \geq 0$, the $G$-linearizations of each differ
(unless, of course, the $G$-linearization of ${\cal O}$ is trivial).

Formally, write
\begin{displaymath}
\bigoplus_{n \geq 0} H^0 ({\bf C}^{4}, {\cal O}^{n}) \: = \: R[z] 
\end{displaymath}
where $z$ is a dummy variable defining the grading.
Assume $z$ has weight $-k \in {\bf Z}$ under $G$, 
defining the $G$-linearization
of ${\cal O}$.
Then we have
\begin{eqnarray*}
\mbox{Proj } \left[ \bigoplus_{n \geq 0} H^0 ( {\bf C}^4, {\cal O}^n )
\right]^G & = & \mbox{Proj } R[z]^{G} \\
& = & \mbox{Proj } \left[ \bigoplus_{n \geq 0} R_{k n} z^n \right] \\
& = & \left\{ \begin{array}{ll}
       \mbox{Proj } \bigoplus_{n \geq 0} R_n z^n & k > 0 \\
       \mbox{Proj } R_0 [z] & k = 0 \\
       \mbox{Proj } \bigoplus_{n \geq 0} R_{-n} z^n  & k < 0 
       \end{array} \right. \\
& = & \left\{ \begin{array}{ll}
       \mbox{Proj } R_{\geq 0} & k > 0 \\
       \mbox{Spec } R_0 & k = 0 \\
       \mbox{Proj } R_{\leq 0} & k < 0 
       \end{array} \right.
\end{eqnarray*}

Clearly, the three possible GIT quotients above are precisely
the conifold and its two small resolutions.  For $k > 0$,
the unstable set on ${\bf C}^4$ is precisely $\{ a_1 = a_2 = 0 \}$,
and for $k < 0$ the unstable set is precisely $\{ b_1 = b_2 = 0 \}$.

Finally, we should mention that GIT quotients are equivalent
to symplectic quotients (see \cite{kirwan} or the appendix to
\cite{git} for more information).
Suppose we are constructing the GIT quotient of a space ${\cal T}$
by the action of a reductive algebraic group $G$ with maximal
compact subgroup $K$.  Let $\iota: {\cal T} \rightarrow {\bf P}^{n}$
denote the projective embedding defined by polarization ${\cal L}$.
Then to define the equivalent symplectic quotient,
we take the symplectic form\footnote{This symplectic form also happens
to be a de Rham representative of the cohomology class
$c_{1}({\cal L})$.  However, not all representatives of $c_{1}({\cal L})$
yield possible symplectic forms.} 
to be the restriction of the K\"ahler form
from the projective embedding, and the moment
map is defined by the composition
\begin{displaymath}
{\cal T} \: \stackrel{ \iota }{\longrightarrow} \:
{\bf P}^{n} \:  \stackrel{ \Phi_1 }{\longrightarrow} \:
pu(n+1)^{*} \: \stackrel{ \Phi_2 }{\longrightarrow} \:
k^{*}
\end{displaymath}
where $\Phi_{1}$ is the moment map for the action of $PU(n+1)$
on ${\bf P}^{n}$, $\Phi_2$ is the transpose of the differential
of the map of $K \subset G$ into $PU(n+1)$, and $k = \mbox{Lie } K$. 
The corresponding symplectic quotient is then
\begin{displaymath}
\frac{ ( \Phi_2 \circ \Phi_1 \circ \iota )^{-1} (0) }{ K }
\end{displaymath}

Note that not all symplectic quotients have equivalent GIT
quotients.  For example, one necessary (but not sufficient)
condition is that the symplectic form be (the image of) an element
of $H^2 ({\bf Z})$.

\section{Filtrations \label{hsfilt}}

One important filtration of a coherent sheaf, often appearing
in the mathematics literature, is called the Harder-Narasimhan
filtration.  Given a torsion-free coherent sheaf ${\cal E}$,
a Harder-Narasimhan filtration of ${\cal E}$ is
a filtration \cite{maruyamaboundedness}
\begin{displaymath}
0 = {\cal E}_{0} \subset {\cal E}_{1} 
\subset \cdots \subset {\cal E}_{n}
= {\cal E}
\end{displaymath}
(where $\subset$ indicates proper subset)
with the properties
\begin{displaymath}
\begin{array}{ll}
(1) & {\cal E}_{i}/{\cal E}_{i-1} \mbox{ is semistable for } 1 \leq i \leq n
\\
(2) & \mu({\cal E}_{i}/{\cal E}_{i-1}) > \mu({\cal E}_{i+1}/{\cal E}_{i})
\mbox{ for } 1 \leq i \leq n-1
\end{array}
\end{displaymath}
For every torsion-free coherent sheaf, a Harder-Narasimhan filtration exists 
and is unique.

Note that the Harder-Narasimhan filtration is trivial (meaning,
of the form $0 = {\cal E}_{0} \subset {\cal E}_{1} = {\cal E}$)
precisely when ${\cal E}$ is semistable.  Intuitively, the
Harder-Narasimhan filtration gives information about the ``instability"
of ${\cal E}$.

It is quite easy to derive the Harder-Narasimhan filtration for
any torsion-free coherent sheaf ${\cal E}$.  Let ${\cal E}_{1}$
be a destabilizing subsheaf (meaning, $\mu({\cal E}_{1}) > \mu({\cal E})$)
of greatest possible slope, such that among all destabilizing subsheaves
of that slope, ${\cal E}_{1}$ has the greatest rank.  Now, let ${\cal
E}_{2}$ be a subsheaf of ${\cal E}$ such that ${\cal E}_{1} \subset
{\cal E}_{2}$, and so that ${\cal E}_{2}/{\cal E}_{1}$ has maximal
slope among subsheaves of ${\cal E}/{\cal E}_{1}$, and such that
among such subsheaves of equal slope, it has maximal rank.
Proceeding in this manner, one can construct each element of
the Harder-Narasimhan filtration.

Another important kind of filtration worth mentioning is a Jordan-H\"{o}lder
filtration.  Just as the Narasimhan-Harder filtration splits any
torsion-free coherent sheaf into semistable factors, a 
Jordan-H\"{o}lder filtration splits a semistable sheaf into stable
factors.

More precisely, let ${\cal E}$ be a semistable sheaf.  A
Jordan-H\"{o}lder filtration of ${\cal E}$ is a filtration
\begin{displaymath}
0 = {\cal E}_{0} \subset {\cal E}_{1} \subset \cdots \subset {\cal
E}_{n} = {\cal E}
\end{displaymath}
such that the factors ${\cal E}_{i}/{\cal E}_{i+1}$ are stable.

Jordan-H\"{o}lder filtrations always exist, but are not necessarily
unique.

Although the Jordan-H\"older filtration is not necessarily unique,
the associated graded filtration defined by $gr_{i}({\cal E}) =  
{\cal E}_i / {\cal  E}_{i+1}$ is unique.  Two semistable
sheaves with isomorphic associated graded filtrations are S-equivalent.

What is S-equivalence?  Recall from the discussion of GIT quotients
that sometimes properly semistable points are identified in forming
the quotient, so only equivalence classes of properly semistable
objects appear in the GIT quotient.  In a moduli space of sheaves,
such an equivalence class of properly semistable sheaves is
said to be an S-equivalence class, and two properly semistable
sheaves belonging to the same S-equivalence class (meaning,
appearing at the same point on the moduli space) are said to be
S-equivalent.  In the language of symplectic quotients,
$S$-equivalence classes occur when one performs symplectic
reduction at a nonregular value of the moment map.

Thus, Jordan-H\"older filtrations are useful for determining
S-equivalence classes of properly semistable sheaves.

\end{document}